\documentclass{comnet}

\DeclareMathAlphabet{\mathcal}{OMS}{cmsy}{m}{n}

\DeclareMathOperator*{\argmin}{arg\,min}

\usepackage{graphicx}
\usepackage{xcolor}
\usepackage{amsmath,amssymb}
\usepackage{amsthm}
\usepackage{amsfonts}
\usepackage{algorithm}
\usepackage{algpseudocode}
\usepackage{hyperref}
\usepackage{xurl}
\usepackage{enumitem}
\usepackage[capitalise]{cleveref}
\usepackage{verbatim}
\usepackage{pifont}
\usepackage{multirow}
\usepackage{multicol}
\usepackage{arydshln}
\usepackage{subcaption}
\usepackage{aliascnt}
\makeatletter

\let\c@proposition\relax

\let\p@proposition\relax

\theoremstyle{definition} 
\newtheorem{proposition}{\sc Proposition}[section]

\crefname{proposition}{Proposition}{Propositions}
\Crefname{proposition}{Proposition}{Propositions}
\makeatother

\newtheorem{assumption}{\sc Assumption}

\newcommand{\E}{\mathbb{E}}
\newcommand{\Reals}{\mathbb{R}}

\newcommand{\cov}{\operatorname{Cov}}

\newcommand{\tr}{\operatorname{tr}}

\newcommand{\guy}[1]{{\color{black} #1}}
\newcommand{\hengxu}[1]{{\color{black} #1}}
\newcommand{\hltwo}[1]{{\color{black} #1}}
\newcommand{\sourav}[1]{{\color{black} #1}}
\newcommand{\guytwo}[1]{{\color{black} #1}}
\newcommand{\guythree}[1]{{\color{black} #1}}
\newcommand{\hlthree}[1]{{\color{black} #1}}
\newcommand{\guyfour}[1]{{\color{black} #1}}
\newcommand{\hlfour}[1]{{\color{black} #1}}
\newcommand{\guyfive}[1]{{\color{black} #1}}

\usepackage{xparse}
\usepackage{xspace}

\newcommand{\sreftxt}[2]{\expandafter\def\csname sref@#1\endcsname{#2}}

\NewDocumentCommand{\sref}{m}{%
  \csname sref@#1\endcsname%
}

\sreftxt{def:gnarid}{Definition~2.1}
\sreftxt{lem:w_lambda_infty}{Lemma~2.1}
\sreftxt{prop:stationary}{Proposition~2.1}
\sreftxt{lem:ma_form}{Lemma~2.2}
\sreftxt{as:stationary}{Assumption~1}
\sreftxt{as:sigma_epsilon}{Assumption~2}
\sreftxt{lem:cov_finite}{Lemma~3.1}
\sreftxt{as:dist}{Assumption~3}
\sreftxt{lem:lin_indep}{Lemma~3.2}
\sreftxt{prop:consistency}{Proposition~3.1}
\sreftxt{cor:cls_min}{Corollary~3.1}
\sreftxt{as:dist_N}{Assumption~4}
\sreftxt{prop:consistency_N}{Proposition~3.2}
\sreftxt{prop:lambda_error}{Proposition~5.1}
\sreftxt{cor:lambda_error}{Corollary~5.1}

\sreftxt{lem:dist_to_indep}{Lemma~C.4}
\sreftxt{lem:norm_finite}{Lemma~C.5}
\sreftxt{lem:cross_norm_finite}{Lemma~C.6}

\begin{document}

\title{Improved Forecasting for Multivariate Time Series
    using GNAR models with General Inverse Distance Weights}

\shorttitle{Improved Forecasting with GNARID} 
\shortauthorlist{Liu et al.} 

\author{
\name{Hengxu Liu$^*$}
\address{Department of Mathematics, Imperial College London, Huxley Building, 180 Queen's Gate, London SW7 2AZ, UK.\email{$^*$Corresponding author: hengxu.liu18@imperial.ac.uk}}
\name{Guy P. Nason}
\address{Department of Mathematics, Imperial College London, Huxley Building, 180 Queen's Gate, London SW7 2AZ, UK.}
\name{Sourav Das}
\address{School of Electrical Engineering, Computing and Mathematical Sciences, Curtin University, Perth, WA 6102, Australia.}
\and
\name{Stephen Grebby}
\address{Terra Motion Limited, Ingenuity Centre, Triumph Road, Nottingham NG7 2TU.}
}

\maketitle

\begin{abstract}
{Many multivariate time series arise on networks where relationships between variables depend on physical or more general distances. Standard vector autoregressive models struggle in such settings due to their quadratic parameter growth, while generalised network autoregressive (GNAR) models offer a parsimonious alternative by explicitly incorporating network structures. Existing GNAR formulations assume fixed weights, typically uniform or reciprocal distances, which may not reflect the true influence of distance in real systems. We introduce a novel model with general inverse distance weights (GNARID) that extends GNAR by introducing a tunable exponent parameter that controls the strength of distance penalisation. We establish theoretical results on the consistency and asymptotic normality of new GNARID estimators, including an extension to the setting where the number of nodes grows. We analyse the effect of misspecifying the exponent on forecasting error, which helps assess how much GNAR models could be improved by our tunable weights. Simulation studies confirm the theoretical properties and show measurable benefits from correct weight specification. Using the Brumadinho ground displacement dataset, we show that proper tuning of the weight exponent improves predictive performance, highlighting the practical value of flexible distance-based weighting in network time series models.}
{network time series; parsimonious model;
    misspecification; Brumadinho}
\end{abstract}

\section{Introduction}
Multivariate time series data arise in many domains, from traffic flows in transportation networks and epidemiological spread over regions to financial interactions among industries. Vector autoregressions (VAR) \citep{quenouille:multiple} are a popular choice
\guytwo{for} modelling multivariate time series, which assume linear dependencies between current and past variates.
\guytwo{A key drawback of VAR models is the quadratic growth in the number of parameters with respect to the number of time series,
often leading to overfitting even with reasonable sample sizes.}
Different methods have been proposed to mitigate this problem, such as the Lasso-regularised VAR \citep{song2011}, the factor-augmented VAR \citep{fvar}, or the Bayesian VAR \citep{BayesianVAR}.

Suppose that \guy{an \mbox{$N$-dimensional}} time series
is accompanied by a network $\mathcal{G}$ with $N$ vertices (nodes). \hlfour{Inspired by earlier work on network time series
\citep{JansenNasonSilverman01, JNS09, NunesKnightNason15}}, over the last decade
a concise yet effective alternative to VAR \guytwo{models}
has been developed: the Generalised Network Autoregressive (GNAR) model of~\cite{gnar_org,GNAR}. \hlfour{See also \cite{zhu2017} for a closely related Network Vector Autoregressive (NAR) model.}
\guytwo{The GNAR model structures dependencies} explicitly through a known network topology rather than imposing generic sparsity through shrinkage or arbitrary zero restrictions, leading to fewer parameters, clearer interpretability, and better scalability to large and complicated systems.

\hlfour{The network time series literature has also developed in several other directions. A general nonlinear NAR framework is developed in \cite{AF23}, while a special case, the Poisson network autoregressive (PNAR) model in \cite{AF24}, is designed for count-valued network time series. The GNARI and NGNAR models in \cite{Liu23} accommodate count data through thinning operations and nonlinear response functions, respectively, while the FCNAR model in \cite{FCNAR} introduces spline-based functional coefficients. Network dependence has also been incorporated into conditional volatility models \cite{ZDPW20,PP24}. 
Further extensions include edge-valued network time series \cite{MCMR23}, long-memory network processes \cite{boetti2025}, community-specific GNAR models \cite{nason2025usa}, and matrix- and tensor-valued network autoregressive models with latent group structures \cite{ren2026gmnar,Ren2024gtnar}.
\guy{Sometimes the time series is not accompanied by a specific
network, but random networks can be used to great effect,
particularly when, e.g., forecasting; see
\cite{leeming,Nason26}}. \hlfour{Another strand treats the network itself as the time series rather than a supporting structure. Time-varying binary adjacency matrices can be modelled by allowing each edge to evolve according to edge-specific transition probabilities; see \cite{jiang2023autoregressive}. This framework is reparameterised in \cite{jiang2025twoway} using two node-specific parameters to capture heterogeneity in tie formation and persistence. A further development in \cite{chang2026} allows the transition probabilities of an edge to depend on the histories of other edges.} Taken together, these works show how network structure can be incorporated into a wide range of time series models. The present work addresses an unresolved issue in network time series modelling: even when the network topology is available, the connection weights governing the relative influence of neighbouring nodes may still be unknown and require estimation.}

The GNAR model \guytwo{uses a} weight matrix that specifies the relative importance of edges in the network. However, such weights are \guytwo{sometimes} unavailable in practice, while a distance matrix, derived from geographical distances, travel times, or any dissimilarity measure applied to exogenous variables of the nodes, is often accessible. This naturally raises the question of how to determine appropriate weights given the available distances. Let $w_{n,m}$ and $d_{n,m}$ denote the weight and `distance' of the edge $(n,m)$ respectively. \hlthree{Weights were proposed in \cite{gnar_org,GNAR} to be constructed using reciprocal distances, that is}
\hlthree{\begin{equation}
    w_{n,m} = d^{-1}_{n,m}/\sum_q d^{-1}_{n,q},
\end{equation}
where the sum is taken over the relevant neighbours $q$ of node $n$.} \guytwo{We introduce} \guy{a more flexible weighting scheme using \guytwo{general}
inverse distance weights \citep{Shepard68,LuWong08} given by
\begin{equation}
    w_{n,m} = d^{-\lambda}_{n,m}/\sum_q d^{-\lambda}_{n,q}, \quad \lambda \geq 0,
\end{equation}
where $\lambda$ is called the exponent in \cite{Shepard68}.
\hlthree{The term `inverse distance weights' was used in \cite{JNS09} and \cite{GNAR}}
to refer to the  $\lambda=1$ situation, but the name is usually
associated with the more general case of some $\lambda \geq 0$.}
\hltwo{General inverse distance weights with parameter $\lambda$ have been used in many other contexts, but to the best of our knowledge, this is the first work to estimate distance-based weights from the data for the class of network autoregressive processes such as GNAR, rather than predefining them as fixed forms such as reciprocal or uniform weights.}

Inverse distance weights generalise the reciprocal distance weighting by introducing a hyperparameter $\lambda$ that controls the strength of penalisation for larger \guytwo{or smaller} distances. Another widely used approach for weight construction
is negative exponential weighting (or inverse softmax),
\begin{equation}
    w_{n,m} \propto \exp(-\lambda d_{n,m}), \quad \lambda \geq 0,
\end{equation}
which arises from the multinomial logit model \citep{mcfadden_conditional_1974}.
\guytwo{These are} essentially the same as inverse distance weights,
if we first  transform the distances \guytwo{exponentially}.

The inclusion of a single modest additional hyperparameter $\lambda$ allows the influence of distance to be tuned more precisely. While this added flexibility increases the complexity of model fitting, such as making the target function non-convex, we will show that $\lambda$ can still be estimated consistently and efficiently. 

Naturally, there are a wide variety of possible different weight functions that could be used. Our inverse distance weighting scheme is one instance from the more general form $w_{n,m} = f(d_{n,m})$ where $f$ is a monotonically decreasing function in $d_{n,m}$.
We  now proceed to explain our method using general inverse-distance
weights and next establish notation.

\subsection{Notation}
We use $A_{i,j}$ to denote the $(i,j)$-th entry of any matrix $A$, \hlthree{$A_{i,:}$} to denote the $i$-th row, \hlthree{$A_{:,j}$} to denote the $j$-th column. Let $I_N$ denote the identity matrix in $\mathbb{R}^{N \times N}$. Let $c$ and $C$ denote arbitrary constants which 
\guytwo{can} vary for each occurrence.
Let $\|x\|$ denote the $2$-norm of a vector $x$. Let \hlthree{$\|A\|_q$,} $\|A\|$ and $\|A\|_F$ denote the \hlthree{induced $q$-norm, }induced $2$-norm and the Frobenius norm of a matrix $A$ respectively. Define $\rho_{\max}(A)$ and $\rho_{\min}(A)$ to be the largest and smallest absolute eigenvalues of $A$ respectively.
We use $\otimes$ to denote the Kronecker product of matrices. $A \succ (\succeq) 0$ means that $A$ is positive definite (positive semi-definite). The Loewner partial order $A \succ (\succeq) B$ means that $A-B \succ (\succeq) 0$. $A \succ (\succeq) c$ means $A \succ (\succeq) cI$. 
\hlthree{Let $ (f_T) $ and $ (g_T) $ be real-valued sequences, and write $ f_T = O(g_T) $ if there exist constants $ C > 0 $ and $ T_0 $ such that $ |f_T| \le C |g_T| $ for all $ T \ge T_0 $. Then $ f_T = \Theta(g_T) \iff f_T = O(g_T)$ and $g_T = O(f_T)$, \guyfour{which means that 
$f_T, g_T$ grow at the same rate
up to constant factors}.}

\section{GNAR with General Inverse Distance Weights}

\guyfour{We define our
multivariate time series by
$X_{i, t}$, where there are
$i=1, \ldots, N$ time series
collected over $t=1, \ldots, T$
time points. The multivariate
time series is a network time series
if it is associated with a graph/network
with $N$ nodes and
$\{ X_{i, t} \}_{t=1}^T$ is associated
with vertex/node $i=1, \ldots, N$ of
the graph.
A GNAR time series model is a network
time series model that
models the dependence of $X_{i, t}$
on past values of $X_{i, s}$ for
$s< t$ and past values of the
neighbours of $X_{i, t}$, where
the neighbourhood is defined in some
way by the graph structure. The way
in which neighbours affect nodes
is controlled by connection weights,
which are often related to distances
between vertices (if they exist) or
some other analogous measure.
The purpose of connection weights
is to permit the parametrisation of
the statistical GNAR model not to
depend on, e.g.\ inter-vertex
distances, and permits the parametrisation
to remain parsimonious,
see Section~3.1 of \cite{gnar_org}.
}
\hlthree{ In the standard GNAR setting these weights are fixed in advance. In this section we extend this construction by allowing the weights to depend on distances through an exponent parameter.
}

Let $\mathcal{X} = (\mathbf{X}_t, \mathcal{G})$ be a
network  time series of length $T$, where $(\mathbf{X}_t)_{t=1}^T$
 is an \mbox{$N$-dimensional} multivariate time series with
 $\mathbf{X}_t = (X_{t,1},\dots,X_{t,N})$ 
and  \guytwo{$\mathcal{G}$ is a}  network with $N$ vertices. Let $\mathcal{K} = \{ 1, \ldots, N\}$ be the set of
all vertices (time series variables) of $\mathcal{G}$. We write $i \leftrightarrow j$ if $i, j$ are connected by an edge.
Let $A\subset \mathcal{K}$ be a subset of vertices.
Define the neighbours of vertices in $A$ by $\mathcal{N}(A) = \{ j \in \mathcal{K}\setminus A: j \leftrightarrow i, \ i\in A \}$.
As in \cite{gnar_org,GNAR} we define the \mbox{$r$-stage} neighbour sets 
$\left\{ \mathcal{N}^{(r)}(n)\right\}_{r\geq 1}$ by
\begin{equation}
    \mathcal{N}^{(r)} (n) = \mathcal{N} \left\{ \mathcal{N}^{(r-1)}(n)\right\} \setminus \left[\left\{\bigcup_{q=1}^{r-1} \mathcal{N}^{(q)}(n)\right\} \cup \{n\} \right],
\end{equation}
for $r= 2, 3, \ldots$ and $\mathcal{N}^{(1)} (n) = \mathcal{N}( \{ n \})$, where $n\in\mathcal{K}$.
Hence, $\mathcal{N}^{(r)}(n)$ is the set of vertices connected to any element of $\mathcal{N}^{(r-1)}(n)$ by an edge that
has not appeared in any previous neighbour set. In other words, the first stage neighbours are $\mathcal{N}^{(1)}(n)$,
which  are the vertices \hlthree{excluding $n$} directly connected to $n$ by an edge.
The second stage neighbours,
$\mathcal{N}^{(2)}(n)$, are any vertices, \guyfour{not already in
$\mathcal{N}^{(1)}(n)$ nor $n$ itself,} connected by an edge to any vertex in $\mathcal{N}^{(1)}(n)$,
and so on.

Given two vertices in the network (equivalently two time series variables) $n, m\in \mathcal{K}$ we are also supplied
with \hlthree{finite strictly positive} distances (or dissimilarities) $d_{n, m}$. If $n, m$ are connected by an edge, then $d_{n, m}$ could be the
dissimilarity or distance conveyed by that edge. If the vertices are not directly connected then the dissimilarities might be
multipath distances; see \cite{GNAR} for more details on how these might be calculated.

For the inverse distance weight exponent $\lambda \geq 0$, \hlthree{for a node $n \in \mathcal{K}$ and a stage $r > 0$, when $\mathcal{N}^{(r)}(n) \neq \emptyset$, for $m \in \mathcal{N}^{(r)}(n)$, }we introduce our general inverse distance
(connection) weights $w_{n, m} (\lambda)$ by
\begin{equation}\label{eq:w}
    w_{n,m} (\lambda) = \frac{d_{n,m}^{-\lambda}}{\sum_{q \in \mathcal{N}^{(r)}(n)} d_{n,q}^{-\lambda}}.
\end{equation}
\hlthree{When $\mathcal{N}^{(r)}(n) = \emptyset$, we set $w_{n,m}(\lambda) = 0$ for all $m$, so that the corresponding stage-$r$ neighbourhood contribution is zero.}
\hlthree{Although the inverse distance interpretation corresponds to $\lambda \geq 0$, under the finite strictly positive condition on the distances, the weight function is smooth for real $\lambda$.}
We now define the GNARID processes.

\begin{definition}[GNARID]\label{def:gnarid}
    
\guytwo{We introduce  the {\em generalised network autoregressive  model 
with inverse distance weights},
GNARID($p,[s_1, \dots, s_p], \lambda$), defined \guytwo{by}
\begin{equation}
\label{eq:gnarid}
X_{t, n} = \hat{X}_{t|t-1,n} + \epsilon_{\guytwo{t,n}} = \sum_{j=1}^p \left\{  \alpha_{j} X_{t-j,n} + \sum_{r=1}^{s_j} \beta_{j, r} \sum_{m \in \mathcal{N}^{(r)}(n)} w_{n, m}(\lambda) X_{t-j,m} \right\} + \epsilon_{t,n},
\end{equation}}%
where $p$ is the maximal lag, $s_j$ is the furthest stage of neighbourhood to include for lag $j$, $\alpha_j$ is the self-autoregressive coefficient for lag $j$, $\beta_{j,r}$ is the autoregressive coefficient acting on stage-$r$ neighbourhood at lag $j$.
The innovations $\mathbf{\epsilon}_{t} = (\epsilon_{t,n})_{n=1}^N$ are i.i.d.\ \hltwo{over time with mean zero and arbitrary 
covariance matrix $\Sigma_\epsilon$, which permits cross-sectional dependence at each time point.}
\guyfour{For the model definition and
theory we have
$\alpha_j, \beta_{j, r}, \lambda \in \Reals$
for 
each $j=1, \ldots, p$.
\hlfour{For the spatial order $[s_1,\dots,s_p]$ to contain no redundant stage neighbourhoods, we require that, for every $r \leq \max_j s_j$, there exists at least one $n$ with $\mathcal{N}^{(r)}(n)$ being non-empty.}
Let $P = \sum_{j=1}^p (1 + s_j)$ and
then the dimension of the parameter
space is $P+1$.}
\end{definition}
\guytwo{The GNARID$(p, [s_1, \ldots, s_p], 1)$ model is equivalent to a GNAR$(p, [s_1, \ldots, s_p])$ model.}
\guyfour{In practice, 
for the inverse-distance weights
function~\eqref{eq:w}
we only consider $\lambda \geq 0$.}
\guy{For a clean exposition,
the GNARID model above is based
on the so-called `global' GNAR model from \cite{gnar_org}. This
is where the $\alpha_j, \beta_{j, r}$ do not depend
on $n$ and are the same across the network. It would be
\hlthree{possible} to extend GNARID to include node-dependence
for the $\alpha_j$ parameter, the `local-$\alpha$' model
from \cite{GNAR}, or indeed the `local-$\alpha\beta$' model
from \cite{Nason26}, which also makes the $\beta_{j, r}$ depend
on $n$.
}

\begin{figure}[!h]
    \centering
        \resizebox{\textwidth}{!}{\includegraphics{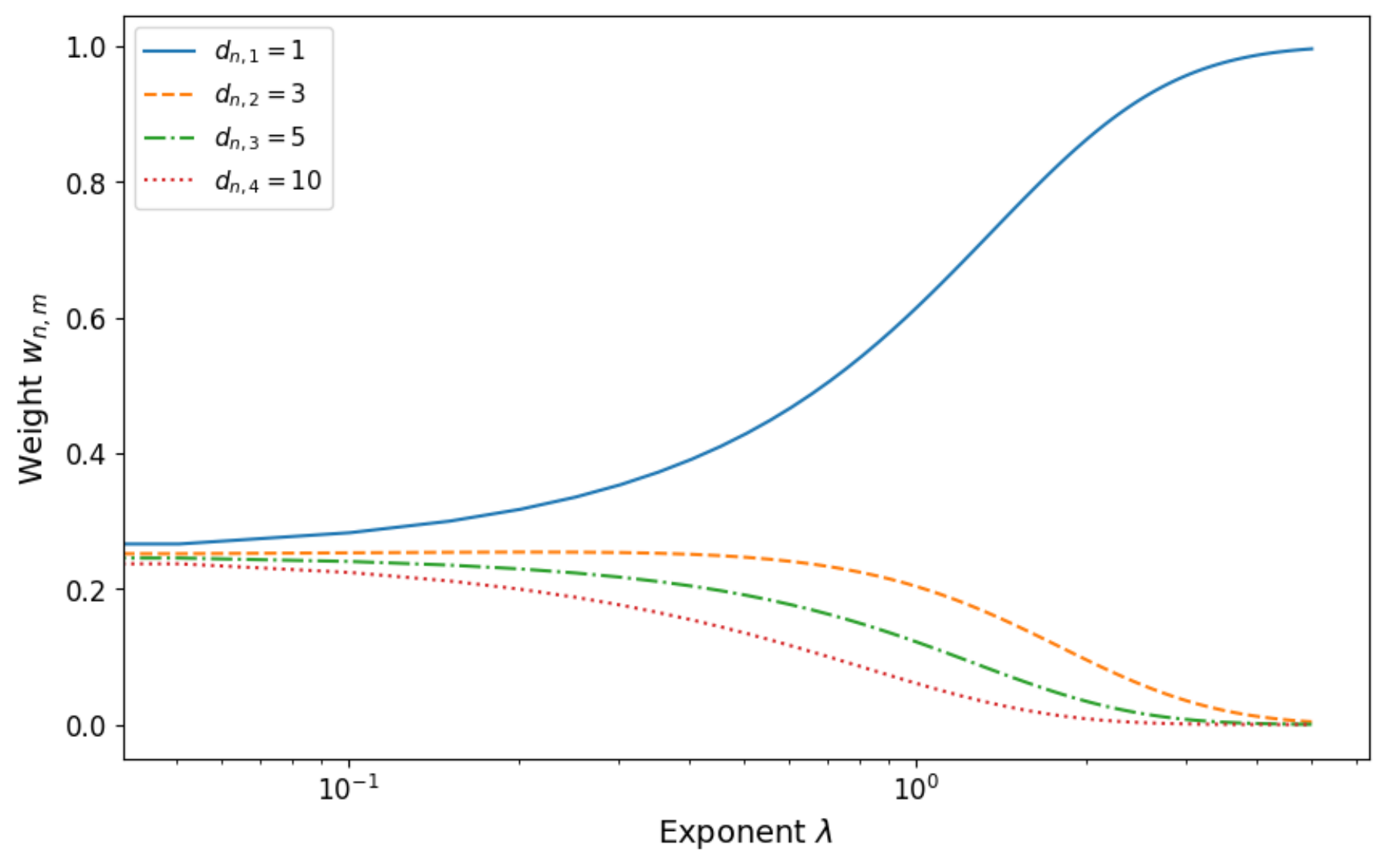}}
        \caption{Plot of weights against $\lambda$ for a given set of distances $\{1,3,5,10\}$.}
        \footnotesize{\textbf{Alt text:} Line plot showing how inverse-distance weights vary with lambda. The smallest distance gains most of the weight as lambda increases, while the larger distances decline toward zero.}
        \label{fig:w_vs_lambda}
\end{figure}

\begin{figure}[!h]
        \centering
        \resizebox{0.6\textwidth}{!}{\includegraphics{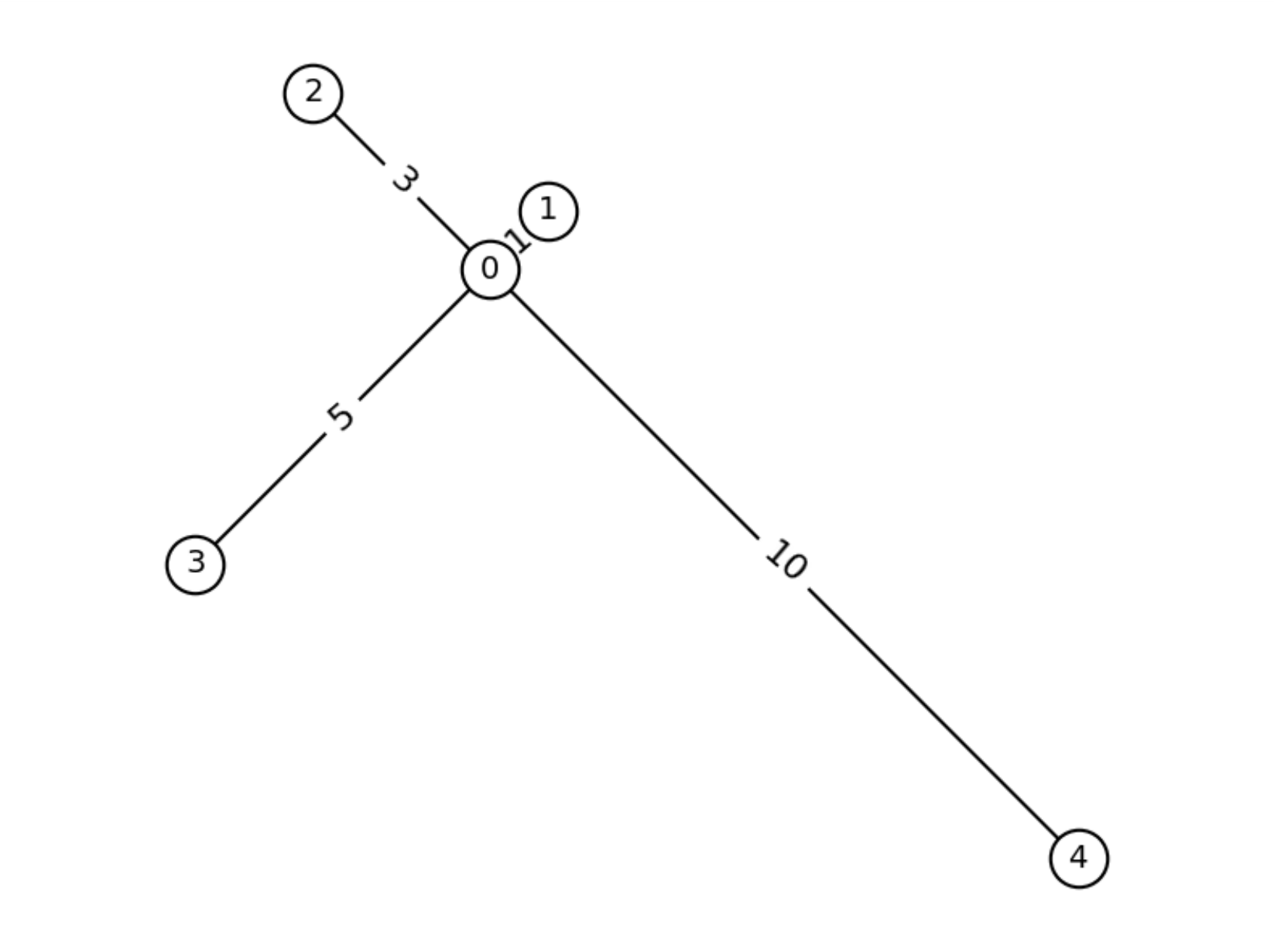}}
        \caption{A star network with distances $\{1,3,5,10\}$.}
        \footnotesize{\textbf{Alt text:} Star network showing a central node connected to four outer nodes. The four edges are labelled by the distances one, three, five, and ten.}
        \label{fig:network_example}
\end{figure}

Compared to the GNAR model, the \guytwo{GNARID model has}  weights \guytwo{that} are now functions of \guy{ the exponent} $\lambda$ and the distances, where $\lambda$ is a parameter to estimate. More specifically, $\lambda$ controls how much we
\guytwo{manipulate (penalise)} the distances. With $\lambda=0$, all edges are treated equally regardless of the distances. With $\lambda \xrightarrow{} \infty$, only the edges corresponding to the minimal distances in their neighbourhoods have non-zero weights. Figure~\ref{fig:w_vs_lambda} \guytwo{illustrates} how $w_{n,m}$ changes with respect to $\lambda$ given a set of distances $\{1,3,5,10\}$. \hlthree{Figure~\ref{fig:network_example} shows a star network whose four edges have these distances.} \hlfour{For node~0 in Figure~\ref{fig:network_example}, $\lambda=0$ assigns uniform weights to all four neighbours, while increasing $\lambda$ progressively shifts more weight towards node~1 which has the shortest distance and less towards the other three neighbours.} Let $(\cdot)^0$ \guytwo{denote} the true parameters. Let $\theta = (\alpha_1,\beta_{1,1},\dots,\alpha_p,\beta_{p,1},\dots,\lambda)'$
\guy{be the vector of parameters}.
    
The following lemma describes the behaviours of $w_{n,m}(\lambda)$ and its derivative as $\lambda \to \infty$, \guy{which is used later}.

\sourav{\begin{lemma}\label{lem:w_lambda_infty}
    Let $x = (x_i)_i$ be a vector with $x_i > 0$. Then, for all $x_i \neq \min (x)$, as $\lambda \rightarrow \infty$,
    \begin{equation}
        w_i(\lambda) = \frac{x_i^{-\lambda}}{\sum_j x_j^{-\lambda}} \rightarrow 0.
    \end{equation}
    Moreover, for all $x_i$, as $\lambda \rightarrow \infty$,
    \begin{equation}
        w_i'(\lambda) = \frac{\partial}{\partial \lambda}\frac{x_i^{-\lambda}}{\sum_j x_j^{-\lambda}} \rightarrow 0.
    \end{equation}
\end{lemma}}

\guytwo{To understand how the weights influence the behaviour of GNARID processes, we introduce weight matrices into the
matrix model specification.}
For all $r = 1, \ldots, \max(s_1, \ldots, s_p)$, let $W^{(r)}(\lambda)$ be the \guytwo{$r$th order weight}
matrix such that, for all $n,m= 1, \ldots, N$
\begin{equation}\label{eq:Wr}
    W^{(r)}(\lambda)_{n,m} = \begin{cases}
        w_{n,m}(\lambda),& \text{if } m \in \mathcal{N}^{(r)}(n), \\
        0,& \text{otherwise.}
    \end{cases}
\end{equation}
\guyfour{As indicated by
Figure~\ref{fig:w_vs_lambda}},
\hlthree{a larger $\lambda$ makes $W^{(r)}(\lambda)$ more concentrated, with more weight assigned to shorter-distance neighbours.} Let $W^{(r)} = W^{(r)}(\lambda^0)$ denote the true stage-$r$ weights. Let \guytwo{matrix} $A_{j}$ be defined as
\begin{equation}
    A_{j} = \alpha_{j}^0 I_N + \sum_{r=1}^{s_j} \beta_{j,r}^0 W^{(r)}.
\end{equation}
Then, we can write the \guy{GNARID} process in \guy{vector autoregressive (VAR) form}
as
\begin{equation}\label{eq:var_form}
    \mathbf{X}_{t} = \sum_{j=1}^p A_{j} \mathbf{X}_{t-j} + \mathbf{\epsilon}_{t}.
\end{equation}
Define the \guytwo{VAR} companion matrix \guytwo{by}
\begin{equation}
    A = \begin{pmatrix}
        A_1 & A_2 & \dots& \dots &A_p\\
        I_N & 0_N & \dots & \dots & 0_N \\
        0_N & I_N \ddots &\dots & \vdots\\
        \vdots & \vdots & \ddots & \ddots & \vdots\\
        0_N & \dots & 0_N & I_N & 0_N
    \end{pmatrix}.
    \label{eq:varcompanion}
\end{equation}
Then, \eqref{eq:var_form} is \guy{equivalent to}
\begin{equation}
    \label{eq:pstacked}
    \bar{\mathbf{X}}_{t} = A\bar{\mathbf{X}}_{t-1} + \bar{\epsilon}_{t}, 
\end{equation}
where $\bar{\mathbf{X}}_{t} = (\mathbf{X}_{t}',\dots,\mathbf{X}_{t-p+1}')'$ and $\bar{\epsilon}_{t} = (\epsilon_{t}',0,\dots,0)$, \guy{where $X'$ denotes
the transpose of $X$. }\hlfour{Having expressed GNARID in VAR form, we can now give a standard stability condition and establish stationarity and ergodicity.}

\begin{assumption}\label{as:stationary}
    $\rho_{\max}(A) < 1.$
\end{assumption}

Since by definition we always have \hlthree{$\sum_{m} w_{n,m} \in \{0,1\}$}, $\sum_{j=1}^p\left( |\alpha_j^0| + \sum_{r=1}^{s_j} |\beta_{j,r}^0| \right) < 1$ guarantees \cref{as:stationary} which is the same as that for the GNAR model. \hlfour{Under \cref{as:stationary}, \cref{lem:ma_form} establishes the moving average representation of the GNARID process, and ultimately leads to stationarity and ergodicity in \cref{prop:stationary}.}

\begin{lemma}\label{lem:ma_form}
    \cref{as:stationary} \guy{implies} that \hlthree{$\mathbf{X}_{t}$} admits the following one-sided MA($\infty$) representation
    \begin{equation}\label{eq:ma_form}
        \mathbf{X}_{t} = \sum_{l=0}^\infty a_l \epsilon_{t-l},
    \end{equation}
    where $a_l$ is a matrix that satisfies
    \begin{equation}\label{eq:ma_bound}
        \sum_{l=0}^\infty \| a_l \|_q< C,
    \end{equation}
    for all $q$.
\end{lemma}

\begin{proposition}\label{prop:stationary}
    \cref{as:stationary} implies that the process \hlthree{$\mathbf{X}_{t}$} is both stationary and ergodic.
\end{proposition}

\hlfour{The above matrix formulation also helps clarify the connection between GNARID and SPDE-based Gaussian Markov random field (GMRF) models, as kindly suggested by a referee.} \hlthree{Our $\lambda$ is analogous to the exponent/smoothness parameter $\nu$ in the Mat\'ern spectral density used in GMRF.
GMRF uses a prespecified mesh or grid, which induces a neighbourhood structure and leads to a sparse precision matrix \cite{lindgren2011}. In this sense, the GMRF construction is similar to network models.
In the standard GMRF construction, $\nu$ is fixed to give the Markov property. However, extensions such as the rational SPDE approach of \cite{bolin2020} allow the smoothness parameter to be estimated from the data, which provides a structural analogy to GNARID, but the mechanisms remain different.} \hlfour{More specifically, for finite $\lambda$, varying $\lambda$ changes the non-zero weights in $W^{(r)}(\lambda)$, whilst
retaining their sparsity pattern, determined by the network, unchanged. By contrast, in the integer-order SPDE-GMRF construction, increasing the smoothness parameter $\nu$ increases the size of the neighbourhood and hence the number of potential non-zero entries in the precision matrix. For example, on a two-dimensional lattice, the limiting case $\nu = 0$ includes only one-step neighbours, while $\nu = 1$ includes also two-step neighbours. }
\section{Coefficient Estimation}

We estimate \guy{GNARID} coefficients using conditional least squares. More specifically, let
\guyfour{$L_T(\theta)$ be our
error criterion, where}
\begin{align}
\begin{split}
    L_T(\theta) = & \frac{1}{T}\sum_{t=1}^T\sum_{n=1}^N (X_{t,n}-\hat{X}_{t|t-1,n})^2\\
    &= \frac{1}{T} \sum_{t=1}^T\sum_{n=1}^N \left\{X_{t, n} - \sum_{j=1}^p \left(  \alpha_{j} X_{t-j,n} + \sum_{r=1}^{s_j} \beta_{j, r} \sum_{m \in \mathcal{N}^{(r)}(n)} \guytwo{w_{n, m}(\lambda)} X_{t-j,m} \right) \right\}^2.
\end{split}
\end{align}
\hlthree{Define the set of global minimisers $G_T = \argmin_{\theta \in \mathbb{R}^{P+1}} L_T(\theta)$ where $P = \sum_{j=1}^p(1+s_j)$ and $P+1$ is the dimension of the parameter space. Any measurable selection $\hat{\theta}_T \in G_T$ is a global estimator.} 

Unlike linear autoregressions, the objective $L_T$ to minimise here is not convex and might have more than one local minimum. This presents some challenges to the estimation which we will discuss later. We first investigate the asymptotic properties of the theoretical estimator $\hat{\theta}_T$. \hltwo{It should be noted that our model, with its dependence on $\lambda$, is no longer linear and hence occupies a different mathematical world than linear VAR and a different kind of nonlinearity than, say, sparse VAR methods. Hence, the results we prove below are new and tailored to this setting, while still drawing on established results where appropriate, since reproving known material would add little value.}

\subsection{Asymptotic Properties}

We will show that
\guyfour{our estimator}
$\hat{\theta}_T$ is strongly consistent, \hlthree{eventually unique} \hlfour{(that is, almost surely, there exists a finite $T_0$, such that the global minimiser is unique for all $T > T_0$)}, and asymptotically normal under some regularity conditions. \hlthree{All three results require positive definiteness of $\cov(\bar{\mathbf{X}}_t,\bar{\mathbf{X}}_t)$ and boundedness of $\cov(\mathbf{X}_t,\mathbf{X}_t)$.
\guyfour{These} are derived in \cref{lem:cov_finite}
\guyfour{using} the same properties of the \guyfour{noise}
covariance matrix  in \cref{as:sigma_epsilon}. To \guyfour{obtain} consistency,
we need global identifiability of the parameters, which is a result of \cref{as:dist}. For uniqueness and asymptotic normality, we additionally require linear independence of the partial derivatives of the conditional mean with respect to the parameters (\cref{lem:lin_indep}), as well as the boundedness of the derivatives up to  third order (\sref{lem:norm_finite} and \sref{lem:cross_norm_finite}) in the appendix.} 

\hlthree{We first state the first-order partial derivatives. For higher-order derivatives, see the appendix.} Let $\hat{\mathbf{X}}_{t|t-1} = (\hat{X}_{t|t-1,n})_n$ be the vector of the conditional mean\guytwo{s} given by \eqref{eq:gnarid}, then
\begin{equation}
    \frac{\partial \hat{X}_{t|t-1,n}}{\partial \alpha_j} = X_{t-j,n},
\end{equation}
\begin{equation}
    \frac{\partial \hat{X}_{t|t-1,n}}{\partial \beta_{j,r}} =  \sum_{m \in \mathcal{N}^{(r)}(n)} w_{n,m}(\lambda) X_{t-j,m},
\end{equation}
\begin{equation}
    \frac{\partial \hat{X}_{t|t-1,n}}{\partial \lambda} = \sum_{j=1}^p\sum_{r=1}^{s_j} \beta_{j,r} \sum_{m \in \mathcal{N}^{(r)}(n)}\frac{\partial w_{n,m}}{\partial \lambda}X_{t-j,m},
\end{equation}
where
\begin{equation}
     \frac{\partial w_{n,m}}{\partial \lambda} = \frac{d_{n,m}^{-\lambda}\sum_{q \in \mathcal{N}^{(r)}(n)}d_{n,q}^{-\lambda}\{\log(d_{n,q})-\log(d_{n,m})\}}{(\sum_{q \in \mathcal{N}^{(r)}(n)} d_{n,q}^{-\lambda})^2}.
\end{equation}

\sourav{In the above derivatives, $\hat X_{t| t-1,n}$ is viewed as a function of the parameter vector $\theta$, while the past observations $\{X_{t-j,m}\}$ and distances $\{d_{n,m}\}$ are conditioned on
and treated here as constants.} 
\hlthree{For boundedness of $\cov(\mathbf{X}_t,\mathbf{X}_t)$ and positive definiteness of $\cov(\bar{\mathbf{X}}_t,\bar{\mathbf{X}}_t)$,} we require that the covariance matrix of the noise
in the GNARID process~\eqref{eq:gnarid}, $\Sigma_\epsilon$, is positive definite and bounded.

\begin{assumption}\label{as:sigma_epsilon}
   We assume that the innovation covariance $\Sigma_\epsilon$ is
   positive definite, $\Sigma_\epsilon \succ c$, and 
   that it has bounded spectral norm, $\|\Sigma_\epsilon\|_2 < C$.
\end{assumption}

\guy{The covariance bounds in}~\cref{lem:cov_finite} are an immediate result of \cref{as:sigma_epsilon}.

\begin{lemma}\label{lem:cov_finite}
    The spectral norm of the
    variance/covariance matrix of $\mathbf{X}_t$ is bounded,
    $\|\cov(\mathbf{X}_t,\mathbf{X}_t)\|_2 < C$, and 
    the \mbox{$p$-stacked} process, $\bar{\mathbf{X}}_t$, 
    from~\eqref{eq:pstacked},
    is positive definite, i.e.\
    $\cov(\bar{\mathbf{X}}_t,\bar{\mathbf{X}}_t) \succ c$.
\end{lemma}

\hlthree{\cref{lem:cov_finite} is  used for two purposes.
\guyfour{First,} it provides a necessary condition for the global
\guyfour{identifiability}
of the parameters, which is used for proving consistency.
\guyfour{Secondly, it} is also a necessary condition for the linear independence of the partial derivatives, as we will show in \cref{lem:lin_indep}. The linear independence guarantees strict local convexity of the loss $L_T$ around $\theta^0$, which is ultimately used to prove uniqueness of the estimator and  asymptotic normality. For the $\alpha$ and $\beta$ directions, both the global
\guyfour{identifiability}
and the linear independence naturally follow from the linear structure of the model and \cref{lem:cov_finite}. For the $\lambda$ direction, we need the following assumption on the distances \guyfour{for identifiability}.}

\hlthree{
\begin{assumption}\label{as:dist}
    There exists at least one pair $(j,r)$ with $r\leq s_j$, such that $\beta^0_{j,r} \neq 0$, and there exists at least one $n=1,\dots,N$ with $m_1,m_2 \in \mathcal{N}^{(r)}(n)$ satisfying $d_{n,m_1} \neq d_{n,m_2}$.
\end{assumption}
}

\cref{as:dist} states that we need at least two edges connecting to the same node to have different distances. \hlthree{The exponent $\lambda$ is only identifiable if changing $\lambda$ changes the weights,
\guyfour{$w_{n, m}(\lambda)$}, in at least one neighbourhood.} For a counter-example, consider a network where each node is connected to only one other node, or a network where the distances are equal in every neighbourhood. \hlthree{In either case, changing $\lambda$ has zero influence on all
weights.}
\guyfour{For example, suppose there were three equal distances
in a neighbourhood $d_{n, m_i} = d$ for $i=1, 2, 3$,
then $w_{n, m_i}(\lambda) = d^{-\lambda}/ (d^{-\lambda} + d^{-\lambda} + d^{-\lambda}) = 1/3$, which does not
depend on $\lambda$.} \hlfour{\cref{as:dist} is only required for identification of the exponent $\lambda$ and is not a criterion for the validity of the GNARID process. If \cref{as:dist} fails and the connection weights do not vary with $\lambda$, the GNARID process remains well defined, but $\lambda$ is redundant and cannot be identified from the data. In this case, the GNARID process is equivalent to a fixed-weight GNAR process, since the conditional mean no longer depends on $\lambda$.}

\hlthree{The identifiability enabled by
\cref{as:dist} allows us to prove consistency of the estimator.} Moreover, under \cref{as:sigma_epsilon}, \cref{as:dist} is a necessary and sufficient condition for the linear independence between the weight function and its derivative, and hence the linear independence of the partial derivatives \hlthree{in \cref{lem:lin_indep}}.

\begin{lemma}\label{lem:lin_indep}
   \guyfour{The derivatives} $\frac{\partial \hat{X}_{t|t-1}}{\partial \alpha_j}$, $\frac{\partial \hat{X}_{t|t-1}}{\partial \beta_{j,r}}$, $\frac{\partial \hat{X}_{t|t-1}}{\partial \lambda}$ are linearly independent in the sense that
    \begin{gather}
    \begin{split}
        \E \left\|\sum_{j=1}^p \left(a_{\alpha_j}\frac{\partial \hat{\mathbf{X}}_{t|t-1}}{\partial \alpha_j}\bigg|_{\theta^0} + \sum_{r=1}^{s_j}a_{\beta_{j,r}}\frac{\partial \hat{\mathbf{X}}_{t|t-1}}{\partial \beta_{j,r}}\bigg|_{\theta^0}\right) + a_\lambda\frac{\partial \hat{\mathbf{X}}_{t|t-1}}{\partial \lambda}\bigg|_{\theta^0} \right\|^2 = 0\\
        \implies a_{\alpha_j} = a_{\beta_{j,r}} = a_\lambda = 0, \quad \forall \quad j=1,\dots,p, \quad r=1,\dots,s_j.
    \end{split}
    \end{gather}
\end{lemma}

\guyfour{The linear independence is
important and feeds into \cref{prop:consistency} via
assumption~C2 of Theorem~3.1
of \cite{estnonlinear}. 
The partial derivatives are with
respect to the parameters and the
linear independence essentially means
that no parameter is redundant in that
the process prediction is influenced
by all of the parameters jointly.
This implies that the information
matrix $U$, next, is strictly positive
definite and, hence, invertible.
Without this, the optimisation/convergence
surface might be flat with hence no
unique minimum for our estimator.
}
\guy{The next result establishes strong consistency, \hlthree{eventual uniqueness} and asymptotic normality
of our estimator $\hat{\theta}_T$, using results from
\cite{cls-asymptotic} and \cite{estnonlinear}.}

\hlthree{\begin{proposition}\label{prop:consistency}
    Under \cref{as:stationary}, \ref{as:sigma_epsilon} and \ref{as:dist}, $|G_T|=1$ for sufficiently large $T$ almost surely. Moreover, any measurable sequence of $G_T=\{\hat{\theta}_T\}$ satisfies $\hat{\theta}_T \overset{a.s.}{\to} \theta^0$. Moreover, let
    \begin{equation}\label{eq:normal_R}
        R = \E \frac{\partial\hat{\mathbf{X}}_{t|t-1}}{\partial \theta}\bigg|_{\theta^0}' \Sigma_\epsilon \frac{\partial\hat{\mathbf{X}}_{t|t-1}}{\partial \theta}\bigg|_{\theta^0},
    \end{equation}
    \begin{equation}\label{eq:normal_U}
        U = \E \frac{\partial\hat{\mathbf{X}}_{t|t-1}}{\partial \theta}\bigg|_{\theta^0}' \frac{\partial\hat{\mathbf{X}}_{t|t-1}}{\partial \theta}\bigg|_{\theta^0},
    \end{equation}
    then $T^{1/2} (\hat{\theta}_T - \theta^0) \overset{d}{\to} N(0,U^{-1}RU^{-1})$.
\end{proposition}}

\hlthree{
\begin{remark}
    \cref{prop:consistency} is stated for the global minimiser of $L_T(\theta)$. $G_T$ being non-empty and a singleton means that for sufficiently large $T$, the global minimum is uniquely attained at a finite real $\lambda$ and not the boundary of the extended reals, $\pm \infty$. Although the objective is non-convex, \cref{prop:consistency} shows that the eventually unique global minimiser converges almost surely to the true $\theta^0$. 

    Moreover, the results in \cref{prop:consistency} apply to the unconstrained estimator. If we minimise instead over the constrained region $\lambda \geq 0$ and the true value is $\lambda^0 = 0$, the consistency conclusion still holds. However, the asymptotic normality on the boundary no longer applies. The limiting distribution for $\hat{\lambda}$ in this case would have a point mass at zero.
\end{remark}}

Suppose we allow the dimensionality of the time series (or network), $N$,
to grow indefinitely. \guy{We would like the norm of the asymptotic variance
of the estimator, $||U^{-1} R U^{-1}||_2$ to
tend to zero, as we are `collecting more data spatially'.}
An assumption stronger than~\cref{as:dist} is needed to ensure asymptotic convergence with respect to $N$.
\hltwo{
\begin{assumption}\label{as:dist_N}
    We require that for all $r =1,\dots,\max_j(s_j)$, $\|W^{(r)}\|_F^2 = \Theta(N)$, and for all $n$, $c< d_{n,m} <C $ for all $m \in \mathcal{N}^{(r)}(n)$. Furthermore, for at least one \hlthree{pair $(j,r)$, $r \leq s_j$ with $\beta^0_{j,r} \neq 0$}, there exists $\mathcal{F}(N)$ where $|\mathcal{F}(N)| = \Theta(N)$ such that for all $n\in \mathcal{F}(N)$, $|\mathcal{N}^{(r)}(n)| = O(1)$ and there exist $m_1,m_2 \in \mathcal{N}^{(r)}(n)$ such that $|d_{n,m_1} - d_{n,m_2}| > c$. In addition, we assume that $\sum_{j=1}^p \|A_j\|_2 < (1+c)^{-1}$. All of the above conditions hold uniformly in $N$.
\end{assumption}
}

\hltwo{\cref{as:dist_N} is only used for the $O(N^{-1})$ decay of the variance for growing $N$ and is not needed for fixed $N$. It guarantees that information continues to accumulate as $N$ grows. $\|W^{(r)}\|_F^2 = \Theta(N)$ ensures that the contribution to $\beta_{j,r}$ direction is of order $N$. The bounded distances ensure that the coupling between $\beta_{j,r}$ direction and $\lambda$ direction is controlled. The set $\mathcal{F}(N)$ which extends \cref{as:dist} ensures that the contribution to the $\lambda$ direction is of order $N$. In particular, it requires that $\Theta(N)$ nodes satisfy \cref{as:dist} with bounded distances and bounded neighbourhood size so that the derivative with respect to $\lambda$ is not degenerate.
The condition $\sum_{j=1}^p \|A_j\|_2 < (1+c)^{-1}$ strengthens the spectral radius constraint in \cref{as:stationary}, since the latter is not enough to guarantee uniform stability for growing $N$. Thus, \cref{as:dist_N} is close to minimal for \cref{prop:consistency_N}. It does not impose full sparsity or globally heterogeneous distances. If we only seek $o(1)$ decay of the variance, then the $\Theta(N)$ conditions can be relaxed to any diverging rate.}

\begin{proposition}\label{prop:consistency_N}
    Let $R$ and $U$ be as in \cref{prop:consistency},
    then $\|U^{-1}RU^{-1}\|_2 = O(N^{-1})$.
\end{proposition}

\subsection{Estimation Methods}
\label{sec:estimationmethods}

Since it is not possible to analytically solve $\argmin_\theta L_T(\theta)$, we must resort to numerical \guyfour{methods}. \hlfour{The choice of numerical optimisation method is an implementation decision and does not form part of the definition of the estimator. Any method that obtains the global minimiser of $L_T(\theta)$ can be used.} {In practice, the choice of numerical method can be affected by the fact that} $L_T(\theta)$ is not convex and it is possible for \guyfour{there to be} more than one minimum. \hlfour{For example, ordinary gradient descent might lead to a non-global minimum.}
However, we do know that, for a fixed $\lambda$, there is a unique solution $\hat{\alpha}_j^{(\lambda)}, \hat{\beta}_{j,r}^{(\lambda)}, \ j=1,\dots,p, \ r = 1,\dots,s_j$ that minimises $L_T(\theta)$
provided that the \guyfour{associated}
least squares linear equation is solvable.
The global minimum is also of the form $(\hat{\alpha}_1^{(\lambda)}, \dots, \hat{\beta}_{1,s_1}^{(\lambda)}, \dots, \hat{\alpha}_p^{(\lambda)},\dots,\hat{\beta}_{p,s_p}^{(\lambda)}, \lambda)$ for some $\lambda$.
\hlthree{\cref{prop:consistency} states that the global minimiser is eventually unique and attained at a finite value of $\lambda$. 
\guyfour{We hence use a modified
grid search, which is
easily parallelisable.
Full details are provided
in Appendix E.}} If we wish to further refine the estimation, we can perform another grid search on a smaller interval or a gradient-based method to a certain level of tolerance. 

\hlthree{
\subsection{More General Weight Functions}\label{sec:general_weight}

\guyfour{We are grateful to a reviewer
who suggested that}
the inverse-power family considered in this paper is not essential for the consistency and asymptotic normality. More generally, one can replace \eqref{eq:w} with the more general form
\begin{equation}
    w_{n,m}(\eta) = \frac{g_{\eta}(d_{n,m})}{\sum_{q\in \mathcal{N}^{(r)}(n)} g_{\eta}(d_{n,q})}, \quad g_{\eta}(d) > 0, \eta \in H \subseteq \mathbb{R}^k.
\end{equation}
Here $H$ is the admissible parameter space for $\eta$. It may be all of $\mathbb{R}^k$, or may be restricted if identifiability or interpretability requires it.

Stationarity, boundedness and positive definiteness of $\cov(\bar{\mathbf{X}}_t, \bar{\mathbf{X}}_t)$ hold as long as \cref{as:stationary} and \cref{as:sigma_epsilon} \guyfour{remain} true.
For consistency, there must be at least one pair $(j,r)$ with $r\leq s_j$ and $\beta^0_{j,r} \neq 0$, such that for all $\eta \in H$, 
\begin{equation}\label{eq:identifiability_eta}
    \eta \neq \eta^0 \implies w_{n,m}(\eta) \neq w_{n,m}(\eta^0) \quad \text{for at least one } n \text{ and at least one } m \in \mathcal{N}^{(r)}(n).
\end{equation}
For asymptotic normality, the analogue of \sref{lem:dist_to_indep} in the appendix must hold to guarantee the linear independence of the partial derivatives. We also need the derivatives up to  third order to be locally bounded. Both \eqref{eq:identifiability_eta} and \sref{lem:dist_to_indep} are guaranteed by \cref{as:dist} for the inverse distance weighting in our case. For certain families, a simple condition similar to \cref{as:dist} is still enough to guarantee these conditions. 

For example, consider the polynomial functions $g_{\eta}(d_{n,m}) = 1 + \eta_1 d_{n,m} + \dots + \eta_k d_{n,m}^k$ with $H$ satisfying $g_{\eta}(d_{n,m}) > 0$ for all relevant $n,m$ and $\eta \in H$. The consistency and asymptotic normality continue to hold if we modify \cref{as:dist} to be such that there exist $m_1,\dots,m_{k+1} \in \mathcal{N}^{(r)}(n)$ such that $d_{n,m_1} ,\dots,d_{n,m_{k+1}}$ are pairwise distinct, i.e., there are $k+1$ distinct distances instead of two within one active neighbourhood. The reason is that a polynomial of degree at most $k$ with $k+1$ distinct zeros is identically zero. Similarly, the same condition also guarantees consistency and asymptotic normality for the family $\exp(\eta_1 d_{n,m} + \dots + \eta_k d_{n,m}^k)$ with $H = \mathbb{R}^k$ for the same reason.

For periodic functions, such as $g_{\eta}(d_{n,m}) = 1+ a \sin (\omega d_{n,m} + \phi)$ with $|a| < 1$, the function is always positive. Unlike the previous examples, having distinct distances itself is not enough to conclude identifiability or linear independence of partial derivatives. In this case, one must check \eqref{eq:identifiability_eta} and the analogue of \sref{lem:dist_to_indep} directly to derive consistency and asymptotic normality.
}

\hlfour{\subsection{Distribution-based Distances Using the Wasserstein Distance}}\label{sec:wasserstein_extension}

\hlfour{Thus far, the distances $d_{n,m}$ have been treated as given quantities describing the dissimilarities between nodes. In some applications, however, suitable distances may not be directly available. Instead, it may be more natural to define distances from the marginal distributions of the network time series themselves. We therefore consider an extension in which the distances form part of the model specification and are constructed from the marginal distributions of the network time series using the Wasserstein distance.}

\hlfour{More precisely, define the set of all relevant pairs of nodes,
\begin{equation}
    \mathcal{P} = \bigcup_{r=1}^{\max_{j=1}^p s_j}\{(n,m)\in \{1,\dots,N\}^2: m \in \mathcal{N}^{(r)}(n)  \}.
\end{equation}
Let $\mathcal{D}$ be the set of all collections of distances $D = \{d_{n,m}: (n,m) \in \mathcal{P} \}$ with $1 \leq d_{n,m} \leq 2$ for all $(n,m) \in \mathcal{P}$. Let
\begin{equation}\label{eq:dist_from_ws_main_text}
    d_{n,m} = 1+ \frac{W_1\{F_n(D),F_m(D)\}}{\max_{(k,l) \in \mathcal{P}} W_1\{F_k(D),F_l(D)\}},
\end{equation}
where $F_n(D)$ is the stationary marginal distribution of $X_{t,n}(D)$ depending on $D$ and 
\begin{equation}
    W_1\{F_n(D),F_m(D)\} = \int_{-\infty}^{\infty} |F_n(D)(x) - F_m(D)(x)| \text{d}x
\end{equation}
is the 1-Wasserstein distance; see \cite[Section 1.2.3]{Panaretos2019}. All other specifications in Definition~\ref{def:gnarid} remain unchanged. The denominator in \eqref{eq:dist_from_ws_main_text} normalises the 1-Wasserstein distances and the offset $1$ is to prevent zero-distances. Although this distance formulation is still a non-random constant and satisfies the GNARID definition, the fact that the stationary distribution itself depends on the distances introduces an apparent circularity to the model. However, we show formally in Section F of the appendix that, under the assumption stated there, such a stationary and ergodic GNARID process exists.}

\hlfour{In practice, the marginal distributions are unknown.}
\hlfour{From observations $\{ \mathbf{X}_t\}_{t=1}^{T}$, we empirically estimate the marginal distribution for node $n$ as 
\begin{equation}
    \hat{F}_{n,T}(x) = \frac{1}{T}\sum_{t=1}^{T} \mathbf{1}_{X_{t,n} \leq x}
\end{equation}
and define the distances used for weight construction as
\begin{equation}\label{eq:bruma_d_nm}
    \hat{d}_{n,m} = 1+ \frac{W_1\{\hat F_{n,T},\hat F_{m,T}\}}{\max_{(k,l) \in \mathcal{P}} W_1\{\hat F_{k,T},\hat F_{l,T}\}}.
\end{equation}
Since $X_{t,n}$ is univariate, $W_1\{\hat F_{n,T},\hat F_{m,T}\}$ can be simplified as
\begin{equation}\label{eq:w_dist_simp}
    W_1\{\hat F_{n,T},\hat F_{m,T}\} = \frac{1}{T}\sum_{t=1}^{T} |X_{(t),n}-X_{(t),m}|,
\end{equation}
where $X_{(t),n}$ is the $t$-th order statistic of $\{X_{t,n}\}_{t=1}^{T}$. Section F of the Appendix also establishes consistency of the empirical Wasserstein distances and shows that the GNARID estimator based on these estimated distances is unique for large $T$ almost surely and is strongly consistent.}

\hlfour{The choice of the 1-Wasserstein distance in the above construction is deliberate. It requires no further tuning parameter and can be directly computed from the observations. In the univariate setting considered here, its empirical version in \eqref{eq:w_dist_simp} is particularly easy to compute. After sorting the observations, it is just the average absolute difference between the order statistics, resulting in a computational cost of roughly $O(T\log T)$. By contrast, MMD typically requires the choice of a kernel and is usually $O(T^2)$, while KL divergence, Hellinger distance and total variation distance typically require density estimation first. The Wasserstein distance is also symmetric, unlike KL divergence. Other cheap and tuning-free choices include the KS and energy distances. However, the KS distance measures only the largest separation between the distribution functions and ignores the actual magnitude of the differences between observations. For example, it assigns the same distance to $(1,1,1,5)$ and $(1,1,1,8)$ as to $(1,1,1,5)$ and $(1,1,1,100)$, whereas the Wasserstein distance clearly distinguishes the two cases. The energy distance is another plausible alternative, but the 1-Wasserstein distance has a particularly direct interpretation. If $X_{t,n} = X_{t,m} + c$ for all $t$, then $W_1\{\hat F_{n,T},\hat F_{m,T}\} = |c|$. Hence, for a pure shift, the 1-Wasserstein distance is exactly the magnitude of that shift, regardless of the observation distributions. The energy distance does not preserve this in general. This combination of no need for tuning parameters, simple computation and direct interpretation makes the 1-Wasserstein distance especially suitable for distance construction in GNARID applications.}

\section{Simulation Study}\label{sec:sim_study}

\begin{figure}[!h]
    \centering
    \resizebox{0.6\textwidth}{!}{\includegraphics{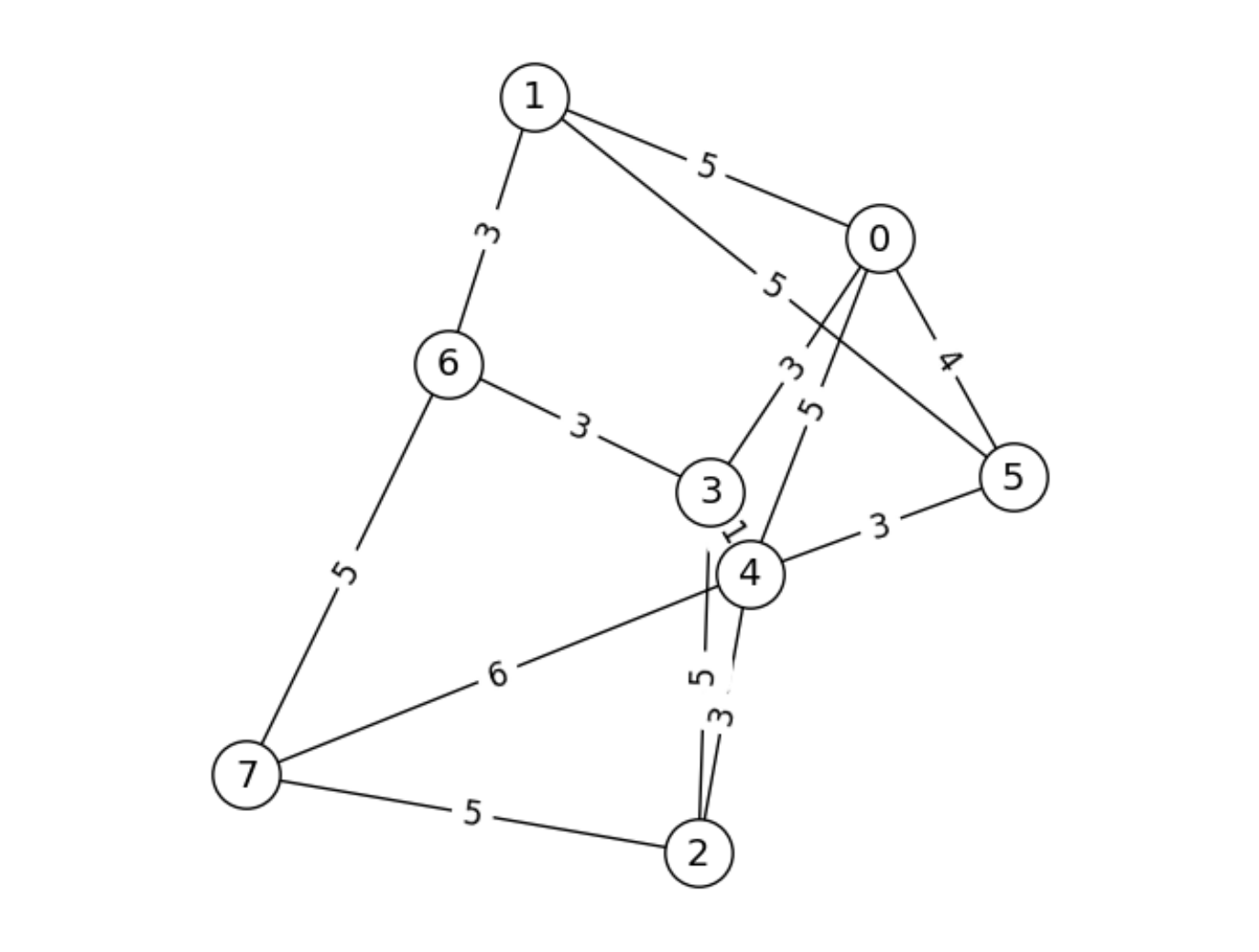}}
    \caption{An $N=8$ network with 14 edges whose connections and distances are specified in \cref{sec:sim_study}.}
    \footnotesize{\textbf{Alt text: }Plot of the network with eight labelled nodes and fourteen edges used in the simulation study. Each edge is annotated by its distance, with the distance values matching those specified in the simulation study.}
    \label{fig:sim_network}
\end{figure}

We now investigate the estimation performance
on simulated GNARID processes. We simulate from the following three model specifications on an $N=8$ network with 14 edges.
The network specification is
0-1(5), 0-3(3), 0-4(5), 0-5(4), 1-5(5), 1-6(3), 2-3(5), 2-4(3), 2-7(5), 3-4(1), 3-6(3), 4-5(3), 4-7(6), 6-7(5), where
$x-y(z)$ indicates an edge exists between vertices $x, y$ and
the length of the edge is $z$. \hlthree{See \cref{fig:sim_network} for an illustration of the network specified above.}

\begin{itemize}
    \item Case A: GNARID($1,[1],\lambda=3$), $\alpha_1 = 0.35$, $\beta_{1,1} = 0.55$.\\
    \item Case B: GNARID($2,[1,1],\lambda =3$), $\alpha_1 = 0.35$, $\beta_{1,1}=0.25$, $\alpha_2= 0.15$, $\beta_{2,1}=-0.20$.\\
    \item Case C: GNARID($2,[1,1],\lambda =0.6$), $\alpha_1 = 0.35$, $\beta_{1,1}=0.25$, $\alpha_2= 0.15$, $\beta_{2,1}=-0.20$.
\end{itemize}

For each case, we simulate 1000 realisations of a $T=1000$ GNARID process. The error terms $\epsilon_{t,n}$ are assumed to be i.i.d. Gaussian with variance 0.1. For each simulation, we estimate $\lambda,\alpha_1,\beta_{1,1}$ \hlthree{for case A, and $\lambda,\alpha_1,\beta_{1,1}, \alpha_2,\beta_{2,1}$ for cases B and C,} by minimising $L_T(\theta)$ via grid search on the interval $[0,10]$. \hltwo{For simulations with higher-dimensional networks and more complex covariance structures, demonstrating the scalability of GNARID, see Appendix D. }\cref{tab:sim} reports the mean
\guy{and standard deviation} of the estimated coefficients over the 1000 realisations in each case.
\begin{table}[h]
    \centering
    \setlength{\tabcolsep}{8pt}
    \begin{tabular}{cccccc}
        Case & $\hat{\lambda}$ & $\hat{\alpha}_1$ & $\hat{\beta}_{1,1}$ & $\hat{\alpha}_2$ & $\hat{\beta}_{2,1}$ \\
        \hline
        A & 3.007(0.21) & 0.350(0.009) & 0.549(0.011)&- &-\\
        \hline
        B & 3.054(0.53)& 0.350(0.010) &  0.250(0.019) & 0.150(0.011) &-0.200(0.018)\\
        \hline
        C & 0.588(0.21)& 0.350(0.011) &  0.250(0.021) & 0.150(0.011) &-0.200(0.021)
    \end{tabular}
    \caption{Mean (standard deviation) of the estimated coefficients over the 1000 realisations for each case.}
    \label{tab:sim}
\end{table}

\cref{tab:sim} shows that in all three cases, the estimation seems consistent. Moreover, we notice that the variance of $\hat{\lambda}$  is significantly larger than that of other estimators in all cases. This aligns with the asymptotic variance in \cref{prop:consistency}. \guy{The simulation assumes} i.i.d.\/
\guy{noise with variance of 0.1} for $\epsilon_{t,n}$,
\guy{which implies that the asymptotic variance $U^{-1}RU^{-1}$
in \cref{prop:consistency} reduces to} $0.1 U^{-1}$. Typically, $\E \|\frac{\partial \hat{X}_{t|t-1}}{\partial \lambda} \|^2$ is likely to be smaller than the
\guy{expected squared derivatives with respect to the other parameters},
which implies that the entry with respect to $\lambda$ in $U^{-1}$ is likely to be larger than the other entries. Indeed, we have also checked the asymptotic 
\guy{standard deviations}
according to \cref{prop:consistency} in each case,
\guy{and they} are roughly the same as those reported in \cref{tab:sim}.

\section{Misspecification of \texorpdfstring{$\lambda$}{lambda}}
\label{sec:misspec}


\subsection{Evaluating Misspecification by Simulation}\label{sec:misspec_sim}

\hlfour{This section assumes that the network is known and focuses specifically on misspecification of the exponent $\lambda$ which determines connection weights. More specifically,} even if the true network is known,
one does not necessarily know the form of the weights $w_{n, m}$, or the sensitivity
of the weights to quantities such as edge distances. 
{\em A priori} there is no reason for a given problem that 
$\lambda=1$
\guyfour{as is often assumed with regular
GNAR and it might be that
the true $\lambda \neq 1$}. This
\guyfour{misspecification} problem has not been
previously studied in the GNAR setting even though, as this section shows, the
choice of $\lambda$ can
influence forecasting performance
significantly. 
We stress that, to our knowledge, all GNAR-type
modelling has either used reciprocal distance weights $(\lambda =1 )$ or
uniform weights. We investigate here what happens if that is not a good
assumption.

\hlthree{
To explore whether misspecification of $\lambda$ can  influence forecasting performance and, if so, by
how much, given the `flatness' of the likelihood in the previous section, we conduct the following simulation experiments. 
\hlfour{We consider three $(T,N)$ settings: $(300,8)$, $(300,16)$ and $(50,100)$.} For each repetition, we first generated a random network with $N$ nodes, with random edges selected with probability $1/2$ and
edge distances distributed as Exp$(1)$ exponential \guyfour{random variables.
(An exponentially distributed
random variable with rate parameter
$\nu>0$ has cumulative
distribution function of
$F(x)=1-\exp(- \nu x)$ for $x\geq 0$,
zero otherwise and denoted
Exp$(\nu)$.)}
\hlfour{We then simulated a GNARID$(1, [1])$ process with $\lambda = \lambda_\text{true} \in \{3,5\}$ and nine pairs of
\begin{align}
\begin{split}
    (\alpha_1,\beta_{1,1}) = (\alpha,\beta) \in \{&(0.75,0.15),(0.65,0.25),(0.55,0.35),(0.45,0.45),(0.35,0.55),\\
    &(0.25,0.65),(0.15,0.75),(0.05,0.85),(0.01,0.89)\}
\end{split}
\end{align}
so that $\alpha + \beta = 0.9$ in each case.}
\guyfour{To be clear, $(1, [1])$ means
$p=1$ and $s_1=1$, $\alpha_1 = \alpha$
and $\beta_{1,1} = \beta$.} The innovation covariance is assumed to be $0.1I_N$, \guyfour{and that is why
the best mean squared predictive errors (MSPE) (scaled by $10^4$) in Figure~\ref{fig:misspecification_mspe} are very
close to 1000.}
This was repeated 1000 times for each setting. Then, we carried out one-step-ahead forecasting using the
misspecified $\lambda=1$ (the default inverse distance weight used in
nearly all GNAR time series research),
our
estimated $\hat{\lambda}$
and the true $\lambda$ --- the oracle
choice.

\begin{figure}[!h]
    \centering

    \begin{subfigure}{0.485\textwidth}
        \centering
        \includegraphics[width=\textwidth]{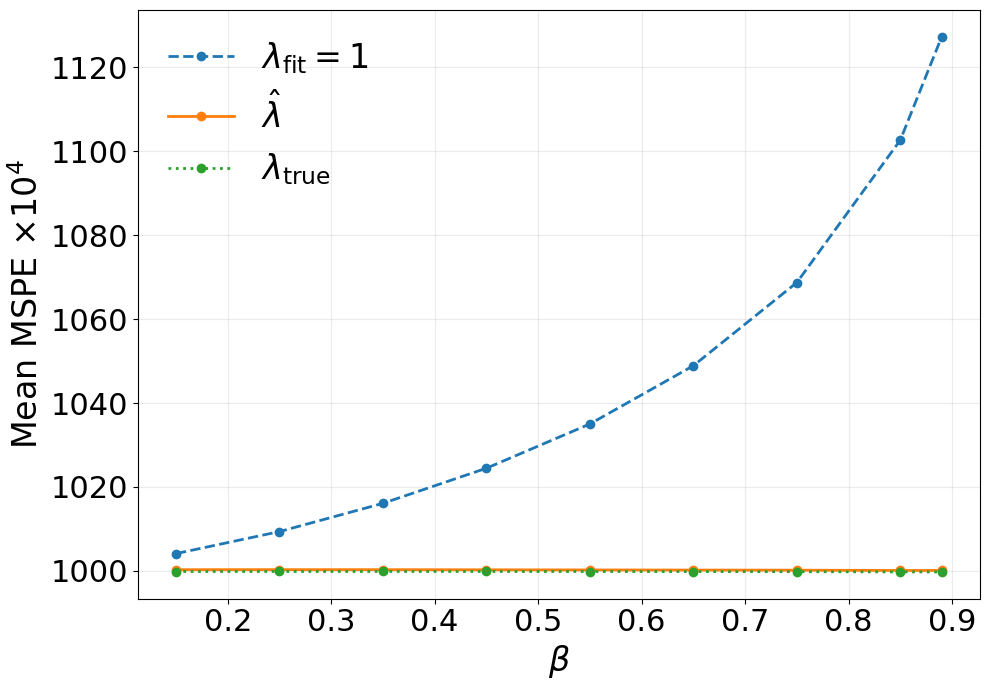}
        \caption{$\lambda_{\text{true}}=3$, $T=300$, $N=8$.}
    \end{subfigure}
    \hfill
    \begin{subfigure}{0.485\textwidth}
        \centering
        \includegraphics[width=\textwidth]{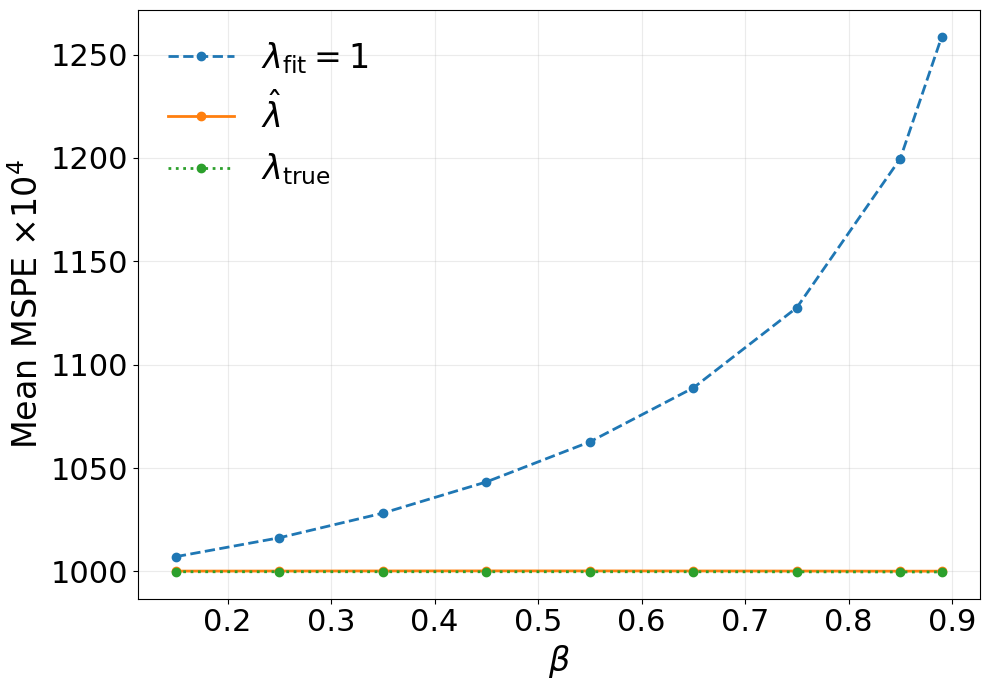}
        \caption{$\lambda_{\text{true}}=5$, $T=300$, $N=8$.}
    \end{subfigure}
\vspace{-.2em}
    \begin{subfigure}{0.485\textwidth}
        \centering
        \includegraphics[width=\textwidth]{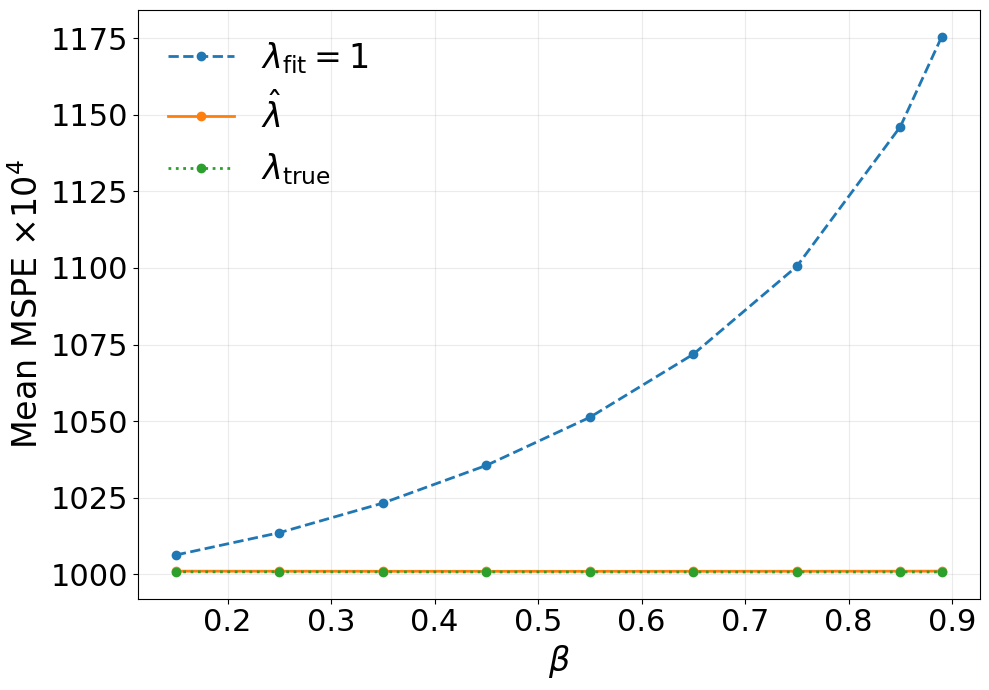}
        \caption{$\lambda_{\text{true}}=3$, $T=300$, $N=16$.}
    \end{subfigure}
    \hfill
    \begin{subfigure}{0.485\textwidth}
        \centering
        \includegraphics[width=\textwidth]{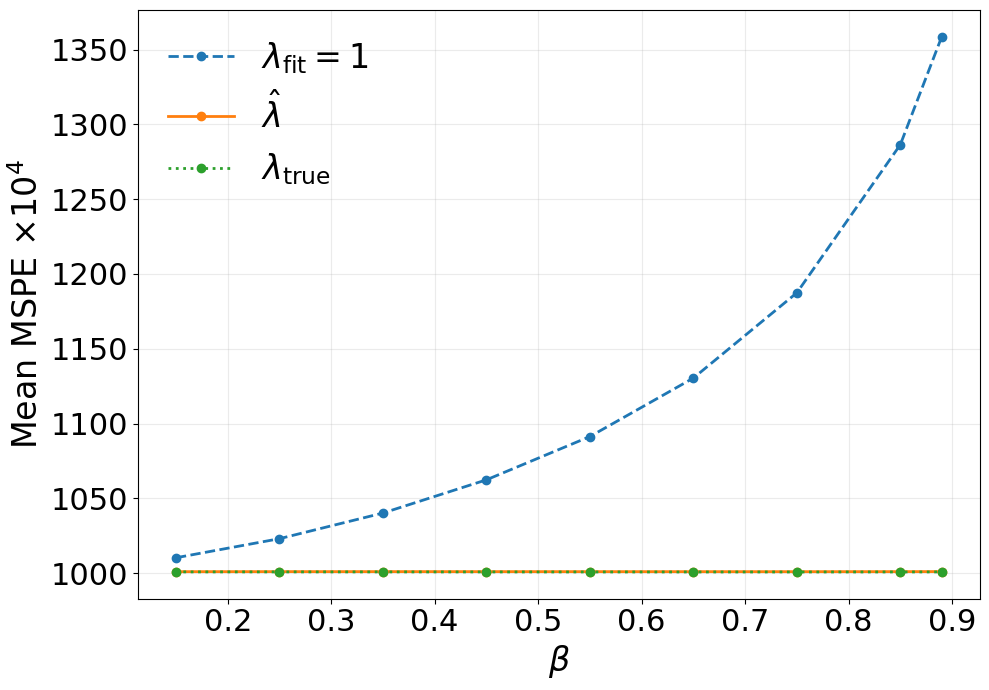}
        \caption{$\lambda_{\text{true}}=5$, $T=300$, $N=16$.}
    \end{subfigure}
\vspace{-.2em}
    \begin{subfigure}{0.485\textwidth}
        \centering
        \includegraphics[width=\textwidth]{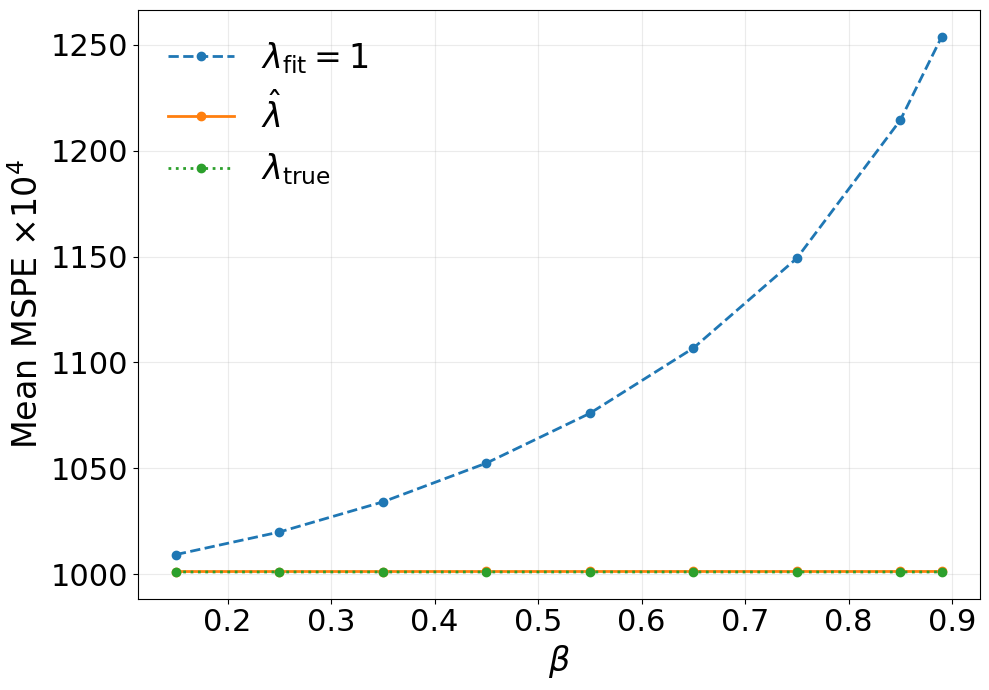}
        \caption{$\lambda_{\text{true}}=3$, $T=50$, $N=100$.}
    \end{subfigure}
    \hfill
    \begin{subfigure}{0.485\textwidth}
        \centering
        \includegraphics[width=\textwidth]{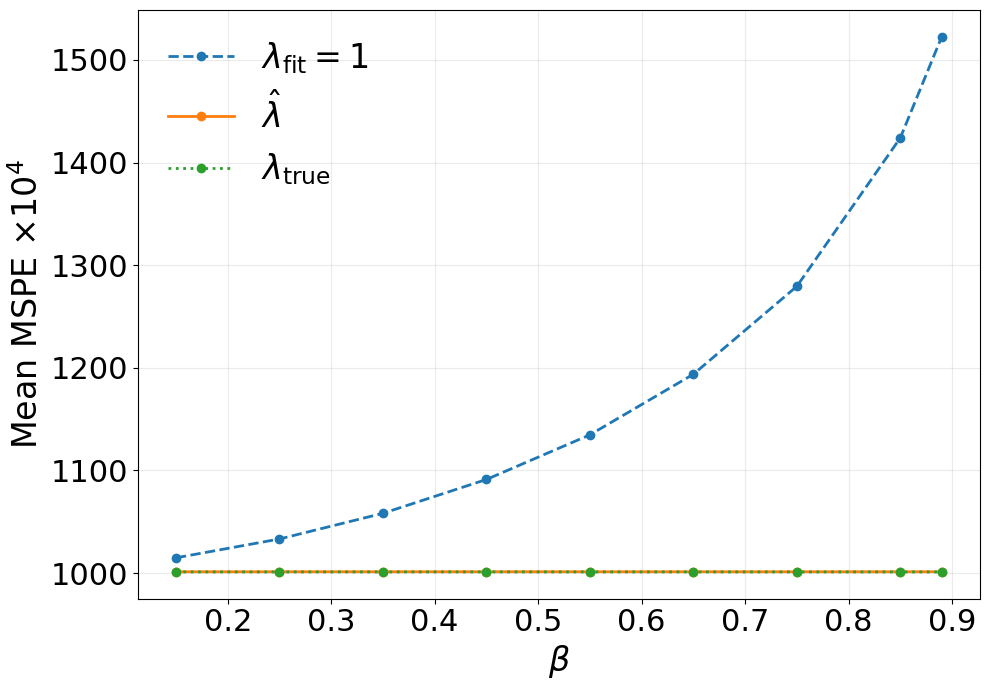}
        \caption{$\lambda_{\text{true}}=5$, $T=50$, $N=100$.}
    \end{subfigure}

    \caption{Mean one-step-ahead MSPE against the network autoregressive coefficient $\beta$ for the misspecified fixed exponent $\lambda_{\text{fit}}=1$, the estimated exponent $\hat{\lambda}$, and the oracle exponent $\lambda_{\text{true}}$, averaged over 1000 realisations.}
    \label{fig:misspecification_mspe}
    \footnotesize{\textbf{Alt text: }Six panel line plot of mean prediction error against network effect for the misspecified, estimated, and oracle exponents. The misspecified error increases substantially with network effect, while the estimated error remains close to the oracle error.}
\end{figure}
\begin{figure}[!h]
    \centering

    \begin{subfigure}{0.485\textwidth}
        \centering
        \includegraphics[width=\textwidth]{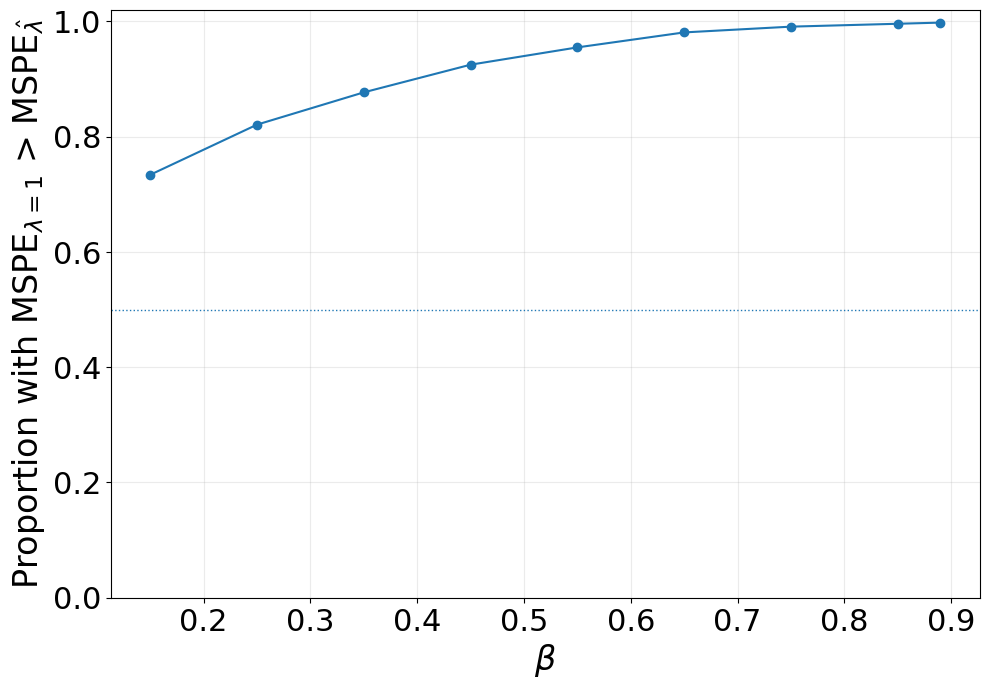}
        \caption{$\lambda_{\text{true}}=3$, $T=300$, $N=8$.}
    \end{subfigure}
    \hfill
    \begin{subfigure}{0.485\textwidth}
        \centering
        \includegraphics[width=\textwidth]{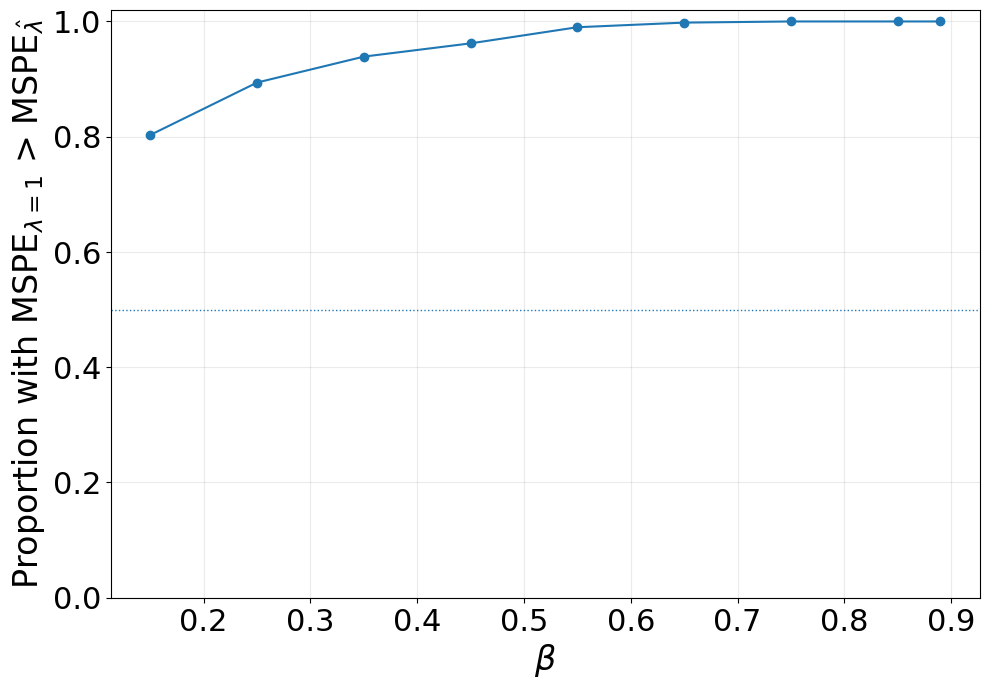}
        \caption{$\lambda_{\text{true}}=5$, $T=300$, $N=8$.}
    \end{subfigure}
\vspace{-.2em}
    \begin{subfigure}{0.485\textwidth}
        \centering
        \includegraphics[width=\textwidth]{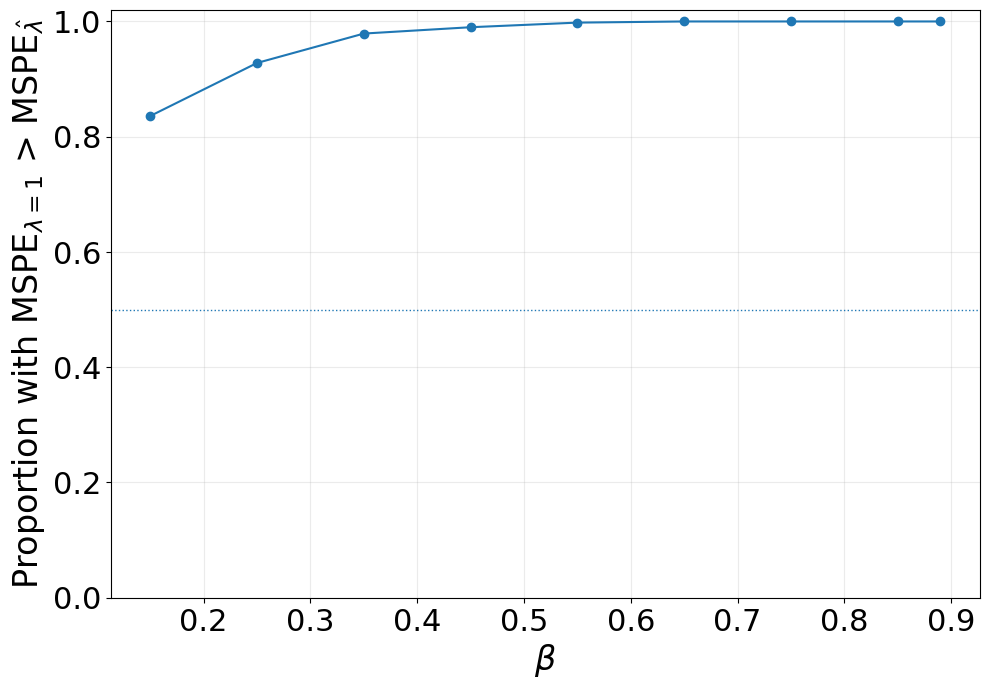}
        \caption{$\lambda_{\text{true}}=3$, $T=300$, $N=16$.}
    \end{subfigure}
    \hfill
    \begin{subfigure}{0.485\textwidth}
        \centering
        \includegraphics[width=\textwidth]{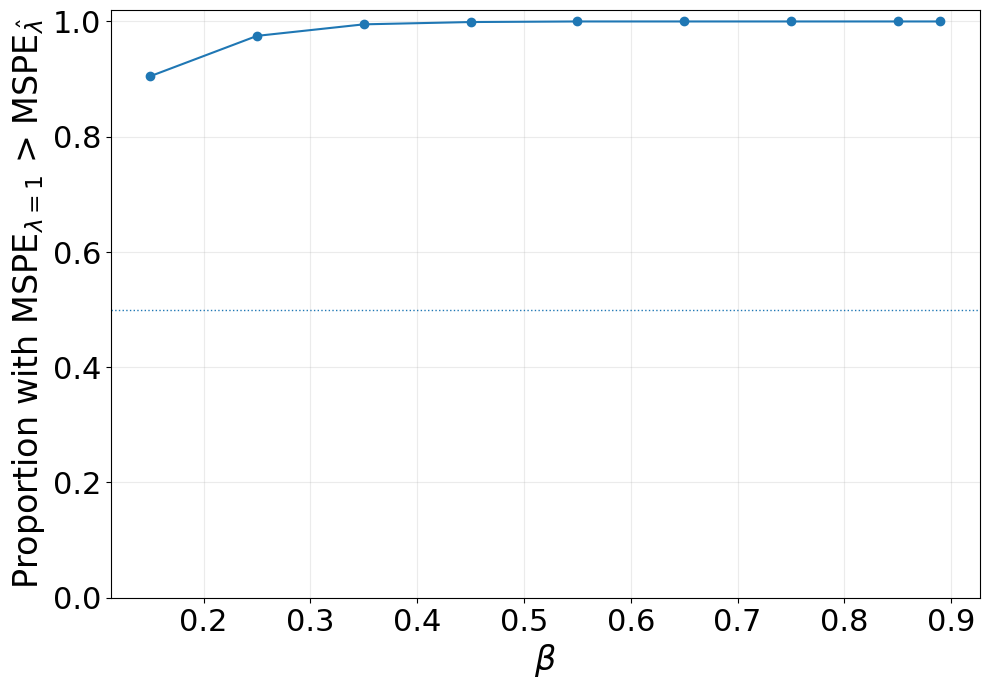}
        \caption{$\lambda_{\text{true}}=5$, $T=300$, $N=16$.}
    \end{subfigure}
\vspace{-.2em}
    \begin{subfigure}{0.485\textwidth}
        \centering
        \includegraphics[width=\textwidth]{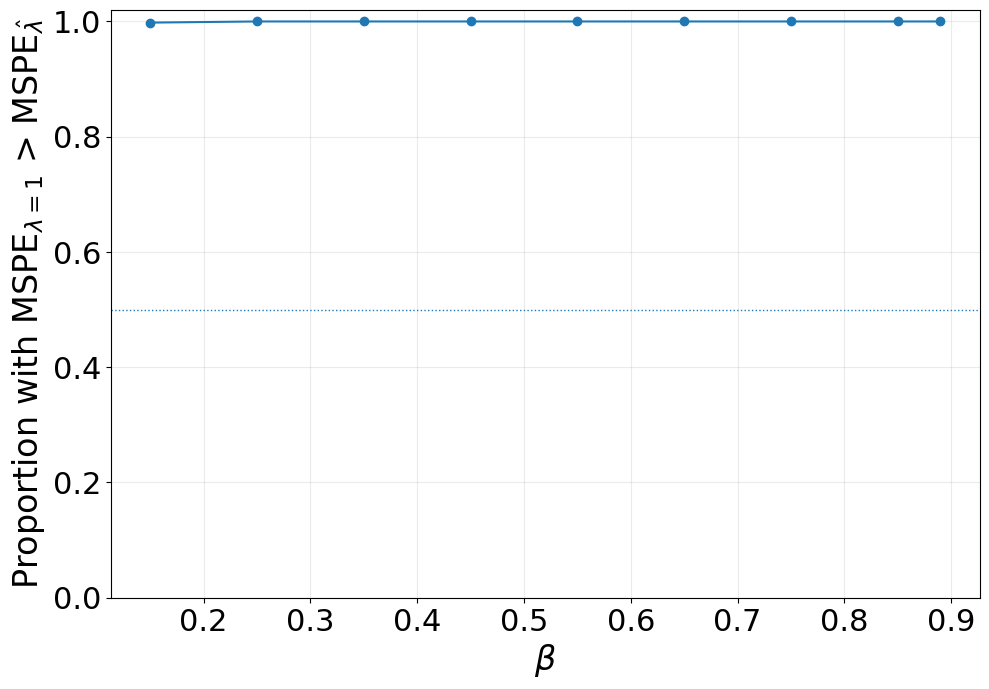}
        \caption{$\lambda_{\text{true}}=3$, $T=50$, $N=100$.}
    \end{subfigure}
    \hfill
    \begin{subfigure}{0.485\textwidth}
        \centering
        \includegraphics[width=\textwidth]{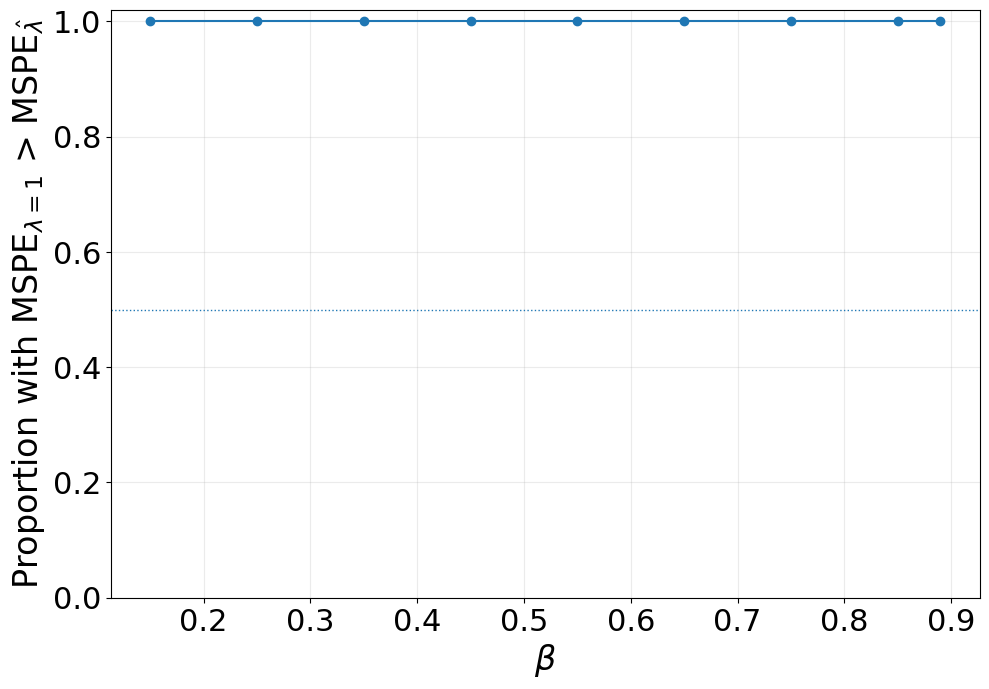}
        \caption{$\lambda_{\text{true}}=5$, $T=50$, $N=100$.}
    \end{subfigure}

    \caption{Proportion of the 1000 realisations for which the MSPE obtained using the misspecified fixed exponent $\lambda_{\text{fit}}=1$ exceeds that obtained using the estimated exponent $\hat{\lambda}$, plotted against the network autoregressive coefficient $\beta$.}
    \label{fig:misspecification_proportion}
    \footnotesize{\textbf{Alt text: }Six panel line plot showing how often estimated exponent achieves a lower prediction error than the misspecified fixed exponent. The trend of the proportion is nondecreasing and goes to one with increasing network effect for each setting.}
\end{figure}
\hlfour{For each setting of $(\lambda_\text{true},T,N)$, \cref{fig:misspecification_mspe} plots the mean one-step MSPE averaged over the 1000 realisations against $\beta$, and \cref{fig:misspecification_proportion} plots the proportion of realisations for which estimating $\lambda=\hat{\lambda}$ produces a smaller MSPE than fixing $\lambda = 1$ against $\beta$.}}
\hlfour{The results show that the forecasting error from fixing $\lambda = 1$ increases markedly as the network coefficient $\beta$ becomes larger. For $\lambda_{\text{true}} = 3$, the percentage increase in mean MSPE relative to estimating $\hat{\lambda}$ rises from $0.38\%$ to $11.28\%$ for $(T,N) = (300,8)$, from $0.53\%$ to $14.84\%$ for $(300,16)$, and from $0.79\%$ to $20.13\%$ for $(50,100)$ as $\beta$ goes from $0.15$ to $0.89$. The corresponding increases for $\lambda_{\text{true}} = 5$ are larger, rising from $0.69\%$ to $20.53\%$, from $0.92\%$ to $26.32\%$, and from $1.34\%$ to $34.23\%$, respectively. Thus, for the two $T=300$ settings the effect of misspecification is stronger for $N=16$ than for $N=8$, while the $(T,N) = (50,100)$ setting gives the largest deterioration. In contrast, the mean MSPE obtained from using $\hat{\lambda}$ remains almost indistinguishable from that of the oracle $\lambda_\text{true}$, with the percentage increases being smaller than $0.05\%$ in all settings.}

\hlfour{The standard deviations of the MSPEs are relatively large compared with the differences in their means, particularly for small $\beta$. For example, when $(T,N)=(300,8)$ and $\lambda_{\text{true}}=3$, at $\beta=0.15$ the difference between the mean MSPEs for $\lambda=1$ and $\hat{\lambda}$ is only $3.83\times10^{-4}$, whereas their standard deviations are $7.27\times10^{-3}$ and $7.19\times10^{-3}$, respectively. In contrast, in the setting with the largest difference, $(T,N)=(50,100)$, $\lambda_{\text{true}}=5$, and $\beta=0.89$, the difference in mean MSPE is $5.21\times10^{-2}$, substantially larger than the corresponding standard deviations of $7.89\times10^{-3}$ and $1.99\times10^{-3}$.}

\hlfour{However, the proportion of realisations for which $\lambda=1$ gives a larger MSPE than $\hat{\lambda}$ provides a more direct measure of how consistently estimating $\lambda$ improves forecasting performance. For $\lambda_{\text{true}}=3$, this proportion increases from $73.4\%$ to $99.8\%$ for $(T,N)=(300,8)$ and from $83.6\%$ to $100\%$ for $(T,N)=(300,16)$ as $\beta$ increases from $0.15$ to $0.89$. For $(T,N)=(50,100)$, it is already $99.8\%$ at $\beta=0.15$ and equals $100\%$ for all larger values of $\beta$. For $\lambda_{\text{true}}=5$, the corresponding proportion increases from $80.3\%$ to $100\%$ for $(T,N)=(300,8)$ and from $90.5\%$ to $100\%$ for $(T,N)=(300,16)$, while for $(T,N)=(50,100)$ it equals $100\%$ for every value of $\beta$ considered.}

\hlfour{Thus, the results in Figures~\ref{fig:misspecification_mspe}~and~\ref{fig:misspecification_proportion} clearly show} the importance of specifying $\lambda$ correctly.
\hlthree{Here, the usual GNAR choice $\lambda_{\text{fit}}=1$ lies well below the true exponent $\lambda_{\text{true}} = 3$ and $5$. This makes the weights too diffuse, assigning relatively too much weight to neighbours of longer distances. This effect becomes more severe as $\lambda_\text{true}$ increases.}
\guy{\hlthree{Moreover},
it is more important to correctly specify $\lambda$ for GNAR processes having
a high `edge parameter' $\beta$, which is reassuring since $\lambda$ helps control
information transmission between nodes linked by edges.}
The more general lesson is that forecasting with GNAR
might be improved by checking a variety of $\lambda$ values or, indeed, other weighting schemes entirely.

\subsection{Evaluating Misspecification by Theory}

We now theoretically investigate the influence of misspecification of $\lambda$ on the forecasting error.
Let $\hat{\mathbf{X}}_{t|t-1}(\lambda)$ be the conditional mean of $\mathbf{X}_t$ where $\alpha_j$ and $\beta_{j,r}$ are set to the true parameters, i.e.,
\begin{equation}
    \hat{\mathbf{X}}_{t|t-1}(\lambda) = \sum_{j=1}^p \left(\alpha_j^0 \mathbf{X}_{t-j} + \sum_{r=1}^{s_j} \beta_{j,r}^0 W^{(r)}(\lambda) \mathbf{X}_{t-j} \right).
\end{equation}
Define
\begin{equation}
    \Delta W^{(r)}(\lambda) = W^{(r)}(\lambda) - W^{(r)},
\end{equation}
where $W^{(r)}(\lambda)$ and $W^{(r)}$ are the weight matrices from~\eqref{eq:Wr},
and define
\begin{equation}
    \Gamma(j) = \E(\mathbf{X}_{t}\mathbf{X}_{t-j}').
\end{equation}
Then, we have
\begin{proposition}\label{prop:lambda_error}
    \begin{equation}
       \E \|\hat{\mathbf{X}}_{t|t-1}(\lambda) - \mathbf{X}_t\|^2 = \sum_{j=1}^p \sum_{i=1}^p \sum_{r=1}^{s_j} \sum_{s=1}^{s_i} \beta_{j,r}^0 \beta_{i,s}^0 \tr\{\Delta W^{(r)}(\lambda) \Gamma(i-j) {\Delta W^{(s)}(\lambda)}'\} + \tr(\Sigma_\epsilon).
    \end{equation}
\end{proposition}

When $\lambda=\lambda^0$, $\E \|\hat{\mathbf{X}}_{t|t-1}(\lambda) - \mathbf{X}_t\|^2 = \tr(\Sigma_\epsilon)$. \guy{For illustrative processes, for the simplest GNAR case,
when $p=1$ and $s_1=1$} we can derive a two-sided bound for $\E \|\hat{\mathbf{X}}_{t|t-1}(\lambda) - \mathbf{X}_t\|^2$ as follows.
\begin{corollary}\label{cor:lambda_error}
    If $p=1$ and $s_1=1$, then
    \begin{equation}
       \beta_{1,1}^2\rho_{\min}\{ \Gamma(0)\}\|\Delta W^{(1)}(\lambda)\|^2_F \leq \E \|\hat{\mathbf{X}}_{t|t-1}(\lambda) - \mathbf{X}_t\|^2 - \tr(\Sigma_\epsilon) \leq \beta_{1,1}^2\rho_{\max}\{ \Gamma(0)\}\|\Delta W^{(1)}(\lambda)\|^2_F
    \end{equation}
\end{corollary}

\begin{figure}[!h]
    \centering
    \resizebox{\textwidth}{!}{\includegraphics{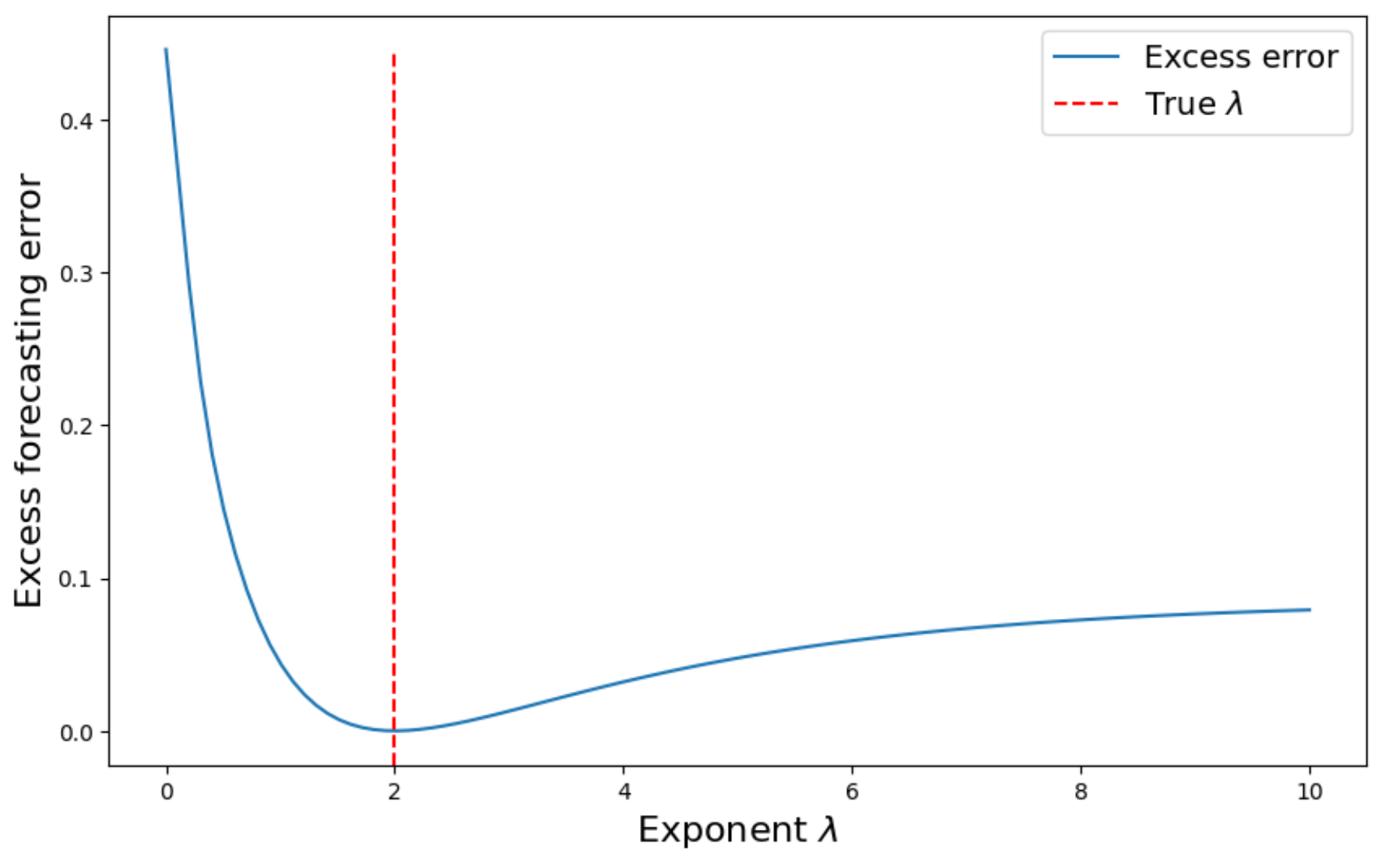}}
    \caption{Plot of $\E \|\hat{\mathbf{X}}_{t|t-1}(\lambda) - \mathbf{X}_t\|^2 - \tr(\Sigma_\epsilon)$ vs $\lambda$ for the GNARID($1,[1]$) processes with $\alpha_{1}^0 = 0.15$, $\beta_{1,1}^0 = 0.75$  on an $N=8$ network. \guy{Red vertical
    dashed line indicates true $\lambda^0 =2$. The label $0.x$ on the $y$-axis
    implies an $x\%$ increase in the forecasting error due to misspecification of $\lambda$.}}
    \footnotesize{\textbf{Alt text: }Line plot of excess forecasting error against lambda. Error is minimised near the true value two, rises sharply below two, and increases more gradually above two.}
    \label{fig:prop:lambda_error}
\end{figure}

\cref{prop:lambda_error} \guy{bounds} the expected sum of squared forecasting error under the assumption that $\alpha_j$ and $\beta_{j,r}$ are fixed to their true values. The exact error with respect to $\lambda$ where $\alpha_j$ and $\beta_{j,r}$ are actually estimated
\guy{would have to be} derived in an asymptotic setting and involves interplay between the coefficients. We can view \cref{prop:lambda_error} as an approximation to the exact forecasting error. \hlthree{It is worth noting that \cref{prop:lambda_error} does not depend particularly on the weight function used. Similar to what we have discussed in \cref{sec:general_weight}, the misspecification error can arise if changing weight parameters from their true values changes the weight matrices. }Figure~\ref{fig:prop:lambda_error} illustrates $\E \|\hat{\mathbf{X}}_{t|t-1}(\lambda) - \mathbf{X}_t\|^2 - \tr(\Sigma_\epsilon)$ for the GNARID($1,[1]$) processes with $\alpha_{1}^0 = 0.15$, $\beta_{1,1}^0 = 0.75$ and $\lambda^0=2$ on the same \hlthree{type of} $N=8$ network used in \cref{sec:misspec_sim}. \guy{The figure indicates that it is better to overestimate
$\lambda$, rather than underestimate it in this case.}

\hlthree{This agrees with the simulation results in \cref{sec:misspec_sim}, where the standard GNAR $\lambda_{\text{fit}}=1$ lies on the underestimation side and the excess forecasting error can increase sharply. More generally, we expect misspecification of $\lambda$ to matter the most when the network coefficients $\beta$ are large and the distances within neighbourhoods are diverse. Conversely, if $\beta$ are small, or if distances within neighbourhoods are nearly uniform, the effect of misspecification would be much weaker.}

\section{Brumadinho Ground Displacement Application}
\guy{On 25th January 2019 a tailings dam near Brumadinho
in Minas Gerais state collapsed, killing 270 people \citep{guardian19,brumadinho22}. \hlthree{It was proposed in \cite{grebby21,Das24}}
that satellite-based monitoring techniques
may help mitigate future catastrophes and suggested
that interferometric synthetic aperture radar (InSAR)
data would be able to monitor the course of events and
potentially prevent a collapse. The data we use here can
be publicly accessed via repositories detailed
in~\cite{grebby21}.}

\begin{figure}[!h]
\centering
\resizebox{\textwidth}{!}{\includegraphics{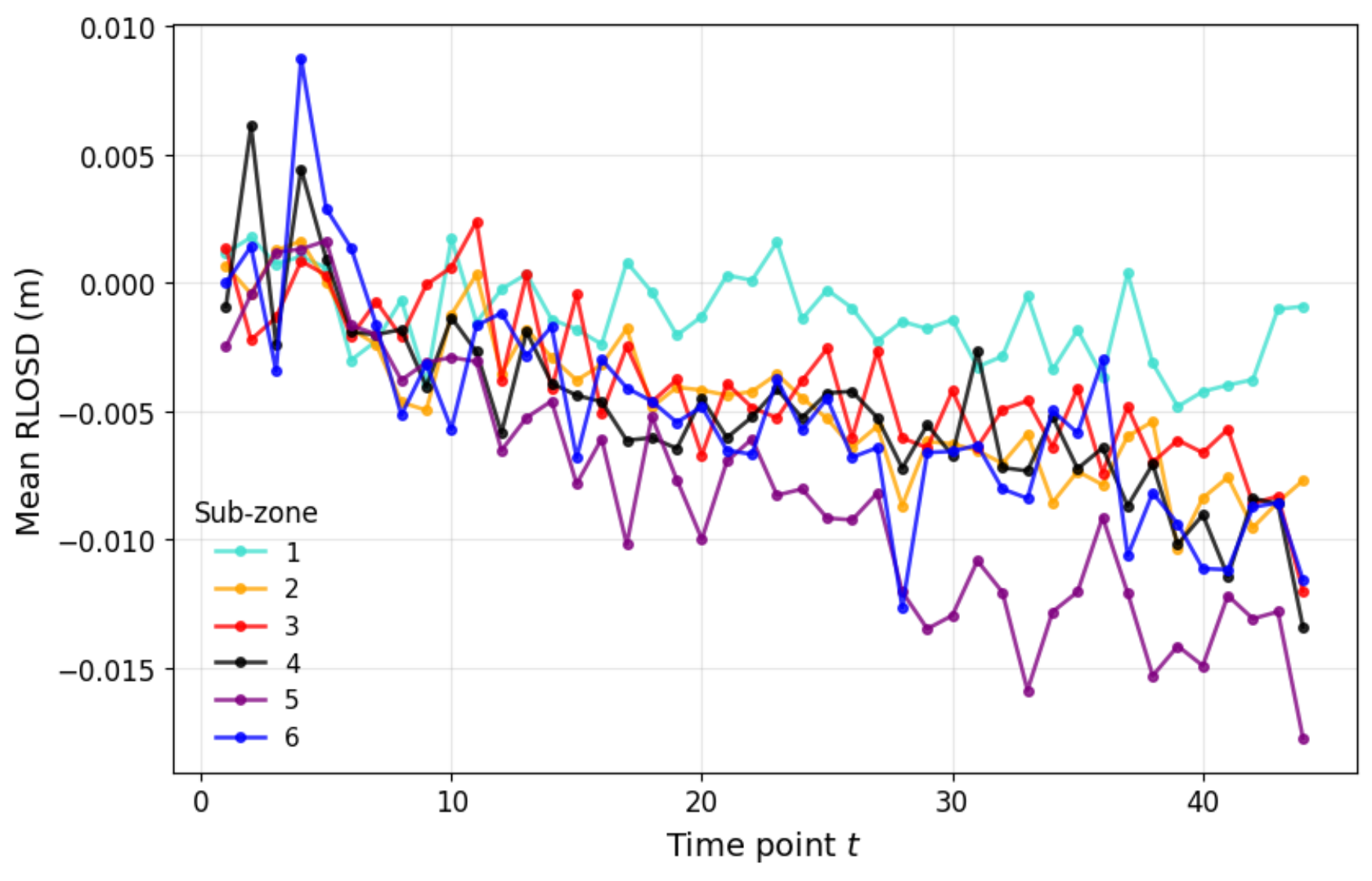}}
\caption{\hlthree{Spatial mean of the time series in each of the six sub-zones against time. The lines are coloured by sub-zone.}}
\footnotesize{\textbf{Alt text: }\hlthree{Line plot showing the spatial mean time series for each of the six sub-zones over forty four time points, with lines coloured by sub-zone. Most sub-zone means show a decreasing trend over time.}}
\label{fig:bruma_subzone_mean}
\end{figure}

The Brumadinho Feij\~{a}o tailings dam data analysed here were obtained by two European Space Agency low earth orbiting satellites that surveyed the area from 19 August 2017 to 17 Jan 2019 (the last acquisition before the dam's failure). The satellites used InSAR technology to sample regularly observed time series of relative line of sight displacement (RLOSD) (in \hlthree{metres}) over 220 gridded pixels, within six geologically partitioned zones.
Within the six zones, 30, 44, 46, 25, 38, 37 pixels were monitored respectively at twelve-day intervals, corresponding to the orbital cycles of the satellites, \hlthree{yielding 44 time points in total}. \cref{fig:bruma_subzone_mean} plots the spatial mean time series for each of the six sub-zones over $44$ time points, with lines coloured by sub-zone. 

%
%
\newpage

\subsection{\hlfour{Forecasting Setup}}

\hlfour{We use rolling forecast origins from $t=34$ to $t=43$ to produce one-step-ahead forecasts for $t=35$ to $t=44$, giving us ten validation time points in total. Starting at the first origin, observations $\{\mathbf{X}_t\}_{t=1}^{34}$ are used to fit the models and $\mathbf{X}_{35}$ is used to validate the one-step-ahead forecasting performance. The training period is then extended by one time point for each subsequent origin, and at the last origin, the models are fitted to $\{\mathbf{X}_t\}_{t=1}^{43}$ to predict $\mathbf{X}_{44}$. All parameters estimated from the time series are recomputed at each origin using the information available only up to this time point. Because the pixels are spatially dependent, we treat each forecast time as a single validation unit, and node-level forecast errors are aggregated into one predictive SSE at each origin. }

\hlfour{\cref{fig:bruma_subzone_mean} indicates that the time series exhibit an approximately linear trend, with the trend differing between sub-zones. We therefore first detrend the series at the initial rolling origin $t=34$. For each sub-zone, we compute its spatial mean time series and fit a linear trend (intercept and slope) to this mean using the available training observations up to $t=34$. The fitted sub-zone trend is then removed from every pixel belonging to that sub-zone. The estimated trend slope for each sub-zone at this first origin is shown in Table~\ref{tab:meanmm}. After detrending, the series fluctuates approximately around zero. We apply pixel-wise KPSS tests \citep{kwiatkowski1992} for stationarity and Newey--West HAC tests \citep{newey1987} for zero mean, with Benjamini--Yekutieli \citep{benjamini2001} correction for multiple testing in both cases. None of the 220 series reject either the zero-mean null hypothesis or the stationarity null hypothesis at the $10\%$ level, and hence also not at the $5\%$ level. At each subsequent rolling origin, the same detrending procedure is repeated using only the training observations available up to that origin. The fitted sub-zone trend at each origin is extrapolated one step ahead and removed from the observation to be predicted. The models are then fitted to these detrended series.}
\begin{table}[h]
\centering
\hlfour{
\begin{tabular}{rr|rr}
\hline
Sub-zone & Est slope ($\times 10^5$) & Sub-zone & Est slope ($\times 10^5$)  \\
\hline
1 & $-6.5$  &  4 & $-22.1$ \\
2 & $-23.4$ &  5 & $-42.4$\\
3 & $-19.6$ &  6 & $-28.6$ \\
\hline
\end{tabular}
\caption{\hlfour{Trend slope estimate for each of the six sub-zones at origin $t=34$.
\label{tab:meanmm}}}}
\end{table}

\subsection{\hlfour{Forecasting Models}}

We fit \hltwo{the following models} \hlfour{at each origin}.
\guyfour{The first three are non-network
competitor methods. Univariate
models build a model and forecast
each time series separately. 
Vector autoregressive (VAR) models jointly model
time series of vectors, where one
component can influence the same
or a different component in the future.
As mentioned in the Introduction,
VAR models are known to have 
 poor data to parameter ratios,
since their dimensionality
grows as $N^2$. VARLasso
works to sparsify VAR and reduce
the number of parameters by using
the Lasso method, see references
mentioned in the list below.}
\hltwo{
\begin{description}
    \item [B-AR-A:] Univariate AR($2$) with same $\alpha$ for all series
    (`global'),
    \item [B-AR-B:] Univariate AR($2$) with different $\alpha$ for each series,
    \item [B-VARLasso:] VARLasso($4$), \hlthree{see \cite{VARlasso}}, 
    \item [B-NAR:] NAR($2$) which is equivalent to GNARID($2,[1,1],\lambda=0$), \hlthree{see \cite{zhu2017}},
    \item [B-GNAR:] GNAR($2,[1,1]$) which is equivalent to GNARID($2,[1,1],\lambda=1$),
    \item [B-GNARID:] ($2,[1,1],\lambda=\hat{\lambda}$).
\end{description}}
\hlfour{For each forecasting origin $b$, from observations $\{ \mathbf{X}_t\}_{t=1}^{b}$, we empirically estimate the distribution-based distances $\hat{d}_{n,m}$ according to \cref{sec:wasserstein_extension}. The construction at each origin uses only observations available at that origin.} 
We use predefined $\lambda$s in model \hltwo{B-NAR and B-GNAR}, but estimate $\lambda$ from the data in model \hltwo{B-GNARID}. As discussed in previous sections, $\lambda=0$ removes the influence of the distances $\hat d_{n,m}$ from the weights, and $\lambda=1$ corresponds to weights reciprocal to $\hat d_{n,m}$. 

\begin{figure}[!h]
\centering
\resizebox{\textwidth}{!}{\includegraphics{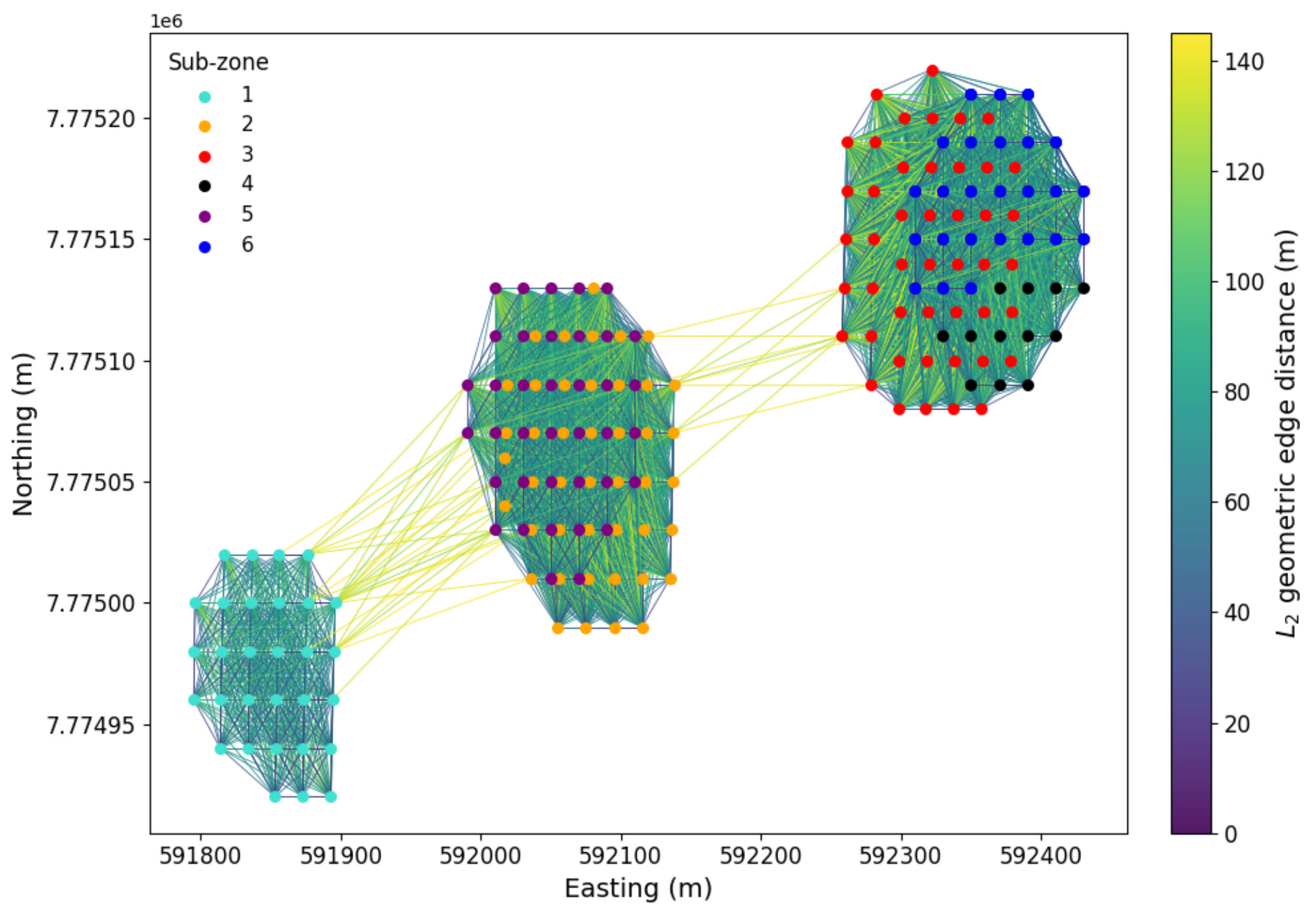}}
\caption{Network connecting any two pixels whose geometric distance is less than \hlfour{145} m. \hlthree{The nodes are coloured by sub-zone, while the edges are coloured by their Euclidean distances.} \guy{The $x$ and $y$ axes are Universal Transverse Mercator coordinates in zone 23S. The units of $x$ (Eastings) are metres (m) and the units of $y$ (Northings) are millions of metres (mM).}}
\footnotesize{\textbf{Alt text: }\hlthree{Network plot of \hlfour{two hundred and twenty} Brumadinho pixels in three spatial clusters, with edges between pixels within \hlfour{one hundred and forty-five} metres. Nodes are coloured by sub-zone and edges are coloured by edge distance.}}
\label{fig:bruma_net}
\end{figure}

\hlfour{The model orders and hyperparameters are chosen at the first origin $t=34$ and fixed for the later origins. For B-GNARID, the maximal lag $p=2$, the furthest stage of neighbourhood $s_j=1, j=1,2$, and the 145 m distance threshold used to construct the network were chosen by minimising the \hltwo{10-fold cross-validated one-step predictive sum of squared errors (SSE) of model B-GNAR on $\{\mathbf{X}_t\}_{t=1}^{34}$.} In particular, with the 145 m distance threshold, two pixels are connected by an edge if and only if their geometric distance is less than 145 m. The resulting adjacency matrix has an edge density of approximately $39\%$. \cref{fig:bruma_net} plots the network constructed above, with nodes coloured by sub-zone, and edges coloured by edge distance. For B-GNAR and B-NAR, the same 10-fold joint cross-validation for model orders and network thresholds results in a furthest stage of neighbourhood of $s_j=0, j=1,2$ for both models regardless of the threshold. With $s_j=0, j = 1,2$, the two models are effectively the same as a univariate AR model with global $\alpha$. Nevertheless, we still include them as part of the comparison and their orders are chosen to be the same as B-GNARID, namely GNAR($2,[1,1]$) and NAR(2), using the same network with threshold $145$ m.} Model \hltwo{B-AR-A} serves as a baseline for assessing the decrease in predictive SSE due to incorporating the network. \hltwo{We therefore used the same $p=2$ for B-AR-A.} The term `global' indicates that the autoregressive coefficients are shared across pixels.
B-AR-B and B-VARLasso are also included for comparison. \hlthree{For VARLasso, the maximal lag \hlfour{$p=4$} and the penalising parameter were chosen by minimising the same 10-fold cross-validation.} \hltwo{If we allowed cross-validation to choose the lag for B-AR-B, it selected $p=1$, but the resulting forecasts were worse, so we instead report $p=2$, which is more comparable to the network-based models and gives B-AR-B slightly better performance. VAR is not included in the comparison because its performance was considerably worse than the other models in this high-dimensional setting.}

\hlfour{At each later origin $b$, we refit the models to $\{\mathbf{X}_t\}_{t=1}^{b}$ using the selected orders. At each origin, each fitted model is then used to compute the one-step predictive SSE on the origin $\mathbf{X}_{b+1}$. We report the average one-step predictive SSE over the ten rolling origins. More specifically, for each model, we report $\frac{1}{10} \sum_{t=35}^{44} \sum_{n=1}^N (X_{t,n} - \hat{X}_{t,n} )^2$, where $\hat{X}_{t,n}$ is the estimated value for node $n$ at time $t$ from the model.}

\hlthree{To assess the uncertainty of the forecasts, \hlfour{for each origin,} we compute 95\% forecast intervals for all models considered above. The forecast intervals are constructed using the symmetric version of
\guyfour{the recent conformal prediction
method from} \cite{Xu2023}~Section 2.B. We evaluate the performance of forecast intervals through three metrics: node-wise coverage, average interval width, and the interval score introduced in (43) of \cite{Gneiting07} which is a special case of the interval loss function in \cite{Winkler72}. Lower scores indicate better performance.}

\subsection{\hlfour{Overall Results}}
\begin{figure}[!h]
    \centering
    \resizebox{\textwidth}{!}{\includegraphics{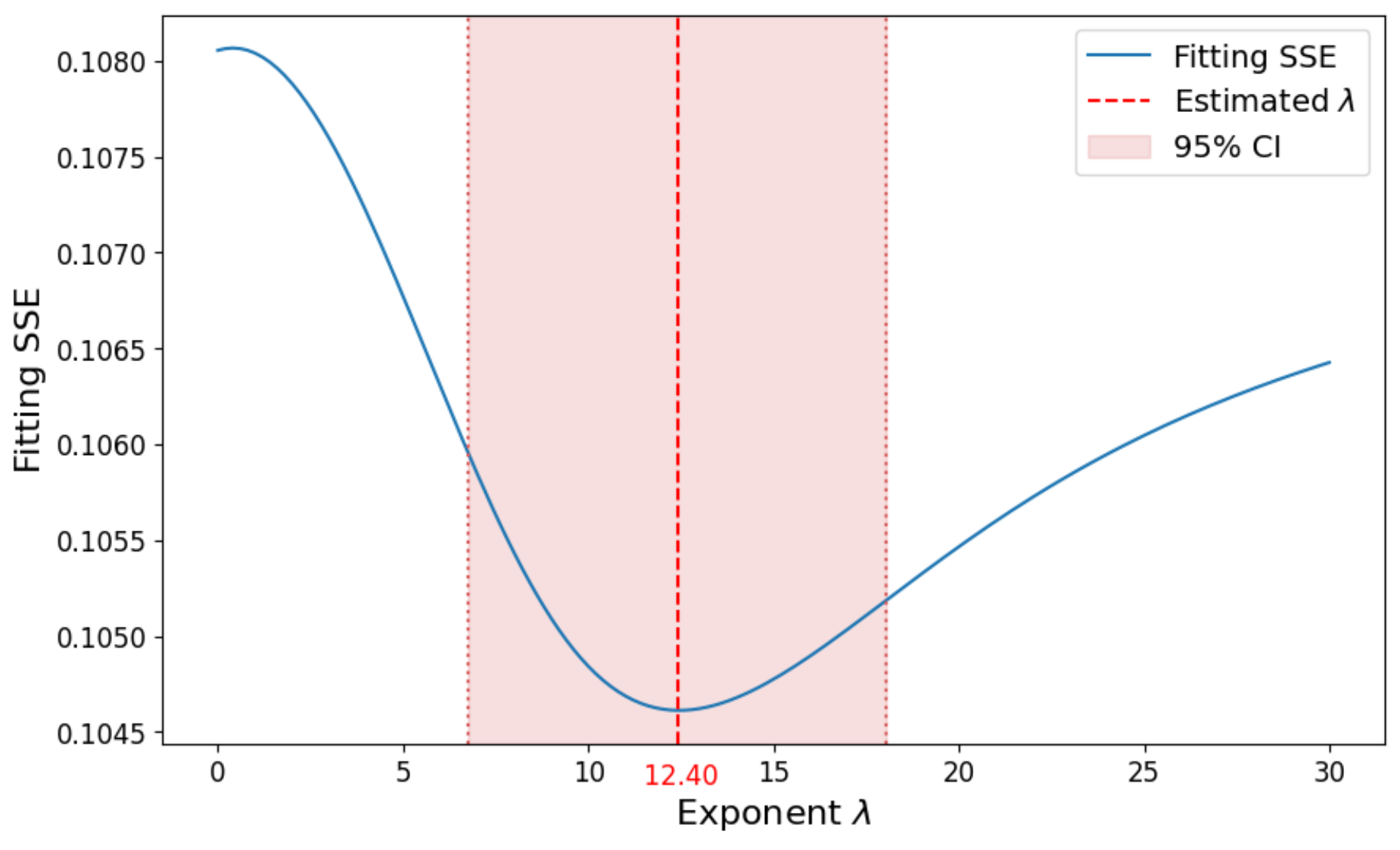}}
    \caption{The fitting SSE, $T L_T(\theta)$, versus $\lambda$ for model \hltwo{B-GNARID} \hlfour{at origin $t=40$}. The optimal exponent is \hlfour{$\hat{\lambda} = 12.4$}, \hlthree{with the shaded region indicating the 95\% confidence interval of $\hat{\lambda}$}.}
    \footnotesize{\textbf{Alt text: }Line plot of fitting sum of squared errors against lambda for the Brumadinho GNARID model \hlfour{fitted to observations before time forty}. The curve is U-shaped and reaches its minimum at lambda \hlfour{twelve point four}.}
    \label{fig:Bruma_SSE}
\end{figure}

Figure~\ref{fig:Bruma_SSE} plots the fitting SSE 
\guythree{for B-GNARID} vs the exponent $\lambda$ \hlfour{at origin $t=40$}. The optimal exponent is \hlfour{$\hat{\lambda} = 12.4$}. \hlthree{The red shaded region shows the 95\% confidence interval of $\hat\lambda$, obtained by applying the empirical form of the asymptotic normal distribution in \cref{prop:consistency}, with $R$ and $U$ in \eqref{eq:normal_R} and \eqref{eq:normal_U} estimated by their sample means \hlfour{and assuming that the distances are known}.
} \hengxu{The corresponding estimated autoregressive coefficients are \hlfour{$(\hat{\alpha}_1, \hat{\beta}_{1,1}, \hat{\alpha}_2, \hat{\beta}_{2,1}) = (0.030, 0.26, 0.023, 0.36)$}}. \hlfour{For the other origins, the estimated optimal exponent ranges from $\hat{\lambda} = 12.0$ to $\hat{\lambda} = 13.0$. The estimated autoregressive coefficients vary but the network effects still dominate.}
\guy{It is  interesting that the best model has strong
network-influencing coefficients, $\hat{\beta}_{1, 1},
\hat{\beta}_{2, 1}$. Quite often, the
self-autoregressive coefficients are larger than the
network ones, but here the opposite is true, and this
shows the utility of the network time series
approach. The other point is that the best model
requires terms from more than one lag.
Much work in the literature
only considers lag-one (and nearest neighbours), but
there is a growing body of work, including this
paper, that shows that this
is just not sufficient for proper statistical modelling,
see, e.g.\ \cite{Nason26}.}

\begin{figure}[!h]
    \centering

    \begin{subfigure}[t]{0.48\textwidth}
        \centering
        \includegraphics[width=\linewidth]{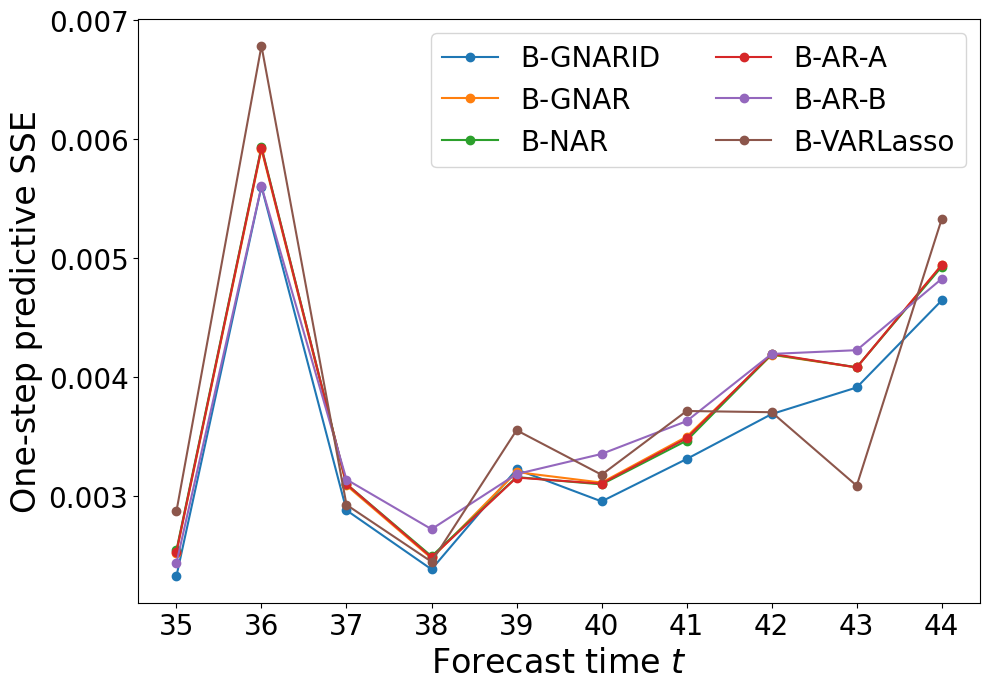}
        \caption{One-step predictive SSE.}
        \label{fig:bruma_pred_sse}
    \end{subfigure}
    \hfill
    \begin{subfigure}[t]{0.48\textwidth}
        \centering
        \includegraphics[width=\linewidth]{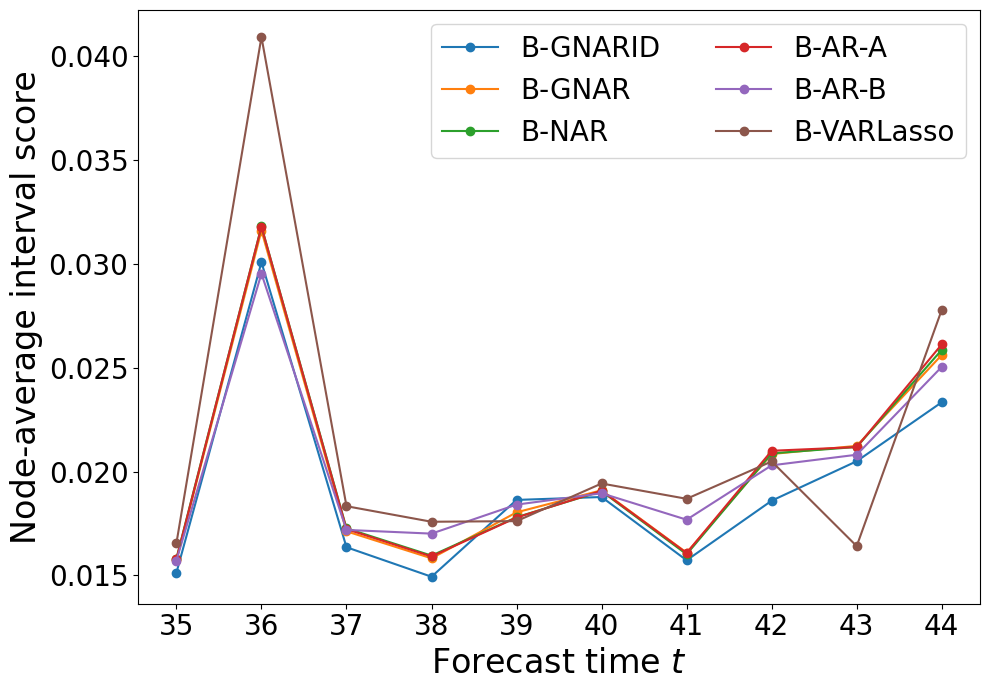}
        \caption{Node-average interval score.}
        \label{fig:bruma_winkler}
    \end{subfigure}

    \vspace{0.5cm}

    \begin{subfigure}[t]{0.48\textwidth}
        \centering
        \includegraphics[width=\linewidth]{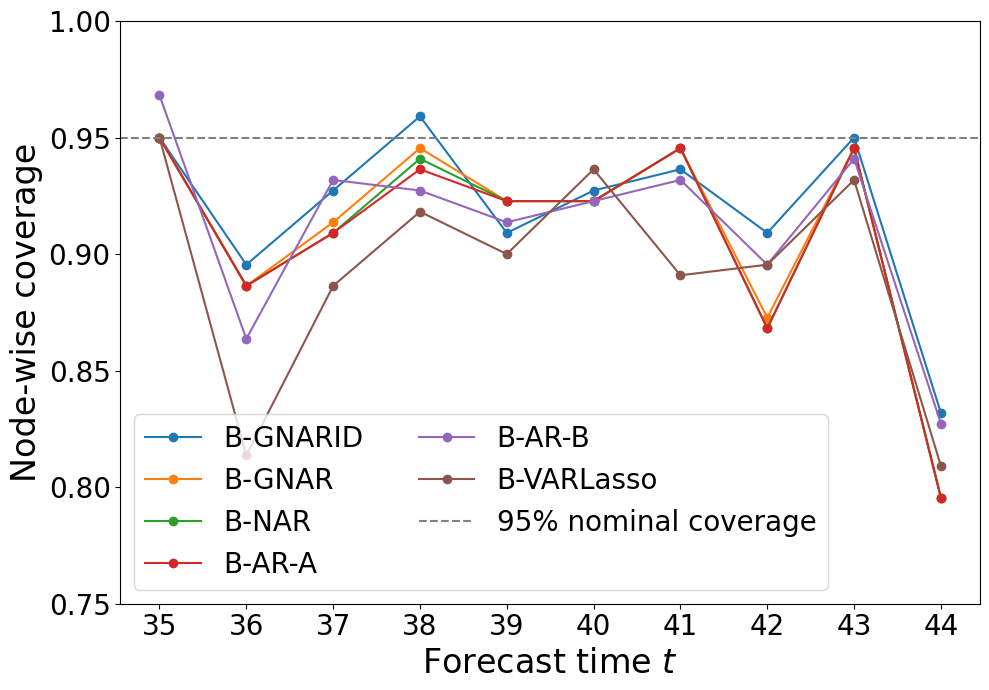}
        \caption{Node-wise coverage.}
        \label{fig:bruma_coverage}
    \end{subfigure}
    \hfill
    \begin{subfigure}[t]{0.48\textwidth}
        \centering
        \includegraphics[width=\linewidth]{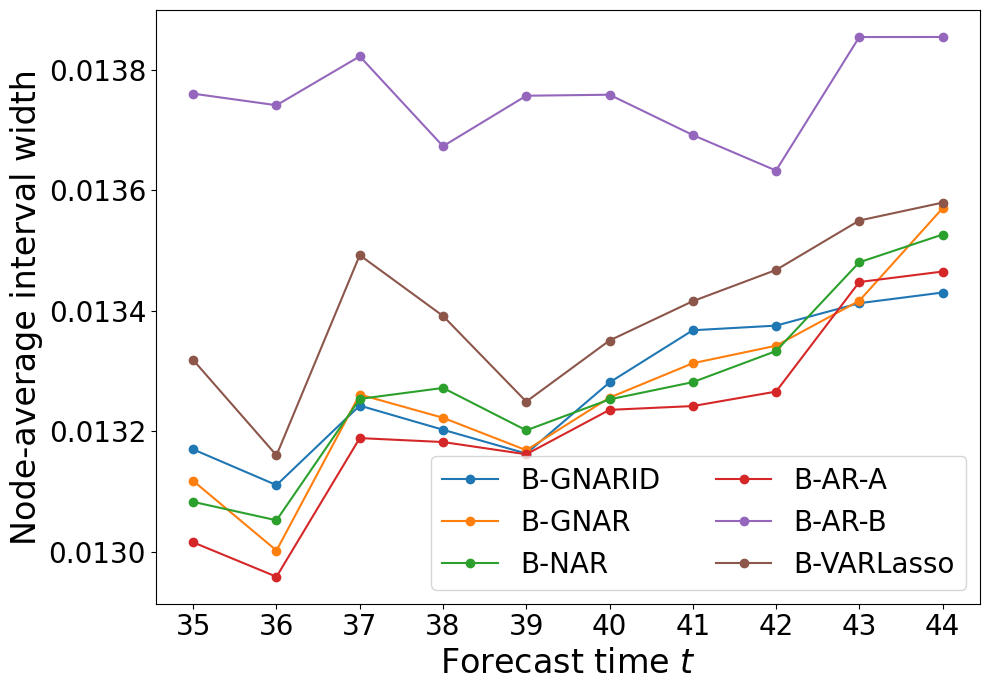}
        \caption{Node-average interval width.}
        \label{fig:bruma_width}
    \end{subfigure}

    \caption{Forecast performance for the Brumadinho ground displacement time series over forecast times $t=35,\dots,44$ for B-AR-A, B-AR-B, B-NAR, B-GNAR, B-GNARID and B-VARLasso. (a) One-step predictive SSE. (b) Node-average interval score. (c) Node-wise coverage, with the dashed line indicating the nominal 95\% coverage. (d) Node-average interval width. Panels (b)--(d) are based on 95\% forecast intervals.}
    \footnotesize{\textbf{Alt text: }Four panel line plot comparing six forecasting models over ten forecast times. Panel a shows one step predictive sum of squared errors, with B-GNARID generally having the lowest values. Panel b shows node average interval scores, with B-GNARID generally among the lowest. Panel c shows node wise coverage, with a horizontal dashed line marking the nominal ninety five percent coverage level. Panel d shows node average interval widths, with B-AR-B producing consistently wider intervals. Panels b to d are based on ninety five percent forecast intervals.}
    \label{fig:bruma_results}
\end{figure}

For each model, \cref{tab:Bruma_SSE} reports \hlthree{the one-step predictive SSE \hlfour{averaged over the ten origins},
\guyfour{percentage increase over the minimum achieved by B-GNARID},
the fitting SSE \hlfour{at the last origin $t=43$}, and node-wise coverage, node-average interval width, and node-average interval score \hlfour{averaged over the ten origins} for $95\%$ forecast intervals}. \hlfour{\cref{fig:bruma_results} plots the one-step predictive SSE, node-wise coverage, node-average interval width, and node-average interval score against the time of prediction $t=35$ to $t=44$ for each model.}
\begin{table}[h]
    \centering\hlfour{
    \setlength{\tabcolsep}{5pt}
    \begin{tabular}{ccrcccc}\hline
        Model& $\lambda$ & \begin{tabular}{c}Pred.\ SSE\\\guyfour{$\mbox{m}^2$}   \ ($\times 10^{-3}$)\end{tabular} & \begin{tabular}{c}Fit.\ SSE\\
        \guyfour{$\mbox{m}^2$}\ ($\times 10^{-2}$)\end{tabular} & Coverage & \begin{tabular}{c}Width, \guyfour{$\mbox{m}$}\\
        ($\times 10^{-2}$) \end{tabular} & \begin{tabular}{c}Score\\($\times 10^{-2}$) \end{tabular}\\
        \hline
        B-AR-A & - & 3.70 \mbox{\small (6\%)} & 12.0 & 90.8\%  &1.32  & 2.02 \\
        B-AR-B & - & 3.73 \mbox{\small (7\%)} & 11.3 & 91.2\% & 1.38 & 2.01\\
        B-VARLasso & - & 3.76 \mbox{\small (8\%)} & 2.35 & 89.3\% & 1.34 & 2.14\\
        B-NAR & 0 & 3.70 \mbox{\small (6\%)} & 12.0 & 90.9\% & 1.33 & 2.02 \\
        B-GNAR & 1 & 3.70 \mbox{\small (6\%)} & 12.0 & 91.0\% & 1.33 & 2.01\\
        B-GNARID&  $\hat{\lambda}=12.4$ & 3.49 \mbox{\small (0\%)} & 11.5 & 92.0\% & 1.33 & 1.92\\\hline
    \end{tabular}}
    \caption{\hlfour{Average one-step predictive SSE, last origin in-sample fitting SSE, and node-wise coverage, average width, and average interval score for $95\%$ forecast intervals from different models fitted to the Brumadinho ground displacement time series.
    The percentages in the Predictive SSE column are the percentage increases
    of the SSE value compared to the smallest for the B-GNARID model.}}
    \label{tab:Bruma_SSE}
\end{table}

\hlfour{In terms of point forecasts, B-AR-A, B-NAR, and B-GNAR have very similar performance. This is due to the fact that, for both B-GNAR and B-NAR, the estimated network autoregressive coefficients, $\hat{\beta}_{1,1}$ and $\hat{\beta}_{2,1}$, range from $-0.059$ to $0.055$, indicating that the spatial information contributes relatively little to the prediction. Hence, B-GNAR and B-NAR achieve similar performance as B-AR-A which does not utilise spatial information at all. B-GNARID gives the lowest prediction error among the models considered. The decrease in average predictive SSE is approximately $5.6\%$ from B-AR-A, B-NAR or B-GNAR to B-GNARID. As we have discussed, in contrast to B-GNAR and B-NAR, the estimated network autoregressive coefficients in B-GNARID are much stronger, showing that B-GNARID can effectively utilise the information from other nodes in the network. \cref{fig:bruma_pred_sse} shows the predictive SSE of each model at each forecast time point. As a summary, GNARID has consistently lower prediction error than each of the other models at nine out of the ten forecast time points. For B-AR-A, B-AR-B, B-NAR and B-GNAR, the time point at which B-GNARID has a larger prediction error is $t=39$. For B-VARLasso, this is $t=43$. It is also worth noting that B-VARLasso achieves a substantially smaller fitting SSE. However, its slightly larger predictive SSE suggests that its lower fitting SSE does not translate into better forecasting performance, and may indicate a modest degree of overfitting even with regularisation.}

\hlfour{In terms of forecast intervals, B-GNARID has the best interval score overall, which is approximately a $4.5\%$ decrease from the second best scores from B-AR-B and B-GNAR. As \cref{fig:bruma_winkler} shows, compared to B-AR-A, B-NAR and B-GNAR, B-GNARID achieves a better interval score at nine out of the ten time points except for $t=39$. Compared to B-AR-B, B-GNARID achieves a better interval score at eight time points except for $t=36$ and $t=39$. Compared to B-VARLasso, GNARID is better at eight time points except for $t=39$ and $t=43$.}

\subsection{\hlfour{Distribution of Weights Analysis}}

\begin{figure}[!h]
    \centering
    \resizebox{\textwidth}{!}{\includegraphics{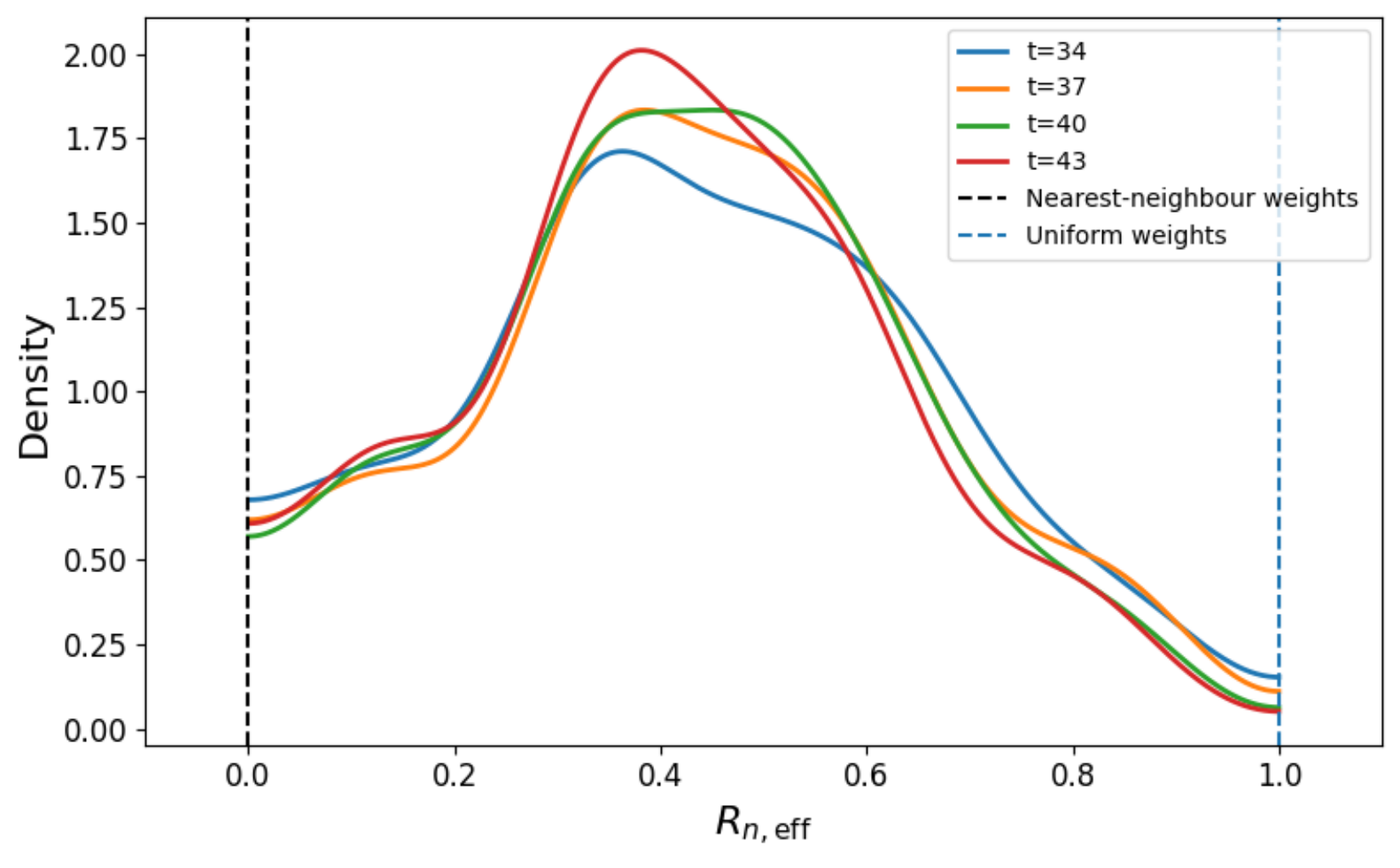}}
    \caption{Gaussian kernel density estimation of the distribution of the scaled effective-number-of-neighbours ratios over the 220 nodes for B-GNARID fitted at four different origins, $t=34,37,40,43$. The two dashed lines indicate two extremes of the ratio, with $R_{n,\text{eff}} = 0$ indicating only one node receives weight one and $R_{n,\text{eff}} = 1$ indicating all nodes received equal weights.}
    \footnotesize{\textbf{Alt text: }Density plot comparing scaled effective neighbour ratios for four forecasting origins. The four curves are similar and concentrated around intermediate values between nearest neighbour weighting and uniform weighting.}
    \label{fig:bruma_eff_neighbours}
\end{figure}

\hlfour{\guyfive{At this point, one might wonder whether $\lambda$ has more than
    a trivial influence on the weights? For example, roughly equal distances
    might just result in equivalence to the GNAR situation or highly
    unequal ones just take weight from one neighbour.} To assess the distribution of weights for each neighbourhood in the B-GNARID model, we compute the effective number of neighbours for each node using the classic order-2 Hill's number in \cite[page 428]{Hill1973},
\begin{equation}
    N_{n,\text{eff}} = \frac{1}{\sum_{m \in \mathcal{N}^{(1)}(n)} w(\hat{\lambda})_{n,m}^2}.
\end{equation}
Then compute the scaled effective-number-of-neighbours ratio \citep[equation (3)]{Jost2010},
\begin{equation}
    R_{n,\text{eff}} = \frac{N_{n,\text{eff}} -1 }{|\mathcal{N}^{(1)}(n)| - 1}.
\end{equation}
The ratio $R_{n,\text{eff}}$ provides a useful way to analyse the distribution of neighbour weights. If all neighbours are assigned uniform weights, then $R_{n,\text{eff}} = 1$. If only one nearest neighbour is assigned weight one while the others are assigned zero weights, then $R_{n,\text{eff}} = 0$. \cref{fig:bruma_eff_neighbours} plots the Gaussian kernel density estimation of the distribution of the ratio $R_{n,\text{eff}}$ over the 220 node neighbourhoods for B-GNARID fitted at origins $t=34,37,40,43$. The distributions at four origins are similar, and show a concentration approximately from $R_{n,\text{eff}} = 0.3$ to $R_{n,\text{eff}} = 0.6$, indicating that the estimated $\hat{\lambda}$ around $12.0$ to $13.0$ is in an intermediate regime. In other words, in many neighbourhoods, closer neighbours receive greater weights, while more distant neighbours do still retain some influence.}

\subsection{\hlfour{Sub-zone Sensitivity Analysis}}

\hlfour{To assess sensitivity on each sub-zone, we also fit the following models at each forecasting origin and validate their forecasting performance using the same metrics: B-AR-A-S, B-NAR-S, B-GNAR-S, B-GNARID-S, B-NAR-L, B-GNAR-L, B-GNARID-L and B-GNARID-$\infty$. The first four, B-AR-A-S, B-NAR-S, B-GNAR-S and B-GNARID-S, fit global-AR($2$), NAR($2$), GNAR($2,[1,1]$) and GNARID($2,[1,1],\lambda=\hat{\lambda}$) models to each sub-zone separately using the network restricted to each sub-zone for the network models. B-NAR-L, B-GNAR-L and B-GNARID-L fit NAR($2$), GNAR($2,[1,1]$) and GNARID($2,[1,1],\lambda=\hat{\lambda}$) models respectively to the full time series but allowing the autoregressive coefficients to be different for each sub-zone. B-GNARID-$\infty$ fits a GNARID($2,[1,1],\lambda=\infty$) model to the full time series. Both (-S) and (-L) models introduce heterogeneity between the sub-zones. However, (-S) models restrict each node to using information in its sub-zone, while in (-L) models, each node can still use information from other sub-zones. B-GNARID-$\infty$ is included here to test the performance of including only the nearest neighbours.}

\hlfour{\cref{tab:Bruma_SSE_sensitivity} reports the one-step-ahead predictive SSE averaged over the ten forecasting origins, in-sample fitting SSE at the last origin, node-wise coverage, node-average interval width, and node-average interval score averaged over the ten origins for each model used in the sensitivity analysis. It shows that accounting for sub-zone heterogeneity has different effects across the models. For B-NAR and B-GNAR, both sub-zone modifications give small improvements in predictive SSE, with fitting separate models to each sub-zone (-S) performing slightly better than allowing sub-zone-local coefficients in the full network (-L). In contrast, B-GNARID-S worsens the predictive SSE of B-GNARID from $3.49 \times 10^{-3}$ to $3.57 \times 10^{-3}$, while B-GNARID-L improves it slightly to $3.45 \times 10^{-3}$. The changes in interval scores follow a similar pattern. This suggests that heterogeneity between sub-zones is potentially useful for GNARID, but that restricting each node to information from its own sub-zone discards useful cross-sub-zone information. B-GNARID-$\infty$, which corresponds to the nearest-neighbour design, increases its predictive SSE to $3.66 \times 10^{-3}$. This is also only slightly better than the other extreme, B-NAR, which assigns equal weights to all neighbours. Together with \cref{fig:bruma_eff_neighbours}, this provides evidence that the fitted distance weighting lies meaningfully between uniform and nearest-neighbour weighting, and that retaining cross-sub-zone information is beneficial. } 

\begin{table}[h]
    \centering\hlfour{
    \setlength{\tabcolsep}{5pt}
    \begin{tabular}{cccccc}\hline
        Model & \begin{tabular}{c}Pred.\ SSE\\\guyfour{$\mbox{m}^2$}   \ ($\times 10^{-3}$)\end{tabular} & \begin{tabular}{c}Fit.\ SSE\\
        \guyfour{$\mbox{m}^2$}\ ($\times 10^{-2}$)\end{tabular} & Coverage & \begin{tabular}{c}Width, \guyfour{$\mbox{m}$}\\
        ($\times 10^{-2}$) \end{tabular} & \begin{tabular}{c}Score\\($\times 10^{-2}$) \end{tabular}\\
        \hline
        B-AR-A-S & 3.68 & 11.9 & 91.4\%  &1.34  & 2.01 \\
        B-NAR-S & 3.67  & 11.8 & 92.0\% & 1.35 & 2.00\\
        B-GNAR-S & 3.66  & 11.8 & 92.0\% & 1.35 & 2.00\\
        B-GNARID-S & 3.57  & 11.5 & 92.5\% & 1.36 & 1.95 \\
        B-NAR-L & 3.69 & 11.8 & 91.4\% & 1.35 & 2.00\\
        B-GNAR-L & 3.69  & 11.8 & 91.3\% & 1.35 & 2.00\\
        B-GNARID-L & 3.45  & 11.3 & 92.1\% & 1.34 & 1.91\\
        B-GNARID-$\infty$ & 3.66  & 11.9 & 90.9\% & 1.32 & 2.01\\
        \hline
    \end{tabular}}
    \caption{\hlfour{Average one-step predictive SSE, last origin in-sample fitting SSE, and node-wise coverage, average width, and average interval score for $95\%$ forecast intervals from the sensitivity analysis models fitted to the Brumadinho ground displacement time series.}}
    \label{tab:Bruma_SSE_sensitivity}
\end{table}

\section{Conclusion and Future Research}
\hengxu{
This \guytwo{article} introduces GNARID, \guytwo{an extension} of GNAR with distance-based, tunable weights. We establish key theoretical properties of the model, including the consistency and asymptotic normality of the estimators with an extension to the case where the number of nodes $N$ grows indefinitely, and validate these results through comprehensive simulation studies. \guy{
The misspecification section~\ref{sec:misspec} clearly shows that in our model setup, choosing $\lambda$ well results in 
better forecasts. More generally, it shows that it is worth considering a variety of weighting schemes so as to improve
forecasting.} We demonstrate the method on the Brumadinho ground displacement dataset and discover that the incorporation of tunable weight functions \hlfour{yields lower prediction errors}. \hlfour{This shows that even with a fixed network topology, the weights are meaningful for  time series prediction, and that allowing the weights to adapt through a tunable parameter can improve the forecasting performance.} Overall, our theoretical results and empirical findings demonstrate that GNARID provides a flexible and effective way to integrate auxiliary information into constructing weights in GNAR models.

The new GNARID model opens the door to further research. For example, a large exponent $\lambda$ effectively maps large distances to nearly zero weights, thus increasing sparsity. A more direct approach for sparsity control would be to remove edges with distances greater than some threshold as in the application. A systematic study of estimating the threshold could be valuable as it involves a more complex discrete loss. We could also allow $\lambda$ to vary across each stage of neighbourhood, offering finer control of penalisation. Another direction is to derive the exact forecasting error due to misspecification of $\lambda$ that accounts for its interaction with the autoregressive coefficients. Furthermore, it would be interesting to broaden the class of functions mapping distances to weights. \hlthree{For example, one could consider the polynomial functions, the exponential-polynomial functions, or the oscillatory functions discussed in \cref{sec:general_weight}. Other possible options are, for example, extending the current inverse distance weighting to its more general form $w_{n,m} \propto (1+\alpha d_{n,m})^{-\lambda}$, taking into account multivariate distances,} or using additive functions that combine multiple components to capture distinct mechanisms (e.g.\ a short range decay term plus a long range baseline term).

\guyfour{Another area for future
research would be to investigate
theoretically and empirically
the improvement in forecasting
error for more than one-step-ahead
forecasts. Such analysis will likely
depend on scrutiny of the eigenstructure
of the VAR companion matrix in~\eqref{eq:varcompanion},
the structure of which would be
considerably simpler for 
parsimonious GNAR(ID) models.
}}

\section*{Funding}
This work was supported by the Engineering and Physical Sciences Research Council NeST Programme Grant [EP/X002195/1 to G.N. and S.D.].

\section*{Acknowledgments}
Nason and Das gratefully acknowledge support from the NeST Programme Grant just mentioned.
Das received an internal Curtin University Grant to visit the U.K.
including Imperial College that supported this work.
Liu gratefully acknowledges support from Imperial College London.
\guyfour{We acknowledge with deep sympathy the loss of life and the profound impacts suffered by the communities affected by the Brumadinho dam collapse. Our aim is to improve monitoring and early-warning capabilities, thereby enhancing risk assessment and helping to prevent  future disasters.}

\section*{Data Availability Statement}

Deidentified data have been made available at the following URL: \hyperlink{https://github.com/Lhx1108/GNARID-python}{https://github.com/Lhx1108/GNARID-python}.


\bibliographystyle{comnet}
\bibliography{Ref}

@string{annstat = {Ann. Statist.}}

@string{cag = {Comput. Geosci.}}

@string{csda = {Comp. Stat. Data Anal.}}

@string{ijf={Int. J. Forecasting}}

@string{jrssb = {J. R. Statist. Soc. {\rm B}}}

@string{jss = {J. Statist. Soft.}}

@string{jts = {J. Time Ser. Anal.}}

@string{qje={Quart.\ J. Econ.}}

@string{statsin = {Stat. Sin.}}

@string{stprap = {Stoch. Process. Appl.}}

@article{GNAR,
 title={Generalized Network Autoregressive Processes and the {GNAR} Package},
 volume={96},
 url={https://www.jstatsoft.org/index.php/jss/article/view/v096i05},
 doi={10.18637/jss.v096.i05},
 number={5},
 journal=jss,
 author={Knight, Marina and Leeming, Kathryn and Nason, G. P. and Nunes, Matthew},
 year={2020},
 pages={1–36}
}

@article{estnonlinear,
title = {Estimation in nonlinear time series models},
journal = stprap,
volume = {21},
number = {2},
pages = {251-273},
year = {1986},
issn = {0304-4149},
doi = {https://doi.org/10.1016/0304-4149(86)90099-2},
url = {https://www.sciencedirect.com/science/article/pii/0304414986900992},
author = {Tjøstheim, D.}
}

@article{cls-asymptotic,
author = {Klimko, L. A. and Nelson, P. I.},
title = {{On Conditional Least Squares Estimation for Stochastic Processes}},
volume = {6},
journal = annstat,
number = {3},
publisher = {Institute of Mathematical Statistics},
pages = {629 -- 642},
year = {1978},
doi = {10.1214/aos/1176344207},
URL = {https://doi.org/10.1214/aos/1176344207}
}

@misc{brumadinho22,
    title="Brazil: Dam collapse in {B}rumadinho marks 3 years with families waiting for reparations, six victims still missing
and no one held responsible",
    year=2022,
    author="{Business and Human Rights Centre}",
    howpublished={\url{https://www.business-humanrights.org/en/latest-news/brazil-dam-collapse-in-brumadinho-marks-3-years-with-families-waiting-for-reparations-six-victims-still-missing-and-no-one-held-responsible/}},
    url="https://www.business-humanrights.org/en/latest-news/brazil-dam-collapse-in-brumadinho-marks-3-years-with-families-waiting-for-reparations-six-victims-still-missing-and-no-one-held-responsible/"}

@misc{guardian19,
    title={`{T}hat's going to burst': {B}razilian dam workers say they warned of disaster},
    author="{The Guardian}",
    year=2019,
    note="{P}ublished on 6th Feb 2019",
    howpublished={\url{https://www.theguardian.com/world/2019/feb/06/brazil-dam-collapse-workers-say-they-warned-owners}},
    url="https://www.theguardian.com/world/2019/feb/06/brazil-dam-collapse-workers-say-they-warned-owners"}

@misc{gnar_org,
    author = {Marina Knight and Guy P. Nason and Matthew Nunes},
    title = {Modelling, Detrending and Decorrelation of Network Time Series}, 
    year={2016},
    note={arXiv:1603.03221}}

@article{fvar,
    author = {Bernanke, Ben S. and Boivin, Jean and Eliasz, Piotr},
    title = {Measuring the Effects of Monetary Policy: A Factor-Augmented Vector Autoregressive ({FAVAR}) Approach},
    journal = qje,
    volume = {120},
    number = {1},
    pages = {387-422},
    year = {2005},
    month = {02},
    issn = {0033-5533},
    doi = {10.1162/0033553053327452},
    url = {https://doi.org/10.1162/0033553053327452},
    eprint = {https://academic.oup.com/qje/article-pdf/120/1/387/5442603/120-1-387.pdf},
}

@ARTICLE{BayesianVAR,
title = {Bayesian approach to evaluate the impact of external shocks on {R}ussian macroeconomics indicators},
author = {A.A. Shevelev},
year = {2017},
journal = {World of economics and management / Vestnik NSU. Series: Social and Economics Sciences},
volume = {17},
number = {1},
pages = {26-40},
url = {https://EconPapers.repec.org/RePEc:nos:wjflnh:2017_1_03e}
}

@incollection{mcfadden_conditional_1974,
	address = {New York},
	title = {Conditional logit analysis of qualitative choice behavior},
	language = {en},
	booktitle = {Frontiers in {Econometrics}},
	publisher = {Academic press},
	author = {McFadden, Daniel},
	editor = {Zarembka, Paul},
	year = {1974},
	pages = {105--142}
}

@article{LuWong08,
    title="An adaptive inverse-distance weighting spatial interpolation technique",
    author="Lu, G. Y. and Wong, D. W.",
    year=2008,
    journal=cag,
    volume=34,
    number=9,
    pages={1044-1055}}

@inproceedings{Shepard68,
    author ="Shepard, D." ,
    title = "A two-dimensional interpolation function for irregularly-spaced data",
    booktitle = "Proceedings of the 1968 23rd ACM national conference",
    year= 1968,
    pages="517--524"}

@article{JNS09,
	Author= {Jansen, M. and Nason, G. P. and Silverman, B. W.},
	Title={Multiscale methods for data on graphs and irregular multidimensional situations},
	Year=2009,
	Journal=jrssb,
	Volume=71,
	PAges="97--126"}

@phdthesis{leeming,
    author="Leeming, K.",
    year=2019,
    title="New methods in time series analysis",
    school="University of Bristol, Bristol, UK"}

@article{grebby21,
    title="Advanced analysis of satellite data reveals ground deformation precursors to the {B}rumadinho {T}ailings {D}am collapse",
    author="Grebby, S. and Sowter, A. and Gluyas, J. and Toll, D. and Gee, D. and Athab, A. and Girindran, R.",
    year=2021,
    journal="Communications Earth and Environment",
    volume=2,
    pages=2}

@article{Das24,
    author="Das, S. and Priyadarshana, A. and Grebby, S.",
    title="Monitoring the risk of a tailings dam collapse through spectral analysis of satellite {InSAR} time-series data.",
    journal="Stoch. Environ. Res. Risk Assess.",
    volume="38",
    pages="2911–2926",
    year="2024"}

@book{Hayashi2000,
  author    = {Hayashi, Fumio},
  title     = {Econometrics},
  year      = {2000},
  publisher = {Princeton University Press},
  address   = {Princeton, NJ},
  isbn      = {9780691010182}
}

@ARTICLE{Xu2023, 
    author={Xu, Chen and Xie, Yao}, 
    journal={IEEE Transactions on Pattern Analysis and Machine Intelligence}, 
    title={Conformal Prediction for Time Series}, 
    year={2023}, 
    volume={45}, 
    number={10}, 
    pages={11575-11587}}

@article{Gneiting07,
author = {Tilmann Gneiting and Adrian E Raftery},
title = {Strictly Proper Scoring Rules, Prediction, and Estimation},
journal = {Journal of the American Statistical Association},
volume = {102},
number = {477},
pages = {359--378},
year = {2007},
publisher = {Taylor \& Francis},
doi = {10.1198/016214506000001437},
URL = {https://doi.org/10.1198/016214506000001437},
eprint = {https://doi.org/10.1198/016214506000001437}
}

@article{Winkler72,
 ISSN = {01621459, 1537274X},
 URL = {http://www.jstor.org/stable/2284720},
 author = {Robert L. Winkler},
 journal = {Journal of the American Statistical Association},
 number = {337},
 pages = {187--191},
 publisher = {[American Statistical Association, Taylor & Francis, Ltd.]},
 title = {A Decision-Theoretic Approach to Interval Estimation},
 urldate = {2026-05-22},
 volume = {67},
 year = {1972}
}

@article{lindgren2011,
  author  = {Lindgren, Finn and Rue, H{\aa}vard and Lindstr{\"o}m, Johan},
  title   = {An explicit link between {Gaussian} fields and {Gaussian} {Markov} random fields: the stochastic partial differential equation approach},
  journal = {Journal of the Royal Statistical Society: Series B (Statistical Methodology)},
  year    = {2011},
  volume  = {73},
  number  = {4},
  pages   = {423--498},
  doi     = {10.1111/j.1467-9868.2011.00777.x}
}

@article{bolin2020,
  author  = {Bolin, David and Kirchner, Kristin},
  title   = {The rational {SPDE} approach for {Gaussian} random fields with general smoothness},
  journal = {Journal of Computational and Graphical Statistics},
  year    = {2020},
  volume  = {29},
  number  = {2},
  pages   = {274--285},
  doi     = {10.1080/10618600.2019.1665537}
}

@article{MCMR23,
author="Mantziou, A. and Cucuringui, M. and Meirinhos, V. and Reinert, G.",
title="The {GNAR}-edge model: a network autoregressive model for networks with time-varying edge weights",
journal="Journal of Complex Networks",
volume=11,
year=2023,
pages="cnad039"}

@article{PP24,
author="Pan, Y. and Pan, J.",
title="Threshold Network {GARCH} model",
journal=jts,
volume=45,
pages="910--930",
year=2024}

@article{AF24,
author="Armillotta, M. and Fokianos, K.",
title="Count Network Autoregression",
journal=jts,
volume=45,
year=2024,
pages="584--612"}

@article{VARlasso,
	author={Hsu, N.-J. and Hung, H.-L. and Chang, Y.-M.},
	year=2008,
	title={Subset selection for vector autoregressive processes using {L}asso},
	journal=csda,
	volume=52,
	pages={3645--3657}}

@inproceedings{JansenNasonSilverman01,
	Author = {Jansen, M. and Nason, G. P. and Silverman, B. W.},
	Booktitle = {Wavelet applications in signal and image processing. Proceedings of SPIE},
	Editor = {Unser, M. and Aldroubi, A.},
	Pages = {87--97},
	Title = {Scattered data smoothing by empirical {B}ayesian shrinkage of second generation wavelet coefficients},
	Volume = 4478,
	Year = 2001}

@incollection{NunesKnightNason15,
	author="Nunes, M. and Knight, M. and Nason, G.P.",
	title="Modelling and prediction of time series arising on a graph.",
	year=2015,
	booktitle={Modeling and Stochastic Learning for Forecasting in High Dimensions},
	Editor={Antoniadis, A. and Poggi, J.-M. and Brossat, X.},
	Pages={183--192},
	publisher={Springer},
	volume={217},
	Series={Lecture Notes in Statistics}}

@article{zhu2017,
author = {Xuening Zhu and Rui Pan and Guodong Li and Yuewen Liu and Hansheng Wang},
title = {{Network vector autoregression}},
volume = {45},
journal = annstat,
number = {3},
publisher = {Institute of Mathematical Statistics},
pages = {1096 -- 1123},
year = {2017},
doi = {10.1214/16-AOS1476},
URL = {https://doi.org/10.1214/16-AOS1476}
}

@book{quenouille:multiple,
author="Quenouille, M. H.",
title="The Analysis of Multiple Time Series",
year=1957,
publisher="Griffin",
address="London"}

@article{Nason26,
title="Forecasting {UK} consumer price inflation with {RaGNAR}: {R}andom generalised network autoregressive processes",
author="Nason, G. P. and Palasciano, H.",
year=2026,
journal=ijf,
volume=42,
pages="181--202"}

@article{ZDPW20,
author="Zhou, J. and Li, D. and Pan, R. and Wang, H.",
title="Network {GARCH} model",
journal=statsin,
volume=30,
year=2020,
pages="1723--1740"}

@article{AF23,
author = {Armillotta, M. and Fokianos, K.},
title = {{Nonlinear network autoregression}},
volume = {51},
journal = {The Annals of Statistics},
number = {6},
publisher = {Institute of Mathematical Statistics},
pages = {2526 -- 2552},
year = {2023},
doi = {10.1214/23-AOS2345},
URL = {https://doi.org/10.1214/23-AOS2345}
}

@misc{Liu23,
      title={New Methods for Network Count Time Series}, 
      author={Liu, H. and Nason, G. P.},
      year={2023},
    note={arXiv:2312.01944},
      eprint={2312.01944},
      archivePrefix={arXiv},
      primaryClass={stat.ME},
      url={https://arxiv.org/abs/2312.01944}, 
}

@article{FCNAR,
  title   = {A Functional Coefficients Network Autoregressive Model},
  author  = {Yin, Hang and Safikhani, Abolfazl and Michailidis, George},
  journal = {Statistica Sinica},
  year    = {2026},
  volume  = {36},
  number  = {1},
  pages   = {305--329},
  doi     = {10.5705/ss.202022.0402}
}

@techreport{song2011,
  author      = {Song, Song and Bickel, Peter J.},
  title       = {Large Vector Auto Regressions},
  institution = {Humboldt University Berlin, Collaborative Research Center 649},
  type        = {SFB 649 Discussion Paper},
  number      = {2011-048},
  year        = {2011},
  url         = {https://arxiv.org/abs/1106.3915}
}

@misc{boetti2025,
      title={Long memory network time series}, 
      author={Chiara Boetti and Matthew A. Nunes and Marina I. Knight},
      year={2025},
      eprint={2512.10446},
      archivePrefix={arXiv},
      primaryClass={stat.ME},
      url={https://arxiv.org/abs/2512.10446}, 
      note={arXiv:2512.10446}
}

@misc{nason2025usa,
      title={Generalized network autoregressive modelling of longitudinal networks with application to presidential elections in the {USA}}, 
      author={Guy P. Nason and Daniel Salnikov and Mario Cortina-Borja},
      year={2025},
      eprint={2503.10433},
      archivePrefix={arXiv},
      primaryClass={stat.ME},
      url={https://arxiv.org/abs/2503.10433},
      note={arXiv:2503.10433}
}

@misc{ren2026gmnar,
      title={Matrix-valued Network Autoregression Model with Latent Group Structure}, 
      author={Yimeng Ren and Xuening Zhu and Ganggang Xu and Yanyuan Ma},
      year={2026},
      eprint={2212.02107},
      archivePrefix={arXiv},
      primaryClass={stat.ME},
      url={https://arxiv.org/abs/2212.02107}, 
      note={arXiv:2212.02107}
}

@article{Ren2024gtnar,
    author = {Yi Ren and Xuening Zhu and Ganggang Xu and Yanyuan Ma},
    title = {Multi-relational Network Autoregression Model with Latent Group Structures},
    journal = {Journal of Machine Learning Research},
    year = 2026,volume = {27},
number = {78},
pages  = {1--135},
}

@article{jiang2025twoway,
  author  = {Jiang, Binyan and Leng, Chenlei and Yan, Ting and
             Yao, Qiwei and Yu, Xinyang},
  title   = {A Two-Way Heterogeneity Model for Dynamic Networks},
  journal = {The Annals of Statistics},
  year    = {2025},
  volume  = {53},
  number  = {6},
  pages   = {2617--2641},
  doi     = {10.1214/25-AOS2557}
}

@article{chang2026,
    author = {Chang, Jinyuan and Fang, Qin and Kolaczyk, Eric D and MacDonald, Peter W and Yao, Qiwei},
    title = {Autoregressive networks with dependent edges},
    journal = {Journal of the Royal Statistical Society Series B: Statistical Methodology},
    pages = {qkag063},
    year = {2026},
    month = {04},
    issn = {1369-7412},
    doi = {10.1093/jrsssb/qkag063},
    url = {https://doi.org/10.1093/jrsssb/qkag063},
    eprint = {https://academic.oup.com/jrsssb/advance-article-pdf/doi/10.1093/jrsssb/qkag063/68094621/qkag063.pdf},
}

@article{jiang2023autoregressive,
  author  = {Jiang, Binyan and Li, Jialiang and Yao, Qiwei},
  title   = {Autoregressive Networks},
  journal = {Journal of Machine Learning Research},
  year    = {2023},
  volume  = {24},
  number  = {227},
  pages   = {1--69}
}

@article{Panaretos2019,
   author = "Panaretos, Victor M. and Zemel, Yoav",
   title = "Statistical Aspects of {Wasserstein} Distances", 
   journal= "Annual Review of Statistics and Its Application",
   year = "2019",
   volume = "6",
   pages = "405-431",
   doi = {10.1146/annurev-statistics-030718-104938},
   url = "https://www.annualreviews.org/content/journals/10.1146/annurev-statistics-030718-104938",
   publisher = "Annual Reviews",
   issn = "2326-831X",
   type = "Journal Article",
  }

@article{Hill1973,
author = {Hill, M. O.},
title = {Diversity and Evenness: A Unifying Notation and Its Consequences},
journal = {Ecology},
volume = {54},
number = {2},
pages = {427-432},
doi = {https://doi.org/10.2307/1934352},
url = {https://esajournals.onlinelibrary.wiley.com/doi/abs/10.2307/1934352},
eprint = {https://esajournals.onlinelibrary.wiley.com/doi/pdf/10.2307/1934352},
year = {1973}
}

@article{kwiatkowski1992,
  title={Testing the Null Hypothesis of Stationarity against the Alternative of a Unit Root: How Sure Are We That Economic Time Series Have a Unit Root?},
  author={Kwiatkowski, Denis and Phillips, Peter C. B. and Schmidt, Peter and Shin, Yongcheol},
  journal={Journal of Econometrics},
  volume={54},
  number={1--3},
  pages={159--178},
  year={1992},
  doi={10.1016/0304-4076(92)90104-Y}
}

@article{newey1987,
  title={A Simple, Positive Semi-Definite, Heteroskedasticity and Autocorrelation Consistent Covariance Matrix},
  author={Newey, Whitney K. and West, Kenneth D.},
  journal={Econometrica},
  volume={55},
  number={3},
  pages={703--708},
  year={1987},
  doi={10.2307/1913610}
}

@article{benjamini2001,
  title={The Control of the False Discovery Rate in Multiple Testing under Dependency},
  author={Benjamini, Yoav and Yekutieli, Daniel},
  journal={The Annals of Statistics},
  volume={29},
  number={4},
  pages={1165--1188},
  year={2001},
  doi={10.1214/aos/1013699998}
}

@Article{Jost2010,
    AUTHOR = {Jost, Lou},
    TITLE = {The Relation between Evenness and Diversity},
    JOURNAL = {Diversity},
    VOLUME = {2},
    YEAR = {2010},
    NUMBER = {2},
    PAGES = {207--232},
    URL = {https://www.mdpi.com/1424-2818/2/2/207},
    ISSN = {1424-2818},
    DOI = {10.3390/d2020207}
}
%








\end{document}


\title{Appendix for ``Improved Forecasting for Multivariate Time Series
    using GNAR models with General Inverse Distance Weights''}

\shorttitle{Improved Forecasting with GNARID Appendix} 
\shortauthorlist{Liu et al.} 

\author{
\name{Hengxu Liu$^*$}
\address{Department of Mathematics, Imperial College London, Huxley Building, 180 Queen's Gate, London SW7 2AZ, UK.\email{$^*$Corresponding author: hengxu.liu18@imperial.ac.uk}}
\name{Guy P. Nason}
\address{Department of Mathematics, Imperial College London, Huxley Building, 180 Queen's Gate, London SW7 2AZ, UK.}
\name{Sourav Das}
\address{School of Electrical Engineering, Computing and Mathematical Sciences, Curtin University, Perth, WA 6102, Australia.}
\and
\name{Stephen Grebby}
\address{Terra Motion Limited, Ingenuity Centre, Triumph Road, Nottingham NG7 2TU.}
}

\maketitle
\appendix
\renewcommand{\thelemma}{\thesection.\arabic{lemma}}
\renewcommand{\theequation}{\thesection.\arabic{equation}}
\section{Proofs}
\begin{proof}[Proof of \sref{lem:w_lambda_infty}]
    For all $x_i \neq \min(x)$,
    \begin{equation}\label{eq:lem:w_lambda_infty_1}
        \frac{x_i^{-\lambda}}{\sum_j x_j^{-\lambda}} <  \frac{x_i^{-\lambda}}{\min(x)^{-\lambda}} = \left(\frac{x_i}{\min(x)}\right)^{-\lambda} \rightarrow 0.
    \end{equation}
    For all $x_i$,
    \begin{equation}
         \frac{\partial}{\partial \lambda}\frac{x_i^{-\lambda}}{\sum_j x_j^{-\lambda}} = \frac{x_i^{-\lambda} \sum_j x_j^{-\lambda} (\log x_j - \log x_i)}{(\sum_j x_j^{-\lambda})^2}. 
    \end{equation}
    For $i$ such that $x_i \neq \min(x)$,
    \begin{equation}
        |w_i'(\lambda)| < w_i(\lambda) \frac{\max_k(|\log x_k- \log x_i|) \sum_j x_j^{-\lambda}}{\sum_j x_j^{-\lambda}} = C w_i(\lambda) \rightarrow 0
    \end{equation}
    by \eqref{eq:lem:w_lambda_infty_1}. Let $A$ be such that if $i \in A$, $x_i = \min(x)$. For $i \in A$, 
    \begin{equation}
        |w_i'(\lambda)| < w_i(\lambda) \frac{\max_k(|\log x_k- \log x_i|) \sum_{j \notin A} x_j^{-\lambda}}{\sum_j x_j^{-\lambda}} = w_i(\lambda) C \sum_{j \notin A} w_j(\lambda) \rightarrow 0
    \end{equation}
    by \eqref{eq:lem:w_lambda_infty_1}.
    
\end{proof}

\begin{proof}[Proof of \sref{lem:ma_form}]
    \hltwo{\sref{as:stationary} states} that $\rho_{\max}(A) < (1+c)^{-1}$ for some $c>0$ which implies that the linear mapping $A(1+c)$ is eventually contracting by \cref{lem:contraction}, i.e., there exists $l_0 > 0$, such that for all $q$,
    \begin{align}
        &\| A^l (1+c)^l \|_q< 1 \ , \ \ \forall \ l>l_0 \\
        \implies &\| A^l \|_q < (1+c)^{-l} \ , \ \ \forall \ l>l_0 \\
        \implies &\| A^l \|_q < C(1+c)^{-l} \ , \ \ \forall \ l,
    \end{align}
    where we for example let $C=\max_l \{ \|A^l\|_q (1+c)^l \}$. Now, 
    \begin{align}\label{eq:companion_ma}
        \bar{\mathbf{X}}_{t} &= A \bar{\mathbf{X}}_{t-1} + \bar{\epsilon}_{t}\\
        &= A (A\bar{\mathbf{X}}_{t-2} + \bar{\epsilon}_{t}) + \bar{\epsilon}_{t}\\
        &=\dots\\
        &= \sum_{l=0}^\infty A^l \bar{\epsilon}_{t-l},
    \end{align}
    or equivalently,
    \begin{align}\label{eq:original_ma}
        \mathbf{X}_{t} &= \sum_{l=0}^\infty [A^l]_{1:N,1:N} \epsilon_{t-l}\\
        &= \sum_{l=0}^\infty a_l \epsilon_{t-l}
    \end{align}
    with
    \begin{equation}
        \sum_{l=0}^\infty \|a_l\|_q < C\sum_{l=0}^\infty (1+c)^{-l} < C.
    \end{equation}
\end{proof}

\hlthree{\begin{proof}[Proof of \sref{prop:stationary}]
    With the causal representation in \sref{lem:ma_form} and the fact that $\epsilon_t$ are i.i.d. over time, $\{\mathbf{X}_t \}$ is strictly stationary and ergodic by Proposition 6.1 (d) in \cite{Hayashi2000}.
    
\end{proof}}

\begin{proof}[Proof of \sref{lem:cov_finite}]
    From \sref{lem:ma_form}, we know that
    \begin{align}
        \cov(\mathbf{X}_t,\mathbf{X}_t) &= \E \left( \sum_{l=0}^\infty a_l \epsilon_{t-l} \sum_{l=0}^\infty (a_l \epsilon_{s-l})' \right)\\
        &= \sum_{l=0}^{\infty} a_{l} \Sigma_\epsilon {a_{l}}'
    \end{align}
    and $\|a_l\|_2 < C(1+c)^{-l}$ for all $l$ for some $c>0$, then
    \begin{equation}
        \|\cov(\mathbf{X}_t,\mathbf{X}_t)\|_2 \leq \sum_{l=0}^{\infty} \|a_{l}\|_2 \|\Sigma_\epsilon\|_2 \|{a_{l}}'\|_2 < C\|\Sigma_\epsilon\|_2 \sum_{l=0}^\infty (1+c)^{-2l} < C.
    \end{equation}
    Let $B \in \mathbb{R}^{pN \times N}$ be 
    \begin{equation}
        B = \begin{pmatrix}
            I_N\\
            \vdots\\
            0
        \end{pmatrix}.
    \end{equation}
    Then $\bar{\epsilon}_t = B\epsilon_t$. \eqref{eq:companion_ma} is then equivalent to
    \begin{equation}
        \bar{\mathbf{X}}_{t} = \sum_{l=0}^\infty A^l B\epsilon_{t-l},
    \end{equation}
    which implies that 
    \begin{equation}
        \cov(\bar{\mathbf{X}}_{t},\bar{\mathbf{X}}_{t}) = \sum_{l=0}^\infty A^l B \Sigma_\epsilon B' {A'}^l
    \end{equation}
    which converges since $\rho_{\max}(A)<1$. For any $v \neq 0 \in \mathbb{R}^{pN}$, write $v$ as $v = (v_1',v_2',\dots,v_p')'$. The form of $A$ implies that $B' {A'}^l v = v_{l+1} + \sum_{i=1}^l D_{l,i} v_i$ for $l=0,\dots,p-1$ where each $D_{l,i}$ is a matrix built from finite sum and product of $A_j$. Take $j \in {1,\dots,p}$ to be such that $v_j \neq 0$ and $v_{j+1} = \dots = v_p = 0$. We can show by induction that $\sum_{l=1}^j \|v_l\|_2 \leq (1+\max_{l,i}\|D_{l,i}\|_2)^p \sum_{l=0}^{j-1}\|B' {A'}^l v \|_2 $. Therefore, $\|v\|_2^2 \leq C \sum_{l=0}^{p-1}\|B' {A'}^l v \|_2^2$. Thus, by \sref{as:sigma_epsilon},
    \begin{align}
        v' \cov(\bar{\mathbf{X}}_{t},\bar{\mathbf{X}}_{t}) v &= \sum_{l=0}^\infty (B' {A'}^l v)' \Sigma_\epsilon B' {A'}^l v \\
        &\geq c \sum_{l=0}^{p-1}\|B' {A'}^l v \|_2^2 \geq c\|v\|_2^2/C.
    \end{align}
    Hence, $\cov(\bar{\mathbf{X}}_{t},\bar{\mathbf{X}}_{t}) \succ c$.
    
\end{proof}


\begin{proof}[Proof of \sref{lem:lin_indep}]
    For a general order ($p,[s_1,\dots,s_p]$) model, suppose
    \begin{equation}
        \E \left\|\sum_{j=1}^p \left(a_{\alpha_j}\frac{\partial \hat{\mathbf{X}}_{t|t-1}}{\partial \alpha_j}\bigg|_{\theta^0} + \sum_{r=1}^{s_j}a_{\beta_{j,r}}\frac{\partial \hat{\mathbf{X}}_{t|t-1}}{\partial \beta_{j,r}}\bigg|_{\theta^0}\right) + a_\lambda\frac{\partial \hat{\mathbf{X}}_{t|t-1}}{\partial \lambda}\bigg|_{\theta^0} \right\|^2 = 0,
    \end{equation}
    where for each entry, this implies for all $n$,
    \begin{align}\label{eq:entry_E_squared=0}
    \begin{split}
        &\E \Bigg|\sum_{j=1}^p \left(a_{\alpha_j} X_{t-j,n} + \sum_{r=1}^{s_j}a_{\beta_{j,r}} \sum_{m \in \mathcal{N}^{(r)}(n)} w_{n,m}(\lambda^0) X_{t-j,m}\right) + \\
        &a_\lambda \sum_{j=1}^p\sum_{r=1}^{s_j} \beta_{j,r}^0 \sum_{m \in \mathcal{N}^{(r)}(n)}\frac{\partial w_{n,m}}{\partial \lambda}\bigg|_{\lambda^0}X_{t-j,m}\Bigg|^2 = 0.
    \end{split}
    \end{align}
    We know from \sref{lem:cov_finite} that $\cov(\bar{\mathbf{X}}_{t-1},\bar{\mathbf{X}}_{t-1}) \succ 0$. Then by \cref{lem:lin_indep_pd}, we have that for all $j=1,\dots,p$, $r=1,\dots,s_j$ and $m \in \mathcal{N}^{(r)}(n)$,
    \begin{gather}
        a_{\alpha_j} = 0,\\
        a_{\beta_{j,r}} w_{n,m}(\lambda^0) +
        a_{\lambda} \beta_{j,r}^0 \frac{\partial w_{n,m}}{\partial \lambda}\bigg|_{\lambda^0} = 0.
    \end{gather}
    Under \sref{as:dist}, \cref{lem:dist_to_indep} implies that $a_{\beta_{j,r}} = a_\lambda = 0$ for at least one set of $j,r$. Substituting $a_\lambda = 0$ into the other equations gives $a_{\alpha_j}=a_{\beta_{j,r}}=a_{\lambda}=0$ for all $j,r$.
    
\end{proof}

\hlthree{\begin{proof}[Proof of \sref{prop:consistency}]
    Define 
     \begin{equation}
        l(\theta) = \E \| \mathbf{X}_t - \hat{\mathbf{X}}_{t|t-1}(\theta) \|_2^2.
    \end{equation}
    Since $\mathbf{X}_t = \hat{\mathbf{X}}_{t|t-1}(\theta^0) + \epsilon_t$ and $\E (\epsilon_t|\mathcal{F}_{t-1})=0$,
    \begin{equation}
        l(\theta) = l(\theta^0) + \E \| \hat{\mathbf{X}}_{t|t-1}(\theta^0) - \hat{\mathbf{X}}_{t|t-1}(\theta) \|_2^2.
    \end{equation}
    Hence $l(\theta) = l(\theta^0)$ only if $\E \| \hat{\mathbf{X}}_{t|t-1}(\theta^0) - \hat{\mathbf{X}}_{t|t-1}(\theta) \|_2^2 = 0$ or equivalently $\hat{\mathbf{X}}_{t|t-1}(\theta^0) = \hat{\mathbf{X}}_{t|t-1}(\theta)$ almost surely. By \sref{lem:cov_finite}, this implies that for each $j$, $A_j(\theta) = A_j(\theta^0)$.

    The diagonal entries give $\alpha_j = \alpha_j^0$. The off-diagonal stage-$r$ entries give $\beta_{j,r}W^{(r)}(\lambda) = \beta_{j,r}^0W^{(r)}(\lambda^0)$. Since the weights are row normalised, this gives $\beta_{j,r} = \beta^0_{j,r}$. For the pair $(j,r)$ in \sref{as:dist}, this implies $W^{(r)}(\lambda) = W^{(r)}(\lambda^0)$. For the node $n$ and neighbours $m_1,m_2$ in \sref{as:dist}, we have for finite $\lambda$,
    \begin{equation}
        \frac{w_{n,m_1}(\lambda)}{w_{n,m_2}(\lambda)} = \frac{w_{n,m_1}(\lambda^0)}{w_{n,m_2}(\lambda^0)} \implies (\frac{d_{n,m_1}}{d_{n,m_2}})^{-\lambda} = (\frac{d_{n,m_1}}{d_{n,m_2}})^{-\lambda^0}. 
    \end{equation}
    As $d_{n,m_1} \neq d_{n,m_2}$, this implies $\lambda = \lambda^0$. Since we require compactness later, $\lambda$ is considered on the extended reals. For the compactified boundary $\lambda = \pm \infty$, at most one of $w_{n,m_1}(\lambda)$ and $w_{n,m_2}(\lambda)$ is non-zero, but $w_{n,m_1}(\lambda^0)$ and $w_{n,m_2}(\lambda^0)$ are all strictly positive. Reversing the logic above gives that $\lambda = \pm \infty$ cannot be a minimiser of $l(\theta)$. Thus $\theta^0$ is the unique minimiser of $l(\theta)$.

    Next, expanding $L_T(\theta)$ gives
    \begin{equation}\label{eq:loss_expansion}
        L_T(\theta) = \frac{1}{T}\sum_{t=1}^T \|\mathbf{X}_t \|_2^2 - 2\tr\{\Tilde{A}(\theta)(\frac{1}{T}\sum_{t=1}^T \mathbf{X_t}\bar{\mathbf {X}}_{t-1}')'\}+ \tr\{\Tilde{A}(\theta) (\frac{1}{T}\sum_{t=1}^T \bar{\mathbf{X}}_{t-1}\bar{\mathbf{X}}_{t-1}') \Tilde{A}(\theta)'\}
    \end{equation}
    where
    \begin{equation}
        \Tilde{A}(\theta) = (A_1(\theta),\dots,A_p(\theta)).
    \end{equation}
    By \sref{lem:cov_finite} and the ergodic theorem, 
    \begin{equation}
        \frac{1}{T}\sum_{t=1}^T \bar{\mathbf{X}}_{t-1}\bar{\mathbf{X}}_{t-1}' \rightarrow \E (\bar{\mathbf{X}}_{t}\bar{\mathbf{X}}_{t}') \succ 0 \quad a.s.
    \end{equation}
    Hence, for sufficiently large $T$, $\frac{1}{T}\sum_{t=1}^T \bar{\mathbf{X}}_{t-1}\bar{\mathbf{X}}_{t-1}' \succ c$ almost surely. Therefore, by Cauchy Schwarz,
    \begin{equation}
        L_T(\theta) \geq \frac{1}{T}\sum_{t=1}^T \|\mathbf{X}_t \|_2^2 - 2\|\Tilde{A}(\theta)\|_F\|\frac{1}{T}\sum_{t=1}^T \mathbf{X_t}\bar{\mathbf {X}}_{t-1}'\|_F + c \|\Tilde{A}(\theta)\|_F^2
    \end{equation}
    By the ergodic theorem again, $\frac{1}{T}\sum_{t=1}^T \mathbf{X_t}\bar{\mathbf {X}}_{t-1}' \rightarrow \E \mathbf{X_t}\bar{\mathbf {X}}_{t-1}'$ almost surely. Therefore, almost surely, with sufficiently large $T$,
    \begin{equation}
        L_T(\theta) \geq -C\|\Tilde{A}(\theta)\|_F + c\|\Tilde{A}(\theta)\|_F^2
    \end{equation}
    with a finite constant $C$. The right-hand side is a quadratic in $\|\Tilde{A}(\theta)\|_F$ with quadratic coefficient $c > 0$. The ergodic theorem also implies that $L_T(\theta^0) \rightarrow l(\theta^0)$ almost surely. Hence, there exists a constant $M'$ such that $\|\Tilde{A}(\theta)\|_F \geq M' \implies L_T(\theta) > L_T(\theta^0)$ for $T$ large enough almost surely. Since $I_N, W^{(r)}(\lambda), r=1,\dots,\max_j(s_j)$ all have disjoint supports,
    \begin{equation}
        \|\Tilde{A}(\theta)\|_F^2 = \sum_{j=1}^p \alpha_j^2 \|I_N\|_F^2 + \sum_{j=1}^p \sum_{r=1}^{s_j} \beta_{j,r}^2\|W^{(r)}(\lambda) \|_F^2
    \end{equation}
    Thus, with the row-normalisation of $W^{(r)}(\lambda)$, $\|\Tilde{A}(\theta)\|_F^2 \geq c'\sum_{j=1}^p(\alpha_j^2 +\sum_{r=1}^{s_j} \beta_{j,r}^2) = c'\|(\alpha,\beta)\|_2^2$ for some $c'>0$ uniformly in $\lambda$, where $\alpha$ and $\beta$ are collection vectors of $\alpha_j$ and $\beta_{j,r}$ respectively. Therefore, there exists a finite constant $M$ such that $\|(\alpha,\beta)\|_2 \geq M \implies L_T(\theta) > L_T(\theta^0)$. Equivalently, for a global minimiser, $\|(\alpha,\beta)\|_2 \leq M$.

    For the $\lambda$ direction, we compactify it with $\lambda = \pm \infty$. Because the distances are finite and strictly positive, the weights have finite limits as $\lambda \rightarrow \pm \infty$. Therefore, functions $\Tilde{A}(\theta)$, $W^{(r)}(\lambda)$, $l(\theta)$, and $L_T(\theta)$ extend continuously to $B = \{\theta: \|(\alpha,\beta)\| \leq M, \lambda \in [-\infty,\infty] \}$ which is compact. Thus, the function $L_T(\theta)$ attains its minimum on $B$ eventually almost surely. 

    Now, on $B$, each entry in $\Tilde{A}(\theta)$ is a bounded continuous function of $\theta$. Then, $L_T(\theta) - l(\theta)$ can be written as a finite sum of terms of the form $f(\theta)(\frac{1}{T}\sum_{t=1}^T X_{t-i,n}X_{t-j,m}-\E X_{t-i,n}X_{t-j,m})$ where $f(\theta)$ is bounded on $B$. By ergodic theorem, we have the uniform convergence
    \begin{equation}\label{eq:prop:consistency:uni_conv_B}
        \sup_{\theta \in B} |L_T(\theta) - l(\theta)| \rightarrow 0 \quad a.s.
    \end{equation}
    Since we have shown that $\lambda = \pm \infty$ cannot minimise $l$, with compactness of $B$ and continuity of $l$,
    \begin{equation}
         \inf_{\theta\in B: \lambda =\pm \infty} l(\theta) > l(\theta^0).
    \end{equation}
    The uniform convergence thus implies that for sufficiently large $T$,
    \begin{equation}
         \inf_{\theta\in B: \lambda =\pm \infty} L_T(\theta) > L_T(\theta^0), \quad a.s.
    \end{equation}
    Since $L_T(\theta)$ attains its minimum on $B$ and not at $\lambda = \pm \infty$, $G_T$ is non-empty for large enough $T$ almost surely.

    Let $d(\theta_1,\theta_2)$ be a distance well defined on $B$ that preserves convergence of the usual Euclidean distance. For example, $d(\theta_1,\theta_2) = \|(\alpha_1,\beta_1) - (\alpha_2,\beta_2) \|_2 + |\arctan(\lambda_1) - \arctan(\lambda_2)|$ with $\arctan(\pm \infty) = \pm \pi/2$.
    Fix $\epsilon > 0$, since we have proved that $l(\theta)$ attains its unique global minimum at $\theta^0$, compactness of $B$ and continuity of $l(\theta)$ give 
    \begin{equation}\label{eq:prop:consistency:population_squeeze}
        \inf_{\theta\in B: d(\theta,\theta^0) \geq \epsilon} l(\theta) > l(\theta^0).
    \end{equation}
    The uniform convergence in \eqref{eq:prop:consistency:uni_conv_B} thus implies that for sufficiently large $T$,
    \begin{equation}
        \inf_{\theta\in B: d(\theta,\theta^0) \geq \epsilon} L_T(\theta) > L_T(\theta^0), \quad a.s.
    \end{equation}
    As $\epsilon$ is arbitrary, we get
    \begin{equation}\label{eq:prop:consistency:consistency}
        \hat{\theta}_T \overset{a.s.}{\to} \theta^0.
    \end{equation}
    We now show that $G_T$ is eventually a singleton. The Hessian of $l(\theta)$ at $\theta^0$ is
    \begin{equation}
        \frac{\partial^2 l(\theta^0)}{\partial \theta \partial \theta'} = 2U.
    \end{equation}
    By \sref{lem:lin_indep}, $U \succ 0$ (see the proof of Theorem 3.1 in \cite{estnonlinear}). By continuity, there exists a neighbourhood $B_0$ of $\theta^0$ with $B_0 \subset \Omega$ in \cref{lem:cross_norm_finite} such that for all $\theta \in B_0$,
    \begin{equation}
        \frac{\partial^2 l(\theta)}{\partial \theta \partial \theta'} \succ \delta
    \end{equation}
    for some $\delta>0$. Each entry in $\frac{\partial^2 L_T(\theta)}{\partial \theta \partial \theta'}$ is of the form 
    \begin{equation}
        \frac{\partial^2 L_T(\theta)}{\partial \theta_1 \partial \theta_2} = 2 \frac{1}{T}\sum_{t=1}^T \{\frac{\partial\hat{\mathbf{X}}_{t|t-1}}{\partial \theta_1}' \frac{\partial\hat{\mathbf{X}}_{t|t-1}}{\partial \theta_2} -  (\mathbf X_t - \hat{\mathbf{X}}_t)' \frac{\partial^2\hat{\mathbf{X}}_{t|t-1}}{\partial \theta_1 \partial \theta_2}\}.
    \end{equation}
    Using the form of the derivatives, \sref{lem:cov_finite} and the ergodic theorem, we have that for each $\theta \in B_0$,
    \begin{equation}\label{eq:prop:consistency:point_conv_B0}
        \frac{\partial^2 L_T(\theta)}{\partial \theta \partial \theta'} \rightarrow \frac{\partial^2 l(\theta)}{\partial \theta \partial \theta'}, \quad a.s.
    \end{equation}
    Each entry of the derivative of $\frac{\partial^2 \|X_t - \hat{X}_{t|t-1}\|^2}{\partial \theta \partial \theta'}$ with respect to each variable is a sum of the two types of terms in \cref{lem:cross_norm_finite}. Using \cref{lem:cross_norm_finite} and the mean value theorem, we have
    \begin{equation}
        \left\|\frac{\partial^2 \|X_t - \hat{X}_{t|t-1}\|^2}{\partial \theta \partial \theta'}\bigg|_{\theta_1} -  \frac{\partial^2 \|X_t - \hat{X}_{t|t-1}\|^2}{\partial \theta \partial \theta'}\bigg|_{\theta_2} \right\| \leq Q(X_t,\dots,X_{t-p})\|\theta_1 -\theta_2 \|
    \end{equation}
    where $\E Q < \infty$. Then,
    \begin{equation}
        \|\frac{\partial^2 L_T(\theta_1)}{\partial \theta \partial \theta'} -  \frac{\partial^2 L_T(\theta_2)}{\partial \theta \partial \theta'} \| \leq \frac{1}{T}\sum_{t=1}^T Q(X_t,\dots,X_{t-p})\|\theta_1 -\theta_2 \|
    \end{equation}
    Similarly,
    \begin{equation}
        \|\frac{\partial^2 l(\theta_1)}{\partial \theta \partial \theta'} -  \frac{\partial^2 l(\theta_2)}{\partial \theta \partial \theta'} \| \leq \E Q(X_t,\dots,X_{t-p}) \|\theta_1 -\theta_2 \|.
    \end{equation}
    Using ergodic theorem,
    \begin{equation}
         \frac{1}{T}\sum_{t=1}^T Q(X_t,\dots,X_{t-p}) \rightarrow \E Q(X_t,\dots,X_{t-p}), \quad a.s.
    \end{equation}
    Then obviously, for sufficiently large $T$,
    \begin{equation}
         \frac{1}{T}\sum_{t=1}^T Q(X_t,\dots,X_{t-p}) \leq \E Q(X_t,\dots,X_{t-p}) +1 \quad a.s.
    \end{equation}
    Take $B_0' \subset B_0$ to be such that for all $\theta \in B_0'$,
    \begin{equation}\label{eq:prop:consistency:after_B0prime}
        \|\theta-\theta^0\| < \frac{\delta}{4\{2\E Q(X_t,\dots,X_{t-p}) +1\}}
    \end{equation}
    Then for all $\theta \in B_0'$,
    \begin{align}
        \|\frac{\partial^2 L_T(\theta)}{\partial \theta \partial \theta'} - \frac{\partial^2 l(\theta)}{\partial \theta \partial \theta'}\|&\leq \|\frac{\partial^2 L_T(\theta)}{\partial \theta \partial \theta'} - \frac{\partial^2 L_T(\theta^0)}{\partial \theta \partial \theta'}\| + \|\frac{\partial^2 L_T(\theta^0)}{\partial \theta \partial \theta'} - \frac{\partial^2 l(\theta^0)}{\partial \theta \partial \theta'}\| +
        \|\frac{\partial^2 l(\theta)}{\partial \theta \partial \theta'} - \frac{\partial^2 l(\theta^0)}{\partial \theta \partial \theta'}\|\\
        &\leq \{2\E Q(X_t,\dots,X_{t-p}) +1 \}\|\theta-\theta^0\| + \|\frac{\partial^2 L_T(\theta)}{\partial \theta \partial \theta'} - \frac{\partial^2 l(\theta^0)}{\partial \theta \partial \theta'}\|
    \end{align}
    Almost surely, the first term is bounded by $\delta/4$ due to the construction of $B_0'$. The second term is eventually bounded by $\delta/4$ due to \eqref{eq:prop:consistency:point_conv_B0}. Hence, almost surely, 
    \begin{equation}
        \|\frac{\partial^2 L_T(\theta)}{\partial \theta \partial \theta'} - \frac{\partial^2 l(\theta)}{\partial \theta \partial \theta'}\| < \delta/2, \quad \theta \in B_0'.
    \end{equation}
    Thus, for $\theta \in B_0'$, $\frac{\partial^2 L_T(\theta)}{\partial \theta \partial \theta'} \succ \delta/2$ for sufficiently large $T$ almost surely, i.e., $L_T$ is eventually strictly convex on $B_0'$ almost surely. Since we just proved any global minimiser $\hat{\theta}_T \overset{a.s.}{\to} \theta^0$ in \eqref{eq:prop:consistency:consistency}, we can easily show that $|G_T| = 1$ by contradiction with the strict convexity.

    It remains to prove asymptotic normality. Theorem 3.2 of \cite{estnonlinear} is stated for the sequence constructed in Theorem 3.1 of \cite{estnonlinear}. However, its proof which references Theorem 2.2 of \cite{cls-asymptotic} does not rely on that particular sequence. The estimator is only required to satisfy (in our notation) $\hat{\theta}_T \overset{a.s.}{\to} \theta^0$ and $\frac{\partial L_T(\hat{\theta}_T)}{\partial \theta} = 0$. Theorem 3.1 of \cite{estnonlinear} is there to supply a sequence with such properties. Here, both are satisfied for our global estimator $\hat{\theta}_T$ (the latter by definition). With \sref{lem:cov_finite}, \sref{lem:lin_indep}, \cref{lem:norm_finite}, \cref{lem:cross_norm_finite}, and the fact that each innovation $\epsilon_t$ is independent of $\mathcal{F}_{t-1}$ in our case, using Corollary 3.1 of \cite{estnonlinear}, the argument for asymptotic normality in Theorem 3.2 of \cite{estnonlinear} applies directly to our case here.
    
\end{proof}}

\begin{proof}[Proof of \sref{prop:consistency_N}]
    \hltwo{First note that \sref{lem:ma_form} still holds for $q=2$ under growing $N$ as $\sum_{j=1}^p\|A_j\|_2 < 1$ by \sref{as:dist_N} so that we can still achieve a geometric bound for $\|A^l\|_2$ by \cref{lem:contraction2}.} Then, \sref{lem:cov_finite} also holds.
    
    Let $dW^{(r)}$ be such that
    \begin{equation}
        dW^{(r)}_{n,m} = \begin{cases}
            w_{n,m}'(\lambda^0),& \text{if } m \in \mathcal{N}^{(r)}(n), \\
            0,& \text{otherwise.}
        \end{cases}
    \end{equation}
    Let 
    \begin{equation}
        U_n = \E \frac{\partial\hat{X}_{t|t-1,n}}{\partial \theta}\bigg|_{\theta^0} \frac{\partial\hat{X}_{t|t-1,n}}{\partial \theta}\bigg|_{\theta^0}',
    \end{equation}
    Let $P=\sum_j(1+s_j)$, For all $a = 1,\dots,P+1$, $\frac{\partial\hat{X}_{t|t-1,n}}{\partial \theta_a}$ can be written as ${v_{a,n}}'\bar{\mathbf{X}}_{t-1}$ where if $\theta_a = \alpha_j$, $v_{a,n} = (0_{(j-1)N}',e_n,0_{(p-j)N}')'$; if $\theta_a = \beta_{j,r}$, $v_{a,n} = (0_{(j-1)N}',W_{n,:}^{(r)},0_{(p-j)N}')'$; if $\theta_a = \lambda$, $v_{a,n} = \sum_{j,r}\beta_{j,r}(0_{(j-1)N}',dW_{n,:}^{(r)},0_{(p-j)N}')' = v_{\lambda,n}$. Then 
    \begin{equation}
        [U_n]_{a,b} = \E ({v_{a,n}}'\bar{\mathbf{X}}_{t-1}) ({v_{b,n}}'\bar{\mathbf{X}}_{t-1}),
    \end{equation}
    or equivalently,
    \begin{equation}
        U_n = \E (V_n'\bar{\mathbf{X}}_{t-1})(V_n'\bar{\mathbf{X}}_{t-1})',
    \end{equation}
    where $V_n \in \mathbb{R}^{pN \times (P+1)}$ whose $a^\text{th}$ column is $v_{a,n}$. By \sref{lem:cov_finite},
    \begin{equation}\label{eq:U_norm}
        \|U^{-1}\|_2 = \|(\sum_n U_n)^{-1}\|_2 = \frac{1}{\rho_{\min}(\sum_n U_n)} \leq \frac{1}{\rho_{\min}(\cov(\bar{\mathbf{X}}_{t-1},\bar{\mathbf{X}}_{t-1}))\rho_{\min}(\sum_n V_n'V_n)} = \frac{1}{c\rho_{\min}(\sum_n V_n'V_n)}.
    \end{equation}
    By construction of $V_n$, $V_n'V_n$ is diagonal up to the last entry. Then, using Schur complement, we have that $\rho_{\min}\Big(\sum_n V_n'V_n\Big)$ is bounded below by
    \begin{align}
    \begin{split}
        \rho_{\min}\Big(\sum_{n=1}^N V_n'V_n\Big) \geq &\min \Big\{\sum_{n=1}^N \|v_{a,n}\|^2 \text{ for } a=1,\dots,P, \text{ and } \sum_{n=1}^N\Big\|v_{\lambda,n}-\sum_{a=1}^P \frac{{v_{a,n}}'v_{\lambda,n}}{\|v_{a,n}\|^2} v_{a,n} \Big\|^2\Big\} \times \\
        &\Big\{1+\sqrt{P}\max_{a=1,\dots,P}\Big(\frac{|\sum_{n=1}^N{v_{a,n}}'v_{\lambda,n}|}{\sum_{n=1}^N\|v_{a,n}\|^{2}} \Big)\Big\}^{-2},
    \end{split}
    \end{align}
    where $\sum_{n=1}^N\Big\|v_{\lambda,n}-\sum_{a=1}^P \frac{{v_{a,n}}'v_{\lambda,n}}{\|v_{a,n}\|^2} v_{a,n} \Big\|^2$ is due to 
    \begin{equation}
        \sum_{n=1}^N\Big\|v_{\lambda,n}-\sum_{a=1}^P \frac{{v_{a,n}}'v_{\lambda,n}}{\|v_{a,n}\|^2} v_{a,n} \Big\|^2\leq \sum_{n=1}^N\Big\|v_{\lambda,n}-\sum_{a=1}^P \frac{\sum_{m}{v_{a,m}}'v_{\lambda,m}}{\sum_m\|v_{a,m}\|^2} v_{a,n} \Big\|^2.
    \end{equation}
    For the first term, for $\theta_a = \alpha_j$, $\sum_{n=1}^N \|v_{a,n}\|^2 = \sum_{n=1}^N 1 = N$. For $\theta_a = \beta_{j,r}$, $\sum_{n=1}^N \|v_{a,n}\|^2 = \Theta(N)$ by \sref{as:dist_N}. For the last item,
    \begin{equation}
        v_{\lambda,n}-\sum_{a=1}^P \frac{{v_{a,n}}'v_{\lambda,n}}{\|v_{a,n}\|^2} v_{a,n} = \sum_{j,r}\beta_{j,r}^0\bigg(dW_{n,:}^{(r)} - \frac{{dW_{n,:}^{(r)}}'W_{n,:}^{(r)}}{\|W_{n,:}^{(r)}\|^2} W_{n,:}^{(r)}\bigg).
    \end{equation}
    Let $z_n$ be as in \eqref{eq:zn}. \hlthree{Fix the pair $(j,r)$ in \sref{as:dist_N} satisfying $\beta^0_{j,r}\neq 0$.} For all $n \in \mathcal{F}(N)$, and $m\in \mathcal{N}^{(r)}(n)$, its \hlthree{$(j-1)N+m$-th entry} is
    \begin{equation}
        \hlthree{\beta_{j,r}^0} \left\{w_{n,m}\{z_n-\log(d_{n,m})\} - \frac{\sum_{q}w_{n,q}^2\{z_n-\log(d_{n,q})\}}{\sum_q w_{n,q}^2} w_{n,m} \right\} = \beta_{j,r}^0 w_{n,m}\bigg\{\frac{\sum_{q}w_{n,q}^2\log(d_{n,q})}{\sum_q w_{n,q}^2}-\log(d_{n,m})\bigg\},
    \end{equation}
    which is $\Theta(1)$ at least for some $m$ due to \hlthree{$\beta^0_{j,r} \neq 0$, }$c< d_{n,m} < C$ and $|d_{n,m_1}-d_{n,m_2}| > c$ in \sref{as:dist_N}. Since $|\mathcal{N}^{(r)}(n)| = O(1)$ and $|\mathcal{F}(N)| = \Theta(N)$ in \sref{as:dist_N}, $\sum_{n=1}^N\|v_{\lambda,n}-\sum_{a=1}^P \frac{{v_{a,n}}'v_{\lambda,n}}{\|v_{a,n}\|^2} v_{a,n} \|^2 = \Theta(N)\Theta(1) = \Theta(N)$.

    For the second term, if $\theta_a = \alpha_j$, $\sum_{n=1}^N{v_{a,n}}'v_{\lambda,n} = 0$. If $\theta_a = \beta_{j,r}$, 
    \begin{equation}\label{eq:beta_lambda_coupling}
        \Big|\sum_{n=1}^N{v_{a,n}}'v_{\lambda,n}\Big| = \Big|\beta_{j,r}\sum_{n=1}^N{dW_{n,:}^{(r)}}'W_{n,:}^{(r)}\Big| = \Big|\beta_{j,r}\sum_{n=1}^N\sum_{q}w_{n,q}^2\{z_n-\log(d_{n,q})\Big| \leq |\beta_{j,r}|N\log\frac{\max_{n,m} d_{n,m}}{\min_{n,m} d_{n,m}} < CN
    \end{equation}
    due to \sref{as:dist_N}. Hence, the second term is $O(1)$.

    Thus, $\rho_{\min}(\sum_{n=1}^N V_n'V_n) = \Theta(N)$ which implies that $\|U^{-1}\|_2 = \Theta(1/N)$ by \eqref{eq:U_norm}.

    For $R$, we have
    \begin{equation}
        \|R\|_2 \leq \bigg\|\E \frac{\partial\hat{\mathbf{X}}_{t|t-1}}{\partial \theta}\bigg|_{\theta^0}' \frac{\partial\hat{\mathbf{X}}_{t|t-1}}{\partial \theta}\bigg|_{\theta^0}\bigg\|_2 \|\Sigma_\epsilon\|_2 = \|U\|_2\|\Sigma_\epsilon\|_2.
    \end{equation}
    Using the same argument in \eqref{eq:beta_lambda_coupling}, we can also show that $\sum_{n} \| v_{\lambda,n}\|^2 = O(N)$. In conjunction with the above bounds, we have that $\|U\|_2 = O(N)$ and from \sref{as:sigma_epsilon} that $\|\Sigma_\epsilon\|_2 < C$. Hence, $\|R\|_2 = O(N)$. The result follows from combining $\|U^{-1}\|_2$ and $\|R\|_2$.
    
\end{proof}

\begin{proof}[Proof of \sref{prop:lambda_error}]
    Since the self-autoregressive part is not dependent on $\lambda$, 
    \begin{equation}
        \hat{\mathbf{X}}_{t|t-1}(\lambda) - \mathbf{X}_t = \sum_{j=1}^p \sum_{r=1}^{s_j} \beta_{j,r}^0 \Delta W^{(r)}(\lambda) \mathbf{X}_{t-j} - \epsilon_t.
    \end{equation}
    Then,
    \begin{equation}
        \E \|\hat{\mathbf{X}}_{t|t-1}(\lambda) - \mathbf{X}_t\|^2 = \sum_{j=1}^p \sum_{i=1}^p \sum_{r=1}^{s_j} \sum_{s=1}^{s_j} \beta_{j,r}^0 \beta_{i,s}^0 \E (\mathbf{X}_{t-i}'\Delta W^{(s)}(\lambda)' {\Delta W^{(r)}(\lambda)}\mathbf{X}_{t-j}) + \E\|\epsilon_t\|^2,
    \end{equation}
    where the cross terms between $\mathbf{X}_{t-j}$ and $\epsilon_t$ have vanished due to independence. Now,
    \begin{align}
        \E (\mathbf{X}_{t-i}'\Delta W^{(s)}(\lambda)' {\Delta W^{(r)}(\lambda)}\mathbf{X}_{t-j}) &= \tr(\Delta W^{(s)}(\lambda)' {\Delta W^{(r)}(\lambda)} \E \mathbf{X}_{t-j} \mathbf{X}_{t-i}')\\
        &=\tr(\Delta W^{(r)}(\lambda) \Gamma(i-j) {\Delta W^{(s)}(\lambda)}')
    \end{align}
    and $\E\|\epsilon_t\|^2 = \tr(\Sigma_\epsilon)$.
    
\end{proof}

\hengxu{
\begin{proof}[Proof of \sref{cor:lambda_error}]
    This follows immediately from \sref{prop:lambda_error}.
    
\end{proof}
}

\section{Higher-order Derivatives of the Conditional Mean}

The second order derivatives of $\hat{X}_{t|t-1,n}$ are
\begin{equation}
    \frac{\partial^2 \hat{X}_{t|t-1,n}}{\partial \beta_{j,r} \partial \lambda} = \sum_{m \in \mathcal{N}^{(r)}(n)}\frac{\partial w_{n,m}}{\partial \lambda}X_{t-j,m},
\end{equation}
\begin{equation}
    \frac{\partial^2 \hat{X}_{t|t-1,n}}{\partial \lambda^2} = \sum_{j=1}^p\sum_{r=1}^{s_j} \beta_{j,r} \sum_{m \in \mathcal{N}^{(r)}(n)} \frac{\partial^2 w_{n,m}}{\partial \lambda^2} X_{t-j,m},
\end{equation}
where
\begin{equation}
\frac{\partial^2 w_{n,m}}{\partial \lambda^2} = d_{n,m}^{-\lambda}\Bigg\{\frac{\log^2 d_{n,m}}{(\sum_{q}d_{n,q}^{-\lambda})} - \frac{\sum_{q}d_{n,q}^{-\lambda}\log^2 d_{n,q}}{(\sum_{q}d_{n,q}^{-\lambda})^2} - 2\frac{\log d_{n,m} \sum_{q} d_{n,q}^{-\lambda}\log d_{n,q}}{(\sum_{q}d_{n,q}^{-\lambda})^2} + 2\frac{(\sum_q d_{n,q}^{-\lambda} \log d_{n,q})^2}{(\sum_{q}d_{n,q}^{-\lambda})^3}\Bigg\},
\end{equation}
and 0 for the others.

Clearly, we have that $|\frac{\partial w_{n,m}}{\partial \lambda}| \leq 2\sup_{n,m} |\log d_{n,m}|$ and $|\frac{\partial^2 w_{n,m}}{\partial \lambda^2}| \leq 6\sup_{n,m} (\log^2 d_{n,m})$ for all $\lambda$.

The third order derivatives of $\hat{X}_{t|t-1,n}$ are
\begin{align}
    \frac{\partial^3 \hat{X}_{t|t-1,n}}{\partial \beta_{j,r} \partial \lambda^2} = \sum_{m \in \mathcal{N}^{(r)}(n)} \frac{\partial^2 w_{n,m}}{\partial \lambda^2} X_{t-j,m},
\end{align}
\begin{equation}
    \frac{\partial^3 \hat{X}_{t|t-1,n}}{\partial \lambda^3} = \sum_{j=1}^p\sum_{r=1}^{s_j} \beta_{j,r} \sum_{m \in \mathcal{N}^{(r)}(n)} \frac{\partial^3 w_{n,m}}{\partial \lambda^3} X_{t-j,m},
\end{equation}
where the exact form of $\frac{\partial^3 w_{n,m}}{\partial \lambda^3}$ is complex, but we can bound it by $|\frac{\partial^3 w_{n,m}}{\partial \lambda^3}| \leq C\sup_{n,m} |\log^3 d_{n,m}|$ for some constant $C>0$.

\section{Lemmas}

\begin{lemma}\label{lem:lin_indep_pd}
    Let $X_1,\dots,X_n$ be random variables with $\E X_k = 0$ and $\E X_k^2 > 0$. Then $\cov((X_1,\dots,X_n)') \succ 0$ implies $\E (\sum_{k=1}^n a_k X_k)^2 = 0 \implies a_k = 0$.
    
\end{lemma}

\begin{proof}
    $\cov((X_1,\dots,X_n)') \succ 0$ implies that for all $a=(a_1,\dots,a_n)'\neq 0$, we have
    \begin{equation}
        a'\cov((X_1,\dots,X_n)')a = \sum_{i,j} a_i \cov(X_i,X_j) a_j = \E (\sum_{k=1}^n a_k X_k)^2 > 0.
    \end{equation}
    
\end{proof}



\begin{lemma}\label{lem:contraction}
    Let $A$ be a square matrix, suppose $\rho_{\max}(A) < 1$, then there exists $l_0$, such that $\|A^l\| < 1$ for all $l > l_0$.
\end{lemma}

\begin{proof}
    From Gelfand's formula, we know that
    \begin{equation}
        \lim_{l \rightarrow \infty} \|A^l\|^{1/l} = \rho_{\max}(A) < 1,
    \end{equation}
    which implies that there exists $l_0$, such that for all $l > l_0$,
    \begin{equation}
        \|A^l\|^{1/l} < 1.
    \end{equation}
\end{proof}

\begin{lemma}\label{lem:contraction2}
    Assume $\sum_{j=1}^p\|A_j\|_q < (1+c')^{-1}$ for some $q=1,\dots,\infty$, then for the companion matrix $A$, $\|A^l\|_q < C(1+c)^{-l}$ for some $C,c>0$ uniformly in $N$.
\end{lemma}

\begin{proof}
    Let $\delta = (1+c')^{1/p} > 1$ be such that $\delta^{p-1}(1+c')^{-1} = \delta^{-1}< 1$ is independent of $N$. For any $x \in \mathbb{R}^{pN}$, write $x$ as $x = (x_1',x_2',\dots,x_p')$. Define $\|x\|_{q,\delta} = \max_{j=1,\dots,p} \delta^{-(j-1)} \|x_j\|_q$. Let $y = Ax$. Then, $y_1 = \sum_{j=1}^p A_j x_j$ and $y_j = x_{j-1}$ for $j=2,\dots,p$.

    Now, 
    \begin{equation}
        \delta^{-(j-1)}\|y_j\|_q = \delta^{-(j-1)}\|x_{j-1}\|_q \leq \delta^{-1} \|x\|_{q,\delta},
    \end{equation}
    and 
    \begin{align}
        \|y_1\|_q &\leq \sum_{j=1}^p \|A_j\|_q \|x_j\|_q \leq \sum_{j=1}^p \|A_j\|_q \delta^{j-1} \|x\|_{q,\delta}\\
        & \leq \delta^{p-1}(\sum_{j=1}^p \|A_j\|_q ) \|x\|_{q,\delta}.
    \end{align}
    Therefore,
    \begin{equation}
        \|Ax\|_{q,\delta} \leq \max\{\delta^{p-1}(1+c')^{-1},\delta^{-1}\} \|x\|_{q,\delta} = (1+c')^{-{1/p}} \|x\|_{q,\delta},
    \end{equation}
    which implies that $\|A^l\|_{q,\delta} \leq (1+c')^{-{l/p}}$. Getting back to the $q$-norm, we have for all $x$, if $q < \infty$,
    \begin{equation}
        \|x\|_q = \Big(\sum_{j=1}^p \|x_j\|_q^q\Big)^{1/q} \leq \Big(\sum_{j=1}^p \delta^{q(j-1)}\Big)^{1/q} \|x\|_{q,\delta},
    \end{equation}
    and if $q=\infty$,
    \begin{equation}
        \|x\|_q \leq \delta^{p-1} \|x\|_{q,\delta}.
    \end{equation}
    In both cases,
    \begin{equation}
        \|x\|_q \geq \|x\|_{q,\delta}
    \end{equation}
    by definition. Then $\|A^l\|_q \leq C (1+c')^{-{l/p}} = C(1+c)^{-l}$ where $C$ is either $\Big(\sum_{j=1}^p \delta^{q(j-1)}\Big)^{1/q}$ or $\delta^{p-1}$ depending on $q$ and $c = (1+c')^{1/p} - 1$.

\end{proof}

\begin{lemma}\label{lem:dist_to_indep}
    \hlthree{Let $j,r$ be as in \sref{as:dist}.} \sref{as:dist} is a necessary and sufficient condition for the following linear system for $a,b$ to have only trivial solutions $a=b=0$:
    \begin{equation}
        a w_{n,m}(\lambda^0) + b \beta_{j,r}^0 \frac{\partial w_{n,m}}{\partial \lambda}\bigg|_{\lambda^0} = 0
    \end{equation}
    for all $n=1,\dots,N$ and $m \in \mathcal{N}^{(r)}(n)$.
\end{lemma}

\begin{proof}
    We can rewrite $\frac{\partial w_{n,m}}{\partial \lambda}$ as 
    \begin{equation}\label{eq:zn}
        \frac{\partial w_{n,m}}{\partial \lambda} = w_{n,m} \bigg\{\frac{\sum_{q \in \mathcal{N}^{(r)}(n)}d_{n,q}^{-\lambda}\log(d_{n,q})}{\sum_{q \in \mathcal{N}^{(r)}(n)} d_{n,q}^{-\lambda}}-\log(d_{n,m})\bigg\} = w_{n,m} (z_{n}-\log(d_{n,m})).
    \end{equation}
    \hlthree{With the fact that $\beta^0_{j,r} \neq 0$,} the linear system is linearly dependent if and only if for all $n$, $z_{n}-\log(d_{n,m})$ is a constant for all $m \in \mathcal{N}^{(r)}(n)$, i.e., $d_{n,m}$ is a constant for all $m \in \mathcal{N}^{(r)}(n)$.
    
\end{proof}

\begin{lemma}\label{lem:norm_finite}
    $\E \| \frac{\partial\hat{X}_{t|t-1}}{\partial \theta_1}|_{\theta^0} \|_2^2 < \infty$, $\E \| \frac{\partial^2 \hat{X}_{t|t-1}}{\partial \theta_1 \partial \theta_2}|_{\theta^0} \|_2^2 < \infty$ for all $\theta_1,\theta_2 = \alpha_1,\beta_{1,1},\dots,\alpha_p,\beta_{p,1},\dots,\lambda$.
\end{lemma}

\begin{proof}
    This follows directly from \sref{lem:cov_finite}.
    
\end{proof}

\begin{lemma}\label{lem:cross_norm_finite}
    For all $\theta \in \Omega$, \hltwo{where $\Omega$ is a bounded open neighbourhood of $\theta^0$}, for all $\theta_1,\theta_2,\theta_3 = \alpha_1,\beta_{1,1},\dots,\alpha_p,\beta_{p,1},\dots,\lambda$, there exist functions $G,H$, such that
    \begin{equation}
        \left| \frac{\partial\hat{\mathbf{X}}_{t|t-1}}{\partial \theta_1}' \frac{\partial^2 \hat{\mathbf{X}}_{t|t-1}}{\partial \theta_2 \partial \theta_3}  \right| \leq G\hlthree{(X_{t-1},\dots,X_{t-p})} , \quad \quad \E G(X_{t-1},\dots,X_{t-p}) < \infty,
    \end{equation}
    \begin{equation}
        \left| (\mathbf X_t-\hat{\mathbf{X}}_{t|t-1})' \frac{\partial^3 \hat{\mathbf X}_{t|t-1}}{\partial \theta_1 \partial \theta_2 \partial \theta_3}\right| \leq H\hlthree{(X_t,X_{t-1},\dots,X_{t-p})} , \quad \quad \E H(X_t,X_{t-1},\dots,X_{t-p}) < \infty.
    \end{equation}
\end{lemma}

\begin{proof}
    We know that \hltwo{$\beta_{j,r}$}, $w_{n,m}$, $\frac{\partial w_{n,m}}{\partial \lambda}$, $\frac{\partial^2 w_{n,m}}{\partial \lambda^2}$, $\frac{\partial^3 w_{n,m}}{\partial \lambda^3}$ are all uniformly bounded \hltwo{in $\Omega$}. The result thus follows from \sref{lem:cov_finite}.
    
\end{proof}

\hltwo{
\section{Estimation Simulation Results}

We simulate 100 realisations from the same three model specifications as in Section 4 with $N=64$ and $T=200$. The network is generated by connecting each pair of nodes with probability $1/2$, with edge distances drawn from an Exp$(1)$ exponential distribution. The covariance matrix is generated at random and scaled so that its trace equals that of $0.1I_{64}$. The total number of data points is 12800, compared with 8000 in Section 4. \cref{tab:sim_appendix} reports the mean and standard deviation of the estimated coefficients over the 100 realisations in each case.

\begin{table}[h]
    \centering
    \setlength{\tabcolsep}{8pt}
    \begin{tabular}{cccccc}
        Case & $\hat{\lambda}$ & $\hat{\alpha}_1$ & $\hat{\beta}_{1,1}$ & $\hat{\alpha}_2$ & $\hat{\beta}_{2,1}$ \\
        \hline
        A & 2.998(0.14) & 0.349 (0.009) & 0.550(0.010)&- &-\\
        \hline
        B & 3.036(0.37)& 0.349(0.012) &  0.250(0.012) & 0.151(0.012) &-0.202(0.011)\\
        \hline
        C & 0.605(0.09)& 0.349(0.012) &  0.254(0.041) & 0.151(0.012) &-0.205(0.035)
    \end{tabular}
    \caption{Mean (standard deviation) of the estimated coefficients over the 100 realisations for each case.}
    \label{tab:sim_appendix}
\end{table}

The estimation is again approximately consistent and the conclusions drawn in Section 4 also hold here, illustrating the scalability of our algorithm.
}

\hlthree{
\section{Upper Cap for Grid Search on $\lambda$}\label{sec:app_lambda_cap}

\hlthree{The purpose of this section is to derive a justified finite data-driven upper cap for the grid search over $\lambda$. We first define some notation to simplify the math.}

Let $\alpha = (\alpha_1,\dots,\alpha_p)$. Let $\beta = (\beta_{1,1}, \dots, \beta_{p,s_p})$
\guyfour{with the same assumptions on these
parameters as specified in the main paper.}
Let $L_T(\lambda) = \min_{\alpha,\beta}L_T\{(\alpha,\beta,\lambda)\}$. Let $V(\lambda)$ be the design matrix for $(\alpha,\beta)$ for each $\lambda$. Then,
\begin{equation}\label{eq:lambda_design}
    TL_T(\lambda) = \min_{\alpha,\beta} \|(X_1,\dots,X_T)' - V(\lambda) (\alpha,\beta)'\|^2.
\end{equation}
\hlthree{ The limit $TL_T(\infty)$ can be easily computed using one least squares fit, which is used as a reference point to control finite $\lambda$. Moreover specifically, we will derive a non-increasing and converging-to-zero bound that controls $T\{L_T(\infty)-L_T(\lambda)\}$ in \cref{lem:lambda_cap}. This way, we can split the search over $\lambda \in [0,\infty]$ into $[0, \lambda_U]$ and $(\lambda_U, \infty]$, where $[0, \lambda_U]$ is grid-searched and $(\lambda_U, \infty]$ is automatically controlled by \cref{lem:lambda_cap}. The bound in \cref{lem:lambda_cap} is constructed based on a non-increasing converging-to-zero entrywise bound $M(\lambda)$ on the absolute difference between the design matrices evaluated at finite $\lambda$ and $\infty$, $|V(\lambda)-V(\infty)|$. To construct $M(\lambda)$, we first compute the difference between the weights evaluated at finite $\lambda$ and $\infty$.}

For each $n,r$, let $d_{n,\min}^{(r)} = \min_{m\in \mathcal{N}^{(r)}(n)} d_{n,m}$. Let $k_{n}^{(r)} = \# \{m\in \mathcal{N}^{(r)}(n): d_{n,m} = d_{n,\min}^{(r)}\}$. If $d_{n,m} = d_{n,\min}^{(r)}$, $w_{n,m}(\infty) = 1/k_{n}^{(r)}$, and
\begin{equation}\label{eq:w_lambda_infty_eq}
    |w_{n,m}(\lambda) - w_{n,m}(\infty)| = \frac{\sum_{q \in \mathcal{N}^{(r)}(n), d_{n,q} > d_{n,\min}^{(r)}} (\frac{d_{n,\min}^{(r)}}{d_{n,q}})^\lambda}{k_{n}^{(r)}\{k_{n}^{(r)} + \sum_{q \in \mathcal{N}^{(r)}(n), d_{n,q} > d_{n,\min}^{(r)}} (\frac{d_{n,\min}^{(r)}}{d_{n,q}})^\lambda \}}.
\end{equation}
Otherwise, $w_{n,m}(\infty) = 0$, and
\begin{equation}\label{eq:w_lambda_infty_neq}
    |w_{n,m}(\lambda) - w_{n,m}(\infty)| \leq \frac{1}{k_{n}^{(r)}} (\frac{d_{n,\min}^{(r)}}{d_{n,m}})^\lambda.
\end{equation}
\hlthree{We are now ready to define $M(\lambda)$.} Let $M(\lambda)$ be a matrix with the same shape as $V(\lambda)$, whose column corresponding to $\alpha_j$ is zero, and entry corresponding to observation $t,n$ and coefficient $\beta_{j,r}$ is
\begin{align}
    M(\lambda)_{(t,n),(j,r)} = &\sum_{m \in \mathcal{N}^{(r)}(n), d_{n,m} = d_{n,\min}^{(r)}} |X_{t-j,m}| \frac{\sum_{q \in \mathcal{N}^{(r)}(n), d_{n,q} > d_{n,\min}^{(r)}} (\frac{d_{n,\min}^{(r)}}{d_{n,q}})^\lambda}{k_{n}^{(r)}\{k_{n}^{(r)} + \sum_{q \in \mathcal{N}^{(r)}(n), d_{n,q} > d_{n,\min}^{(r)}} (\frac{d_{n,\min}^{(r)}}{d_{n,q}})^\lambda \}}\\
    &+\sum_{m \in \mathcal{N}^{(r)}(n), d_{n,m} \neq d_{n,\min}^{(r)}} |X_{t-j,m}|\frac{1}{k_{n}^{(r)}} (\frac{d_{n,\min}^{(r)}}{d_{n,m}})^\lambda.
\end{align}
Since the column in $V(\lambda)$ corresponding to $\alpha_j$ does not depend on $\lambda$, using \eqref{eq:w_lambda_infty_eq} and \eqref{eq:w_lambda_infty_neq}, we have for each entry $a,b$ of $V(\lambda)$,
\begin{equation}
    |V(\lambda)_{a,b} - V(\infty)_{a,b}| \leq M(\lambda)_{a,b}.
\end{equation}
Moreover, we can easily see that $M(\lambda)_{a,b}$ is non-increasing in $\lambda$, and that $M(\lambda) \rightarrow 0$ as $\lambda \rightarrow \infty$.

Let $(\hat{\alpha}(\infty),\hat{\beta}(\infty))'$ be the solution to \eqref{eq:lambda_design} at $\lambda = \infty$, and let \begin{equation}
    \hat{\epsilon}(\infty) = (X_1,\dots,X_T)' - V(\infty) (\hat{\alpha}(\infty),\hat{\beta}(\infty))'.
\end{equation}
Then, $V(\infty)'\hat{\epsilon}(\infty) = 0$ and 
\begin{equation}\label{eq:TL_T_infty}
    TL_T(\infty) = \| \hat{\epsilon}(\infty)\|^2.
\end{equation}
\hlthree{ The following lemma lower bounds $TL_T(\lambda)$ using $M(\lambda)$ and the reference point $\lambda=\infty$.}

\begin{lemma}\label{lem:lambda_cap}
    If $\|M(\lambda)\|_2 < \sigma_{\min}\{V(\infty)\}$ where $\sigma_{\min}\{V(\infty)\}$ is the smallest singular value of $V(\infty)$, then
    \begin{align}
    \begin{split}
        &T\{L_T(\infty)-L_T(\lambda)\} \leq 2|\hat{\epsilon}(\infty)|'M(\lambda)|(\hat{\alpha}(\infty),\hat{\beta}(\infty))'|\\
        &+ \frac{\{\|M(\lambda)'|\hat{\epsilon}(\infty)|\| + \| |V(\infty)|'M(\lambda)|(\hat{\alpha}(\infty),\hat{\beta}(\infty))'| \| + \| M(\lambda)'M(\lambda)|(\hat{\alpha}(\infty),\hat{\beta}(\infty))'|\|\}^2}{\{\sigma_{\min}\{V(\infty)\} - \|M(\lambda)\|_2\}^2}.
    \end{split}
    \end{align}
\end{lemma}

\begin{proof}
    Let $\Delta V(\lambda) = V(\lambda) - V(\infty)$. We know that $|\Delta V(\lambda)| \leq M(\lambda)$ entrywise. The residuals for $V(\lambda)$ and $(\hat{\alpha}(\infty),\hat{\beta}(\infty))'$ are
    \begin{equation}
        (X_1,\dots,X_T)' - V(\lambda) (\hat{\alpha}(\infty),\hat{\beta}(\infty))' = \hat{\epsilon}(\infty) - \Delta V(\lambda)(\hat{\alpha}(\infty),\hat{\beta}(\infty))'.
    \end{equation}
    The actual least squares equation with $V(\lambda)$ is $TL_T(\lambda) = \min_{\alpha,\beta} \|(X_1,\dots,X_T)' - V(\lambda) (\alpha,\beta)'\|^2$. Equivalently,
    \begin{equation}
        TL_T(\lambda) = \min_{x} \|\hat{\epsilon}(\infty) - \Delta V(\lambda)(\hat{\alpha}(\infty),\hat{\beta}(\infty))' - V(\lambda)x\|^2.
    \end{equation}
    Solving for $x$ gives us
    \begin{align}\label{eq:TL_T_lambda}
    \begin{split}
        TL_T(\lambda) = &\| \hat{\epsilon}(\infty) - \Delta V(\lambda)(\hat{\alpha}(\infty),\hat{\beta}(\infty))'\|^2\\
        &- [V(\lambda)'\{\hat{\epsilon}(\infty) - \Delta V(\lambda)(\hat{\alpha}(\infty),\hat{\beta}(\infty))' \}]' \\
        &\times \{ V(\lambda)'V(\lambda)\}^{-1}\\
        &\times [V(\lambda)'\{\hat{\epsilon}(\infty) - \Delta V(\lambda)(\hat{\alpha}(\infty),\hat{\beta}(\infty))' \}]
    \end{split}
    \end{align}
    Since $V(\infty)'\hat{\epsilon}(\infty) = 0$,
    \begin{align}
    \begin{split}
        &V(\lambda)'\{\hat{\epsilon}(\infty) - \Delta V(\lambda)(\hat{\alpha}(\infty),\hat{\beta}(\infty))' \}\\ 
        = &\Delta V(\lambda)' \hat{\epsilon}(\infty) - V(\infty)' \Delta V(\lambda) (\hat{\alpha}(\infty),\hat{\beta}(\infty))' - \Delta V(\lambda)'\Delta V(\lambda) (\hat{\alpha}(\infty),\hat{\beta}(\infty))'.
    \end{split}
    \end{align}
    Combining \eqref{eq:TL_T_infty} and \eqref{eq:TL_T_lambda}, and dropping the non-positive term $-\|\Delta V(\lambda) (\hat{\alpha}(\infty),\hat{\beta}(\infty))' \|^2$, we have
    \begin{equation}
        T\{L_T(\infty)-L_T(\lambda)\} \leq 2 |\hat{\epsilon}(\infty)' \Delta V(\lambda) (\hat{\alpha}(\infty),\hat{\beta}(\infty))'| + \frac{\|V(\lambda)'\{\hat{\epsilon}(\infty) - \Delta V(\lambda)(\hat{\alpha}(\infty),\hat{\beta}(\infty))' \}\|^2}{\sigma_{\min}^2\{V(\lambda)\}}
    \end{equation}
    For the first term,
    \begin{equation}
        |\hat{\epsilon}(\infty)' \Delta V(\lambda) (\hat{\alpha}(\infty),\hat{\beta}(\infty))'| \leq |\hat{\epsilon}(\infty)|'M(\lambda)|(\hat{\alpha}(\infty),\hat{\beta}(\infty))'|.
    \end{equation}
    For the second term,
    \begin{align}
    \begin{split}
        &\|V(\lambda)'\{\hat{\epsilon}(\infty) - \Delta V(\lambda)(\hat{\alpha}(\infty),\hat{\beta}(\infty))' \}\|\\ 
        \leq &\|M(\lambda)'|\hat{\epsilon}(\infty)|\| + \| |V(\infty)|'M(\lambda)|(\hat{\alpha}(\infty),\hat{\beta}(\infty))'| \| + \| M(\lambda)'M(\lambda)|(\hat{\alpha}(\infty),\hat{\beta}(\infty))'|\|,
    \end{split}
    \end{align}
    and 
    \begin{equation}
        \sigma_{\min}^2\{V(\lambda)\} \geq \{\sigma_{\min}\{V(\infty)\} - \|M(\lambda)\|_2\}^2.
    \end{equation}
\end{proof}

The bound in \cref{lem:lambda_cap} is non-increasing and converges to 0 in $\lambda$. \hlthree{ We now describe how to find an upper cap $\lambda_U$ in practice so that we can control $\lambda > \lambda_U$ automatically using \cref{lem:lambda_cap}.}

The grid search algorithm is:

\begin{enumerate}
\item Fix an initial upper cap for searching $\lambda$, $\lambda_1$. Let $\lambda_0 = 0$. 

\item Solve the least squares equation for $\lambda = \infty$ to obtain $\hat{\alpha}(\infty), \hat{\beta}(\infty)$ and $T L_T(\infty)$.

\item For $\text{i} = 1,\dots$, grid search on $[\lambda_{i-1},\lambda_i]$. If the minimum $\tilde{\lambda}$ is obtained not at $\lambda_i$, store $\hat{\alpha}(\tilde{\lambda}), \hat{\beta}(\tilde{\lambda})$ and $T L_T(\tilde{\lambda})$, set $\lambda_L = \lambda_i$, and break. Otherwise, set $\lambda_{i+1} = 2\lambda_{i}$ until the first non-boundary minimum $\tilde{\lambda}$ is obtained. If $\lambda_{i+1}$ exceeds a prespecified upper cap $\lambda_{m1}$ which does not have to be big (we will compute the final upper cap for search later), set an artificial $L = (1- \text{rtol})T L_T(\infty)$ with a prespecified relative tolerance rtol, set $\lambda_L = \lambda_{i}$, break.

\item Suppose that the above procedure gives a non-boundary minimum. If $T L_T(\tilde{\lambda}) < T L_T(\infty)$, set $L = T L_T(\tilde{\lambda})$. Otherwise, set again an artificial $L = (1- \text{rtol})T L_T(\infty)$.

\item We now find an upper cap $\lambda_U$ such that the condition in \cref{lem:lambda_cap} is satisfied and that the bound in \cref{lem:lambda_cap} is smaller than $T L_T(\infty) - L$. In other words, all $\lambda > \lambda_U$ must have $TL_T(\lambda) \geq L$, which can be safely ignored, unless the fallback in the next step is triggered.

\item To do this, we test for each $\lambda_j = 2^j \lambda_L$, $j=1,2,\dots$ with a prespecified upper cap $\lambda_{m2}$ for $\lambda_j$. $\lambda_{m2}$ can be notably larger than $\lambda_{m1}$ since the bound in \cref{lem:lambda_cap} is much cheaper than least squares. If for $\lambda_j$, the condition in \cref{lem:lambda_cap} is satisfied and the bound in \cref{lem:lambda_cap} is below $T L_T(\infty) - L$, break. Set $\lambda_u = \lambda_j$ and $\lambda_l = \lambda_{j}/2$. Otherwise, continue. If no $\lambda_j \leq \lambda_{m2}$ satisfies the above, set $\lambda_U = \lambda_{m2}$ and skip the next step.

\item For $k=1,\dots$, check the condition and the bound in \cref{lem:lambda_cap} for $\frac{\lambda_l + \lambda_u}{2}$. If both are satisfied, set $\lambda_u = \frac{\lambda_l + \lambda_u}{2}$. Otherwise, set $\lambda_l = \frac{\lambda_l + \lambda_u}{2}$. Continue the refinement until $\lambda_u - \lambda_l$ is smaller than a prespecified tolerance, and set $\lambda_U = \lambda_u$.

\item Finally, grid search on the region $[\lambda_L, \lambda_U]$, where $\lambda_L$ is the largest $\lambda$ checked previously. For the minimum $\check{\lambda}$ attained here, if $TL_T(\check{\lambda}) < L$, set $\hat{\lambda} = \check{\lambda}$. Otherwise, set $\hat{\lambda} = \tilde{\lambda}$ if $L = T L_T(\tilde{\lambda})$, and $\hat{\lambda} = \infty$ if $L = (1- \text{rtol})T L_T(\infty)$.
\end{enumerate}

The size of the grid $s$ can be chosen such that $|L_T(\hat{\lambda}) - L_T(\hat{\lambda} \pm s)|/L_T(\hat{\lambda}) \leq \text{rtol2}$ for some prespecified relevant tolerance rtol2. If this is not satisfied, refine the grid search locally around $\hat{\lambda}$.

\cref{fig:Bruma_SSE_lambda_cap} illustrates the upper cap of the search region obtained from the above algorithm for the Brumadinho application \hlfour{at origin $t=40$} in the main text.

\begin{figure}[h]
    \centering
    \includegraphics[width=0.75\linewidth]{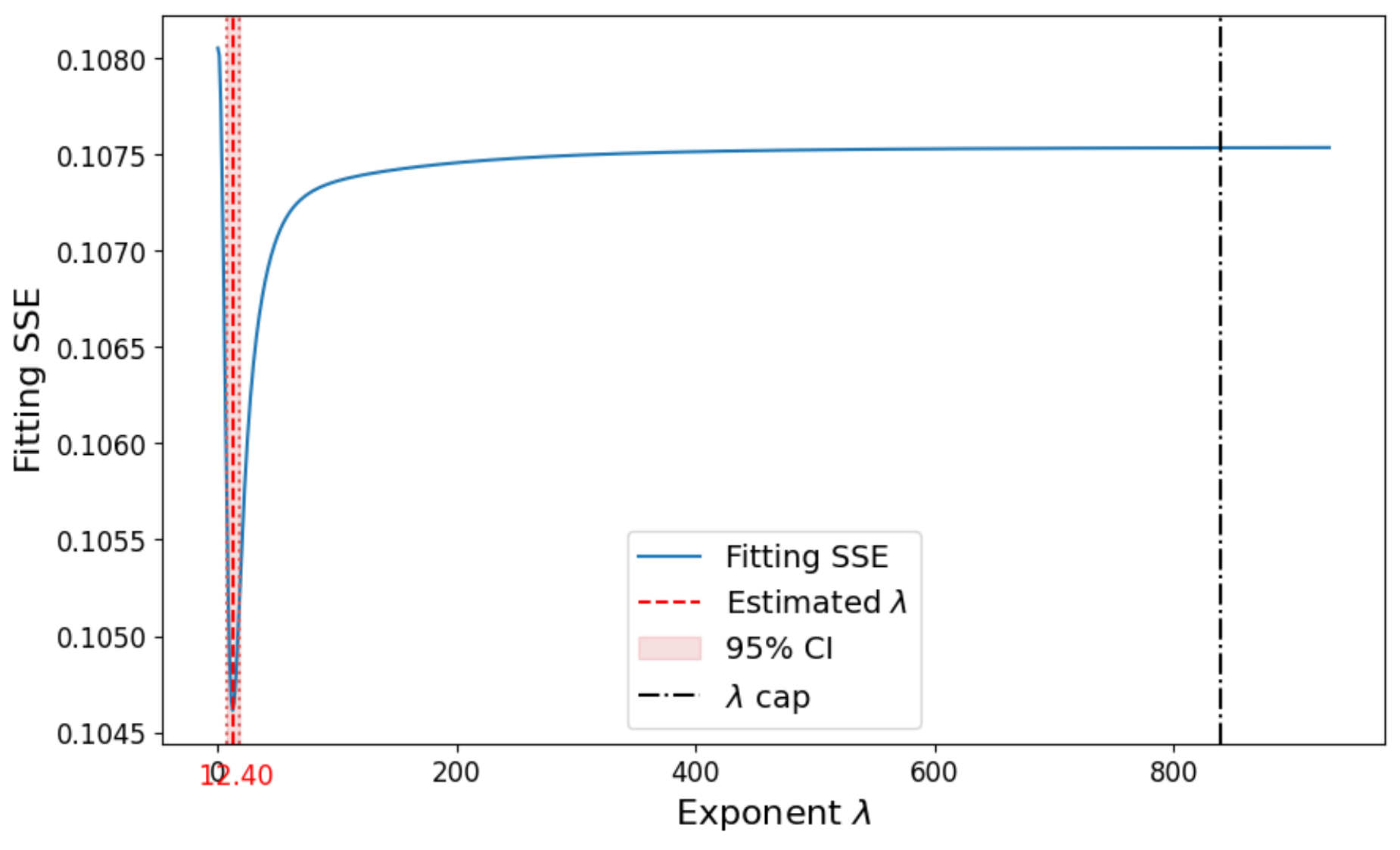}
    \caption{The fitting SSE, $T L_T(\theta)$, versus $\lambda$ for model B-GNARID. The optimal exponent is $\hat{\lambda} = 12.4$, with the shaded region indicating the 95\% confidence interval of $\hat{\lambda}$. The black line, $\lambda = 839$, is the upper cap for the search region obtained using the algorithm in \cref{sec:app_lambda_cap}. All $\lambda$ greater than this can be safely ignored according to \cref{lem:lambda_cap}.}
    \footnotesize{\textbf{Alt text: }Line plot of fitting sum of squared errors against lambda for the Brumadinho GNARID model at origin forty. The line is V-shaped for small lambda and flat for larger lambda. It reaches its minimum at lambda twelve point four. A vertical black line at lambda eight hundred thirty-nine marks the upper cap of the search region; lambda values greater than this are safely ignored.}
    \label{fig:Bruma_SSE_lambda_cap}
\end{figure}
}

\hlfour{
\section{Consistency with estimated Wasserstein distances}

In this section, we consider distances constructed from the marginal distributions of the network time series. For ease of reading, we restate here the distribution-based distance construction given in Section 3.4. Define the set of all relevant pairs of nodes,
\begin{equation}
    \mathcal{P} = \bigcup_{r=1}^{\max_{j=1}^p s_j}\{(n,m)\in \{1,\dots,N\}^2: m \in \mathcal{N}^{(r)}(n)  \}.
\end{equation}
Let $\mathcal{D}$ be the set of all collections of distances $D = \{d_{n,m}: (n,m) \in \mathcal{P} \}$ with $1 \leq d_{n,m} \leq 2$ for all $(n,m) \in \mathcal{P}$. Let
\begin{equation}\label{eq:dist_from_ws}
    d_{n,m} = 1+ \frac{W_1\{F_n(D),F_m(D)\}}{\max_{(k,l) \in \mathcal{P}} W_1\{F_k(D),F_l(D)\}},
\end{equation}
where $F_n(D)$ is the stationary marginal distribution of $X_{t,n}(D)$ depending on $D$ and 
\begin{equation}
    W_1\{F_n(D),F_m(D)\} = \int_{-\infty}^{\infty} |F_n(D)(x) - F_m(D)(x)| \text{d}x
\end{equation}
is the 1-Wasserstein distance; see \cite[Section 1.2.3]{Panaretos2019}. All other specifications in \sref{def:gnarid} remain unchanged. The denominator in \eqref{eq:dist_from_ws} normalises the 1-Wasserstein distances and the offset $1$ is to prevent zero-distances. Although this distance formulation is still a non-random constant and satisfies the GNARID definition, the fact that the stationary distribution itself depends on the distances introduces an apparent circularity to the model. However, we will show that \eqref{eq:dist_from_ws} is well defined and such a GNARID model exists and ultimately that using an empirical estimate of $d_{n,m}$ still leads to a consistent $\hat{\theta}_T$.

We first impose an assumption on the innovation covariance so that $\max_{(k,l)\in \mathcal{P}} W_1\{F_k(D),F_l(D)\}$ is always strictly positive. 

\begingroup
\renewcommand{\theassumption}{F.1}
\begin{assumption}\label{as:ws_max_neq0}
    Let $\gamma = \sum_{j=1}^p\left( |\alpha_j^0| + \sum_{r=1}^{s_j} |\beta_{j,r}^0| \right)$ and suppose that the stationarity condition after \sref{as:stationary} holds, so that $\gamma < 1$. In addition, we assume that there exists a pair of nodes $(n,m) \in \mathcal{P}$ such that $[\Sigma_\epsilon]_{n,n} < (1-\gamma^2) [\Sigma_\epsilon]_{m,m}$.
\end{assumption}
\endgroup

\cref{as:ws_max_neq0} imposes a condition on the marginal variances of the innovations, stating that there exists one pair of nodes with enough variance separation relative to the sum of absolute coefficients. The following Lemma builds on \cref{as:ws_max_neq0} and proves the existence of a stationary GNARID process with distances satisfying \eqref{eq:dist_from_ws}.

\begin{lemma}\label{lem:ws_existence}
    Under \cref{as:ws_max_neq0}, there exists at least one stationary and ergodic GNARID process whose distances satisfy \eqref{eq:dist_from_ws}.
\end{lemma}

\begin{proof}
     Fix a $\theta^0$ satisfying \cref{as:ws_max_neq0}. For any $D \in \mathcal{D}$, as we have shown before, \cref{as:ws_max_neq0} guarantees a stationary and ergodic GNARID process. Let $F_n(D)$ denote the marginal stationary distribution of its node $n$. We first show that $\max_{(k,l) \in \mathcal{P}} W_1\{F_k(D),F_l(D)\} > 0$ for any $D \in \mathcal{D}$.

    Suppose, to the contrary, that there exists a $D \in \mathcal{D}$ such that $\max_{(k,l) \in \mathcal{P}} W_1\{F_k(D),F_l(D)\} = 0$. Then, $F_{n}(D) = F_m(D)$ for all $(n,m) \in \mathcal{P}$. In particular, for all $n$, $\var (X_{t,m}) = \var (X_{t,n}) = v_n$ for all $m \in \mathcal{N}^{(r)}(n)$ for all $r=1,\dots,\max_{j}s_j$ for some $v_n > 0$. Since $\mathbf{\epsilon}_t$ is independent of the past, we have
    \begin{equation}
        v_n = \var (\hat{X}_{t|t-1,n}) + [\Sigma_\epsilon]_{n,n}.
    \end{equation}
    Now, the conditional mean $\hat{X}_{t|t-1,n}$ is itself a linear combination of random variables of variances $v_n$, with coefficients whose absolute sum is at most $\gamma$. Then, simply expanding $\var(\hat{X}_{t|t-1,n})$ gives us $\var(\hat{X}_{t|t-1,n}) \leq \gamma^2 v_n$ and therefore $v_n \leq \gamma^2 v_n + [\Sigma_\epsilon]_{n,n}$. Hence,
    \begin{equation}
        (1-\gamma^2) v_n \leq  [\Sigma_\epsilon]_{n,n} \leq v_n.
    \end{equation} 
    Using the same argument for any $m \in \mathcal{N}^{r}(n)$ for any $r$, we get
    \begin{equation}
        (1-\gamma^2) v_n \leq  [\Sigma_\epsilon]_{m,m} \leq v_n.
    \end{equation}
    Therefore, for any $(n,m) \in \mathcal{P}$,
    \begin{equation}
        [\Sigma_\epsilon]_{n,n} \geq (1-\gamma^2) [\Sigma_\epsilon]_{m,m},
    \end{equation}
    contradicting the assumption, which proves $\max_{(k,l) \in \mathcal{P}} W_1\{F_k(D),F_l(D)\} > 0$.

    Fix $D \in \mathcal{D}$. Consider any sequence $D_q \rightarrow D$. By continuity of the weighting function $w$ with respect to $D$, we immediately have that for the companion matrix $A$, $A(D_q) \rightarrow A(D)$ and therefore $a_l(D_q) \rightarrow a_l(D)$ for every fixed $l$; see \eqref{eq:original_ma}. Recall that we have already established the fact that $\|a_l(D)\| < C(1+c)^{-l}$ for some $c,C > 0$ uniformly in $l$ and independent of $D$ in the proof of \sref{lem:ma_form}. Now, using the same sequence $\{\epsilon_t\}$, 
    \begin{equation}
        \mathbf{X}_t(D_q) - \mathbf{X}_t(D) = \sum_{l=0}^\infty \{a_l(D_q) - a_l(D)\} \epsilon_{t-l},
    \end{equation}
    which implies that 
    \begin{equation}
        \E \|\mathbf{X}_t(D_q) - \mathbf{X}_t(D)\| \leq \E \| \epsilon_{t}\| \sum_{l=0}^\infty \|a_l(D_q) - a_l(D)\|.
    \end{equation}
    For every fixed $l$, $a_l(D_q) \rightarrow a_l(D)$ implies that $\|a_l(D_q) - a_l(D)\| \rightarrow 0$. $\|a_l(D_q)\| < C(1+c)^{-l}$ implies that $\|a_l(D_q) - a_l(D)\| \leq 2C(1+c)^{-l}$ uniformly in $l$. Since $\sum_{l=0}^\infty2C(1+c)^{-l}$ is clearly finite, we have by the dominated convergence theorem that
    \begin{equation}
        \lim_{q \rightarrow \infty} \sum_{l=0}^\infty \|a_l(D_q) - a_l(D)\| =  \sum_{l=0}^\infty \lim_{q \rightarrow \infty} \|a_l(D_q) - a_l(D)\| = 0.
    \end{equation}
    Hence,
    \begin{equation}
        \E \|\mathbf{X}_t(D_q) - \mathbf{X}_t(D)\| \rightarrow 0.
    \end{equation}
    Thus, for every $n$,
    \begin{align}
        W_1\{F_n(D_q),F_n(D)\} &= \int_{-\infty}^{\infty} |F_n(D_q)(x) - F_n(D)(x)| \text{d}x\\
        &= \int_{-\infty}^{\infty} |\E (\mathbf{1}_{X_{t,n}(D_q) \leq x} - \mathbf{1}_{X_{t,n}(D) \leq x})| \text{d}x\\
        &\leq \int_{-\infty}^{\infty} \E | \mathbf{1}_{X_{t,n}(D_q) \leq x} - \mathbf{1}_{X_{t,n}(D) \leq x}| \text{d}x\\
        &= \E \int_{-\infty}^{\infty}  | \mathbf{1}_{X_{t,n}(D_q) \leq x} - \mathbf{1}_{X_{t,n}(D) \leq x}| \text{d}x\\
        &= \E | X_{t,n}(D_q) - X_{t,n}(D) | \rightarrow 0.
    \end{align}
    Using the triangle inequality, we have
    \begin{equation}
        W_1\{F_n(D_q),F_m(D_q)\} \rightarrow W_1\{F_n(D),F_m(D)\}.
    \end{equation}
    Thus, the distance mapping
    \begin{equation}
        D \mapsto \{1+ \frac{W_1\{F_n(D),F_m(D)\}}{\max_{(k,l) \in \mathcal{P}} W_1\{F_k(D),F_l(D)\}}: (n,m) \in \mathcal{P} \}
    \end{equation}
    is continuous in $D$ and clearly lands in $\mathcal{D}$. Since $\mathcal{D}$ is non-empty, convex and compact, by Brouwer's fixed point theorem, \eqref{eq:dist_from_ws} has at least one fixed point, thus proving the lemma.
\end{proof}

Now, fix such a fixed point $D$ and the corresponding stationary and ergodic process. For simplicity, denote its marginal distribution for node $n$ as $F_n$. From observations $\{ \mathbf{X}_t\}_{t=1}^T$, we empirically estimate $F_n$ as 
\begin{equation}
    \hat{F}_{n,T}(x) = \frac{1}{T}\sum_{t=1}^T \mathbf{1}_{X_{t,n} \leq x}
\end{equation}
and define, for $(n,m) \in \mathcal{P}$,
\begin{equation}
    \hat{d}_{n,m} = 1+ \frac{W_1\{\hat F_{n,T},\hat F_{m,T}\}}{\max_{(k,l) \in \mathcal{P}} W_1\{\hat F_{k,T},\hat F_{l,T}\}}.
\end{equation}
In practice, $W_1\{\hat F_{n,T},\hat F_{m,T}\}$ can be simplified as
\begin{equation}
    W_1\{\hat F_{n,T},\hat F_{m,T}\} = \frac{1}{T}\sum_{t=1}^T |X_{(t),n}-X_{(t),m}|,
\end{equation}
where $X_{(t),n}$ is the $t$-th order statistic of $\{X_{t,n}\}_{t=1}^T$.

The next lemma establishes the consistency of the estimator $\hat{d}_{n,m}$.

\begin{lemma}\label{lem:est_d_consistency}
    For fixed $N$, $\max_{(n,m) \in \mathcal{P}} |\hat{d}_{n,m} - d_{n,m}| \xrightarrow{a.s.}0 $
\end{lemma}

\begin{proof}
    Because $\mathbf{X}_t$ is stationary ergodic, by the ergodic theorem, we have 
    \begin{equation}
        \hat{F}_{n,T}(x) \xrightarrow{a.s.} F_{n}(x)
    \end{equation}
    for every continuity point $x$ of $F_n$ (no jump at $x$). In other words, $\prob (E_x) = 1$ for every continuity point $x$, where $E_x = \{\omega: \hat{F}_{n,T}(x) \xrightarrow{} F_{n}(x) \}$. Since a CDF can have only countably many discontinuity points, we can always find a countable set of continuity points, $\{x_j\}_{j=1}^\infty$, dense in $\mathbb{R}$. Then, $P (\bigcap_{j=1}^\infty E_{x_j}) = 1$. One way to achieve this is to consider the set of all open intervals $(a,b)$ in $\mathbb{R}$ with rational $a,b$. Then pick a continuity point in each interval. This is possible since discontinuity points are countable, while points between every $(a,b)$ are uncountable. This leads to a countable set of continuity points dense in $\mathbb{R}$. On $\bigcap_{j=1}^\infty E_{x_j}$, for every continuity point $x$, there always exist two sequences in $\{x_j\}$, $x_{j_k} \uparrow x$ and $x_{j_l} \downarrow x$. Moreover, $x$ being a continuity point leads to $F_n(x_{j_k}) \rightarrow F_n(x)$ and $F_n(x_{j_l}) \rightarrow F_n(x)$. For each $j_k$ and $j_l$, $\hat{F}_{n,T}(x_{j_k}) \rightarrow F_{n}(x_{j_k})$ and $\hat{F}_{n,T}(x_{j_l}) \rightarrow F_{n}(x_{j_l})$. Since CDF is non-decreasing, we have as $T \rightarrow \infty$,
    \begin{equation}
        F_{n}(x_{j_k}) \leq \liminf_{T \rightarrow \infty} \hat{F}_{n,T}(x) \leq \limsup_{T \rightarrow \infty} \hat{F}_{n,T}(x) \leq F_{n}(x_{j_l}).
    \end{equation}
    The two sides both converge to $F_n(x)$ as $k \rightarrow \infty$ and $l \rightarrow \infty$. Thus, we have $\hat{F}_{n,T}(x) \xrightarrow{} F_{n}(x)$ by squeezing. Let $\hat{Z}_{n,T}$ be a random variable with CDF $\hat{F}_{n,T}$. Then, on the event $\bigcap_{j=1}^\infty E_{x_j}$ with probability 1, the convergence in continuity points in one-dimensional space is equivalent to $\hat{Z}_{n,T} \xrightarrow{d} X_{t,n}$. 
    
    Next, by finite first absolute moment, $\E |X_{t,n}| < \infty$, and the ergodic theorem, we have convergence in the first absolute moment,
    \begin{equation}
        \E_{\hat{F}_{n,T}}|\hat{Z}_{n,T}| = \frac{1}{T}\sum_{t=1}^T |X_{t,n}| \rightarrow \E_{F_n} |X_{t,n}|
    \end{equation}
    on an event $E$ with probability 1. Thus, using the fact that convergence in distribution and convergence in the first absolute moment lead to convergence in $W_1$ \citep[Section 1.2]{Panaretos2019}, we have
    \begin{equation}
        W_1(\hat{F}_{n,T}, F_n) \xrightarrow{a.s.} 0,
    \end{equation}
    using the intersection of events $\bigcap_{j=1}^\infty E_{x_j}$ and $E$.

    By triangle inequality,
    \begin{equation}
        |W_1(\hat{F}_{n,T}, \hat{F}_{m,T}) - W_1(F_{n}, F_{m})| \leq W_1(\hat{F}_{n,T}, F_n) + W_1(\hat{F}_{m,T}, F_m) \xrightarrow{a.s.} 0.
    \end{equation}
    Then, for each $N$, $\max_{(k,l) \in \mathcal{P}} W_1\{\hat F_{k,T},\hat F_{l,T}\} \xrightarrow{a.s.} \max_{(k,l) \in \mathcal{P}} W_1\{F_{k},F_{l}\}$. As we have shown that $\max_{(k,l) \in \mathcal{P}} W_1\{F_{k},F_{l}\}$ is strictly positive, the denominator $\max_{(k,l) \in \mathcal{P}} W_1\{\hat F_{k,T},\hat F_{l,T}\}$ is strictly positive eventually almost surely. Therefore, for each $(n,m) \in \mathcal{P}$,
    \begin{equation}
        |\hat{d}_{n,m} - d_{n,m}| \xrightarrow{a.s.}0.
    \end{equation}
    Taking the maximum for fixed $N$ proves the lemma.
\end{proof}

Let $\hat{D}$ be the estimated version of $D$, i.e., $\hat{D} = \{\hat{d}_{n,m}:(n,m) \in \mathcal{P}\}$.
Let $L_T(\theta;\hat{D})$ and $L_T(\theta;D)$ be the loss with respect to estimated distances and true distances respectively. Let $G_T(\hat{D}) = \argmin_{\theta \in \mathbb{R}^{P+1}} L_T(\theta;\hat{D})$ be the set of all global minimisers of $L_T(\theta;\hat{D})$. Any $\hat{\theta}_T(\hat{D}) \in G_T(\hat{D})$ is a global minimiser of $L_T(\theta;\hat{D})$. We are now ready to show that the estimator $\hat{\theta}_T(\hat{D})$ is also eventually unique as $T \rightarrow \infty$ and consistent.

\begin{proposition}\label{prop:wd_consistency}
    Under \sref{as:stationary}, \sref{as:sigma_epsilon}, \sref{as:dist} and \cref{as:ws_max_neq0}, $|G_T(\hat{D})|=1$ for sufficiently large $T$ almost surely. Moreover, any measurable sequence of $G_T(\hat{D})=\{\hat{\theta}_T(\hat{D})\}$ satisfies $\hat{\theta}_T(\hat{D}) \overset{a.s.}{\to} \theta^0$.
\end{proposition}

\begin{proof}
    The proof follows the same strategy as that of \sref{prop:consistency}. We will omit the arguments already stated in the previous proof and focus on genuinely new logics. We still use $l(\theta)$ to denote the population loss with respect to the true distances. Recall that in the proof of \sref{prop:consistency}, we have shown that $\theta^0$ is the unique minimiser of $l(\theta)$. 

    Using \cref{lem:est_d_consistency}, the loss expansion in \eqref{eq:loss_expansion} and the ergodic theorem, and aggregating the terms back, we have
    \begin{equation}
        L_{T}(\theta^0;\hat{D}) \xrightarrow{a.s.} l(\theta^0).
    \end{equation}
    As the weights from estimated distances are still row-normalised and non-negative, the arguments leading to $\|(\hat\alpha_T(\hat{D}),\hat\beta_T(\hat{D}))\|_2$ being eventually bounded by some finite constant $M$ almost surely for the $(\alpha,\beta)$ component of any global minimiser remain unchanged. 
    
    For the $\lambda$ direction, we regard $\lambda$ as taking values on the extended real line $[-\infty,\infty]$, with the weights defined by their finite limits as $\lambda \rightarrow \pm \infty$. Suppose that on an event $E$ with non-zero probability there exists a subsequence $\hat\lambda_{T_k} \rightarrow \infty$. Since the weights from $\hat{D}$ are row-normalised and non-negative, and $\|(\hat\alpha_{T_k},\hat \beta_{T_k})\|_2 \leq M$ eventually almost surely on $E$, we have $\tilde{A}(\hat\theta_{T_k};\hat{D})$ eventually bounded almost surely on $E$. Then, almost surely on $E$, there exists a further sub-subsequence $(\hat\alpha_{T_{k_l}},\hat \beta_{T_{k_l}}) \rightarrow (\alpha^*, \beta^*)$ and $\tilde{A}(\hat\theta_{T_{k_l}};\hat{D}) \rightarrow \tilde{A}^*$ for some finite $(\alpha^*, \beta^*)$ and $\tilde{A}^*$. 
    
    Let $j,r,n,m_1,m_2$ be those from \sref{as:dist} such that $\beta_{j,r}^0 \neq 0$ and $d_{n,m_1} \neq d_{n,m_2}$. Without loss of generality, assume that $d_{n,m_1} < d_{n,m_2}$. By \cref{lem:est_d_consistency}, almost surely on $E$,
    \begin{equation}
        \frac{\hat d_{n,m_1}}{\hat d_{n,m_2}} \rightarrow \frac{d_{n,m_1}}{d_{n,m_2}} < 1.
    \end{equation}
    Let $q=\frac{d_{n,m_1}}{2d_{n,m_2}} + \frac{1}{2} < 1$. For all $R>0$, we can also find $k_0$ large enough, such that, if $k > k_0$, $\hat{\lambda}_{T_k}$ is either $\hat{\lambda}_{T_k} = +\infty$ or $R < \hat{\lambda}_{T_k} < \infty$, and $\frac{\hat d_{n,m_1}}{\hat d_{n,m_2}} < q$. If $\hat\lambda_{T_k} = + \infty$, then almost surely on $E$, by definition, $w_{n,m_2}(\hat\lambda_{T_k};\hat{D}) = 0$. Otherwise, almost surely on $E$,
    \begin{equation}
        \frac{w_{n,m_2}(\hat\lambda_{T_k};\hat{D})}{w_{n,m_1}(\hat\lambda_{T_k};\hat{D})} = \left( \frac{\hat d_{n,m_1}}{\hat d_{n,m_2}}\right)^{\hat\lambda_{T_k}} <  q^R.
    \end{equation}
    Since $w_{n,m_1}(\hat\lambda_{T_k};\hat{D})$ is at most 1, $w_{n,m_2}(\hat\lambda_{T_k};\hat{D}) <  q^R$ almost surely on $E$. We can then conclude that $w_{n,m_2}(\hat\lambda_{T_k};\hat{D}) \rightarrow 0$ almost surely on $E$. As $\hat{\beta}_{j,r}{}_{T_k}(\hat{D})$ is eventually bounded almost surely on $E$, 
    \begin{equation}
        \hat{\beta}_{j,r}{}_{T_k}(\hat{D}) w_{n,m_2}(\hat\lambda_{T_{k}};\hat{D}) \rightarrow 0
    \end{equation}
    almost surely on $E$. However, at the true parameters,
    \begin{equation}
        \beta_{j,r}^0 w_{n,m_2}(\lambda^0;D) \neq 0.
    \end{equation}
    Thus, the sub-subsequence limit $\tilde{A}^* \neq \tilde{A}(\theta^0;D)$.  

    Now, using the loss expansion in \eqref{eq:loss_expansion} and the ergodic theorem, and aggregating the terms back into squared norm, we have
    \begin{equation}
        L_{T_{k_l}}(\hat{\theta}_{T_{k_l}};\hat{D}) \rightarrow \E \|\mathbf{X}_t - \tilde{A}^* \bar{\mathbf{X}}_{t-1} \|_2^2
    \end{equation}
    almost surely on $E$. By \sref{lem:cov_finite}, we have $\cov(\bar{\mathbf{X}}_{t}) \succ 0$. Then if $\tilde{A}^* \neq \tilde{A}(\theta^0;D)$ almost surely on $E$, we get 
    \begin{equation}
        \E \|\mathbf{X}_t - \tilde{A}^* \bar{\mathbf{X}}_{t-1} \|_2^2 - l(\theta^0) = \E \| (\tilde{A}(\theta^0;D) - \tilde{A}^*) \bar{\mathbf{X}}_{t-1} \|_2^2 > 0
    \end{equation}
    almost surely on $E$, where $\tilde{A}^*$ inside the expectations is held fixed. We have shown above that
    \begin{equation}
        L_{T_{k_l}}(\theta^0;\hat{D}) \rightarrow l(\theta^0).
    \end{equation}
    almost surely on $E$. However, by definition, $L_T(\hat\theta_T;\hat{D}) \leq L_{T}(\theta^0;\hat{D})$ for any $T$ and $\omega$, a contradiction. Hence, $E$ has probability 0. The case for $\hat\lambda_{T_k} \rightarrow -\infty$ on an event with non-zero probability is analogous. Then, no subsequence of $|\hat{\lambda}_T|$ can diverge to $\infty$ almost surely. In other words, $\hat{\lambda}_T$ remains eventually bounded almost surely.

    Using the same compactification argument as in the proof of \sref{prop:consistency} with $D$ replaced by $\hat{D}$, we can show that all global minimisers of $L_T(\theta;\hat{D})$ eventually lie in a compact set $B= \{\theta: \|(\alpha,\beta)\| \leq M, \lambda \in [-\infty,\infty]\}$ almost surely. We have now shown that $L_T(\theta;\hat{D})$ attains its minimum on $B$ and not at $\lambda = \pm \infty$ eventually almost surely. Then, $G_T(\hat{D})$ is eventually non-empty almost surely. Moreover, we can also construct a random compact set $\mathcal{B} = \{\theta: \|(\alpha,\beta)\| \leq M, |\lambda| \leq M_\lambda\}$ where $M_\lambda$ is a random variable depending on the path, such that the global minimisers eventually lie in $\mathcal{B}$ almost surely. This is to ensure the uniform continuity leading to the convergence in \eqref{eq:prop:wd_consistency:w_conv}. On this compact set $\mathcal{B}$, using the triangle inequality,
    \begin{equation}
        \sup_{\theta \in \mathcal{B}} |L_T(\theta;\hat{D}) - l(\theta)| \leq \sup_{\theta \in \mathcal{B}}|L_T(\theta;\hat{D}) - L_T(\theta;D)| + \sup_{\theta \in \mathcal{B}}|L_T(\theta;D) - l(\theta)|.
    \end{equation}
    For the second term, we have shown in \cref{eq:prop:consistency:uni_conv_B} that $\sup_{\theta \in B}|L_T(\theta;D) - l(\theta)| \xrightarrow{a.s.} 0$. For the first term, by \cref{lem:est_d_consistency} and the fact that the weights are uniformly continuous jointly in $\lambda$ and $D$ over the compact set $[-M_\lambda, M_\lambda] \times \mathcal{D}$, 
    \begin{equation}\label{eq:prop:wd_consistency:w_conv}
        \sup_{|\lambda| \leq M_{\lambda}} \| W^{(r)}(\lambda;\hat{D}) - W^{(r)}(\lambda;D)\|_F \xrightarrow{a.s.} 0.
    \end{equation}
    Expanding $A_j(\theta;\hat{D})$ gives
    \begin{equation}
        \sup_{\theta \in \mathcal{B}} \| \tilde{A}(\theta;\hat{D}) - \tilde{A}(\theta;D)\|_F \xrightarrow{a.s.} 0.
    \end{equation}
    Using the loss expansion, the ergodic theorem and \sref{lem:cov_finite}, we thus obtain that
    \begin{equation}
        \sup_{\theta \in \mathcal{B}}|L_T(\theta;\hat{D}) - L_T(\theta;D)| \xrightarrow{a.s.} 0.
    \end{equation}
    Following the exact same argument from \eqref{eq:prop:consistency:population_squeeze} to \eqref{eq:prop:consistency:consistency}, we have $\hat{\theta}_T(\hat{D}) \xrightarrow{a.s.} \theta^0$. 

    The eventual uniqueness, $|G_T(\hat{D})| = 1$ follows from the same argument as in the proof of \sref{prop:consistency}. The only new ingredient here is that, for the set $B_0'$ defined before \eqref{eq:prop:consistency:after_B0prime},
    \begin{equation}
        \sup_{\theta \in B_0'}\left\| \frac{\partial^2 L_T(\theta;\hat{D})}{\partial\theta \partial \theta'} - \frac{\partial^2 L_T(\theta;D)}{\partial\theta \partial \theta'} \right\| \xrightarrow{a.s.}0,
    \end{equation}
    which follows from \cref{lem:est_d_consistency} and the same argument from the proof of \sref{prop:consistency} using the ergodic theorem and \cref{lem:cross_norm_finite}.
\end{proof}
}
\bibliographystyle{comnet}
\bibliography{Ref}
%






